\documentclass[3p]{elsarticle}

\journal{Information Sciences}

\usepackage{hyperref}
\hypersetup{colorlinks,urlcolor=blue}
\usepackage{amsmath}
\usepackage{amssymb}
\usepackage{multirow}
\usepackage[noend]{algpseudocode}
\usepackage[Algorithm]{algorithm}
\usepackage{graphicx} 
\usepackage{caption,subcaption}
\usepackage{enumitem}

\begin{document}
\date{}
\begin{frontmatter}



\title{Mining Weighted Sequential Patterns in Incremental Uncertain Databases}

\author[firstaddress]{Kashob Kumar Roy}
\ead{kashobroy@gmail.com}
\author[firstaddress]{Md Hasibul Haque Moon}
\ead{hasibulhq.moon@gmail.com}
\author[firstaddress]{Md Mahmudur Rahman}
\ead{mahmudur@cse.du.ac.bd, mahmudur@du.ac.bd}
\author[firstaddress]{Chowdhury Farhan Ahmed\corref{mycorrespondingauthor}}
\ead{farhan@du.ac.bd, farhan@cse.du.ac.bd}
\author[secondaddress]{Carson Kai-Sang Leung}
\ead{kleung@cs.umanitoba.ca}
\cortext[mycorrespondingauthor]{Corresponding author at: Department of Computer Science and Engineering, University of Dhaka, Dhaka-1000, Bangladesh.}


\address[firstaddress]{Department of Computer Science and Engineering, University of Dhaka, Bangladesh}
\address[secondaddress]{Department of Computer Science, University of Manitoba, Canada}

\begin{abstract}
Due to the rapid development of science and technology, the importance of imprecise, noisy, and uncertain data is increasing at an exponential rate. Thus, mining patterns in uncertain databases have drawn the attention of researchers. Moreover, frequent sequences of items from these databases need to be discovered for meaningful knowledge with great impact. In many real cases, weights of items and patterns are introduced to find interesting sequences as a measure of importance. Hence, a constraint of weight needs to be handled while mining sequential patterns. Besides, due to the dynamic nature of databases, mining important information has become more challenging. Instead of mining patterns from scratch after each increment, incremental mining algorithms utilize previously mined information to update the result immediately. Several algorithms exist to mine frequent patterns and weighted sequences from incremental databases. However, these algorithms are confined to mine the precise ones. Therefore, we have developed an algorithm to mine frequent sequences in an uncertain database in this work.
Furthermore, we have proposed two new techniques for mining when the database is incremental. Extensive experiments have been conducted for performance evaluation. The analysis showed the efficiency of our proposed framework.
\end{abstract}


\end{frontmatter}


\section{Introduction} 
The way toward mining concealed information from the massive extent of data is known as data mining.
This information can be frequent patterns, irregular patterns, correlated patterns, association rules among events, etc. 
Researchers have concentrated mostly on mining frequent patterns, which can be a set of items or a sequence of itemsets or any substructure that frequently occurs in a given database and a minimum support threshold. A plethora of mining algorithms have shown their superior performance in different applications of data.

\textit{Apriori}~\cite{agrawal1994fast_apriori} is the first algorithm based on candidate generation and testing paradigm in the field of data mining which finds frequent itemsets and association rules among them. A well-known anti-monotone (downward closure) property was introduced in~\cite{agrawal1994fast_apriori} to reduce the search space for mining frequent itemsets. Due to the huge memory and time requirements of \textit{Apriori}~\cite{agrawal1994fast_apriori} like algorithms, pattern growth-based approaches~\cite{han2004mining} have drawn great attention for this task. Consequently, different extensions of frequent itemset mining problems were introduced i.e., handling weight constraint \cite{ishita2018efficient_WINCSPAN,yun2008new_WSPAN} to mine interesting patterns,
and utility constraint \cite{gan2018survey,gan2017mining,lin2020high,lin2020incrementally} to mine high-utility patterns.

In many applications, the order or sequence of the items in databases is essential and must be maintained while mining.
Thus, the problem of sequential pattern mining was introduced first in \cite{srikant1996mining_GSP}.~\textit{GSP} \cite{srikant1996mining_GSP} is a generalized solution based on a candidate generation and testing approach. It generates a lot of false-positive sequences and requires multiple scans of the database to obtain true-positive ones. Afterward,
\textit{PrefixSpan}~\cite{pei2004mining_prefixSpan} was proposed to overcome this limitation in the mining of sequential patterns. 
It is based on a pattern-growth approach and follows a depth-first search strategy.
Approaches like \textit{PrefixSpan} employ different efficient pruning techniques to reduce the search space while growing larger patterns. Rizvee et al.~\cite{rizvee2020tree} proposed a tree-based approach to mine sequential patterns and demonstrated their efficient performance.

However, uncertainty is inherent in real-world data due to the noise and inaccuracy of many data sources. With the increasing use of uncertain data in modern technologies such as sensor data, environment surveillance, medical diagnosis, security, and manufacturing systems, uncertain databases are growing larger.
Uncertainty of an item in a database makes both the itemset mining and sequential pattern mining difficult.
Several algorithms developed in \cite{aggarwal2009frequent,lee2015uncertainty,leung2012fast,leung2013puf} are able to mine frequent itemsets in uncertain databases.
Uncertain itemset mining with weight constraints is addressed in \cite{ahmed2016mining,leung2014reducing,lin2016weighted}.
Following \textit{PrefixSpan}~\cite{pei2004mining_prefixSpan}, there are several algorithms proposed to mine sequential patterns in uncertain databases~\cite{muzammal2011mining_uSeq1st, rahman2019mining_uWSeq, zhao2013mining_uncertainSeq}. 

Moreover, many real-life applications reflect that all the frequent sequences are not equally important. 
Hence, different weights are assigned to different items of a database corresponding to the significance of an item.
As a result, each sequence gets a weight, and thus interesting sequential patterns can be mined against different weight and support thresholds.
Sequential pattern mining with weight constraints has been explored in \cite{yun2008new_WSPAN}. To maintain the anti-monotone property, the upper bound of weight is commonly set as the maximal weight of all items in the database~\cite{ishita2018efficient_WINCSPAN,yun2008new_WSPAN}.
\textit{uWSequence}~\cite{rahman2019mining_uWSeq} can handle weight constraints in mining uncertain sequence databases. It follows the expected support based mining process. To incorporate anti-monotone property in the mining process, it has proposed an upper-bound measure of expected support. Besides, it uses a weighting filter separately to handle the weight constraint in mining sequences. Thus,  it can efficiently mine only those sequences that have high frequencies with  high weights. Generally, in weighted pattern mining, sequences that have low frequencies with high weights or high frequencies with low weights are also important in several real-life applications. Again, it is necessary to design sophisticated pruning upper bounds to mine weighted patterns efficiently while maintaining the anti-monotone property as well as limiting false positive pattern generation. 

Therefore, we propose multiple novel pruning upper bounds that are theoretically tightened instead of respective upper bounds already introduced in the literature.  Besides, efficient maintenance of candidate patterns is required to develop a faster method for computing expected supports of patterns. Hence, we utilize a hierarchical index structure to maintain candidate patterns in a space-efficient way that leads to a way faster support computation method than state-of-art methods. 

Furthermore, most real-life databases are dynamic in nature. In recent years, a large amount of research has been conducted in incremental mining, i.e.,~\cite{ahmed2012single,gan2018survey,lee2018single,lin2020incrementally,nam2020efficient,wu2020incrementally} to frequent itemsets with/without different constraints such as weight, utility etc.,~\cite{ cheng2004incspan_INCSPAN,ishita2018efficient_WINCSPAN,lin2015incrementally} to find the updated set of sequential patterns, or weighted sequential patterns or high utility sequential patterns from dynamic databases. Note that all above incremental algorithms are confined into precise databases. Nonetheless, all of these sequential pattern mining algorithms for uncertain databases are not efficient to handle the dynamic nature of data. 
Because, running a \textit{batch algorithm} like \textit{PrefixSpan} or \textit{uWSequence} from scratch for each increment requires a huge amount of time and memory.
Thus the lack of efficient methods on incremental mining of weighted sequential patterns in uncertain databases and its importance has stimulated us to explore this field.  

\subsection{Motivation}
\label{subsec:motivation}
The use of uncertain data in the modern world is increasing day by day. In most cases, the database is not static. New increments are added to the database gradually; hence the set of frequent sequences may change. After each increment, running existing algorithms from scratch, which can mine frequent sequences in static uncertain databases, is very expensive in terms of time and memory. Therefore it has become inevitable to design an efficient technique that can maintain and update frequent sequences when the database grows incrementally. A few scenarios described below reflect the importance of finding frequent weighted sequences from the incremental uncertain database.

\subsubsection{Example One}
Frequent pattern mining is being widely applied in medical data.  It is beneficial to discover hidden knowledge or extract important patterns from massive data on patient symptoms. Health workers/organizations can use these discovered patterns to give patients proper treatment at the right time or observe disease behavior during an outbreak. For instance, the novel coronavirus SARS-CoV-2 causes the coronavirus disease of 2019 (COVID-19), which was first seen in China in late 2019. Researchers from the whole world are working on the COVID-19 outbreak as the world is expecting to face a huge economic recession with losing a lot of people.  If we see the nature of this disease's symptoms, then it is clear that the symptoms have the nature of sequential occurrences. Again, there is uncertainty in the patient data inherently. Because the same physiological index
corresponds to different symptom association probabilities for patients due to their different
physical conditions. 

Let us consider a symptom sequence of a COVID-19 patient e.g., $\{(fever: 0.4), (cough: 0.5), (\textit{\text{sore throat}}: 0.7, breathing\ problem: 0.9)\}$ where each real value denotes the association probability for the corresponding symptom. Furthermore, all symptoms may not have the same significance to diagnose a disease. Some symptoms might be severe, but others might not be, e.g., shortness of breath, chest pain, etc., are severe symptoms for COVID-19; in contrast, dry cough, tiredness, etc., are mild symptoms. Consequently, the weights of symptoms should play a crucial role in mining important symptom sequences from patient data.  
So, considering the nature of the symptoms, it can be said that finding weighted frequent sequences of the symptoms of the corona patients can be helpful to predict whether someone is infected or even the current stage of infection. Besides, it will be possible to take proper treatment for an infected patient by predicting the next stage of infection based on discovered frequent patterns of symptoms. For example, at a particular stage of infection, a patient may have mild pain. From the database of symptoms, it has been seen that a patient having mild pain would have severe breathing problems afterward. So it will be helpful to take proper precautions for the patient.

Moreover, the spreading of the virus is so rapid that the amount of patient data is growing every moment; its nature is changing over time. Even the nature of COVID-19 is also varying from place to place. Thus, due to rapidly growing patient data, non-incremental mining algorithms are not efficient enough to find frequent sequences within a short time; it requires a massive amount of time to run the mining algorithm on the whole database from scratch every time we get new data. Consequently, an efficient algorithm to find weighted frequent sequences from incremental databases is very much needed. Therefore, incrementally mined patterns from massive patient data flow are essential to determine how the symptoms change with time or vary from one region/country to another.

\subsubsection{Example Two} 
    Mining social network behavioral patterns can be another example of uncertain data. These patterns can be discovered from users' activities in social networks. If we observe a user's activities over a while, we can estimate how similar a user is with a student, a photographer, and so on. In other words, we can assign a probability associated to each category, e.g., $\{(student: 0.9), (photographer: 0.7, cyclist: 0.3), (tourist: 0.7)\}$ where it indicates that the probability of being a student for a particular user is estimated after analyzing his activities over a period of time in social networks, as a photographer or a cyclist after analyzing his activities over the next period, and so on. This period could be a couple of minutes or hours or even days or weeks. Again, users are using their social networks at every moment while new users are joining social networks. As a result, a huge amount of data is being produced at every moment, which means that the database of users' activities is growing rapidly. Therefore, an efficient incremental algorithm is required to mine user behavioral patterns on the fly. Nowadays, these discovered patterns are beneficial in a wide range of applications. For instance, in this era of social network-based digital marketing, the study of consumer activities is most important for marketers to understand consumers' behavioral patterns for their customized advertisements. If we discover the patterns on the fly, it is possible to customize advertisements on a user's newsfeed based on his latest activities over social networks. Besides,  different patterns discovered from social network data are also crucial in influencer marketing, trend analysis, social event detection, social spam analysis, etc.
    
    Furthermore, anomaly or fraud analysis over users' activities requires observing the behavioral patterns over a period of time. The behavioral patterns of a fraud user show a deviation from that of regular users over a while. Thus, to detect fraud users from its social network activities, it needs to mine the behavioral patterns incrementally for a certain period of time. Therefore, an efficient incremental mining algorithm can play an essential role in discovering anomalies in the patterns on the fly because it is essential to detect fraud users or other anomalies, e.g., rumors, within the shortest possible time.

\subsubsection{Example Three}
    The traffic management system is being automated to identify the behavior of vehicles and drivers. Several methods like Automatic Number Plate Recognition, Speed Recognition, Vehicle Type Recognition, and Trajectory Analysis produce uncertain data. We can mine different patterns like "Going to road A first and then road B is a frequent behavior of 10\% of the cars". These patterns are generally sequential. In an intelligent transportation system (ITS), different roads/junctions may carry different significance. Again, all data collected through sensor/GPS are associated with some certainty values.
    Consequently, it is to be said that weighted sequential patterns can play an important role in traffic automation, such as planning and monitoring traffic routes. However, different patterns can be seen in different parts of a day, i.e., patterns at the office/school opening time are quite the opposite of office/school closing time in a day. Again, weekday patterns are different from weekend patterns. Patterns even vary between different periods in a year. Note that patterns may change over different periods of a day, a week, a month, or even a year. Therefore, these changes in daily patterns, weekly patterns, or yearly patterns can be mined from incrementally growing traffic databases. Patterns mined from different database increments are helpful to draw the pattern trends across different periods of a day or even a year. These pattern trends are required for analyzing the seasonal behavior of traffic in ITS. 

A few other applications are TNFR (Tumor Necrosis Factor Receptor) disease analysis, DNA sequencing (micro-level information with uncertainty), mining in crime data, weather data, fashion trend \cite{rahman2019mining_uWSeq}, and vehicle recognition data \cite{muzammal2011mining_uSeq1st}; WSN (Wireless Sensor Network) data monitoring \cite{zhao2013mining_uncertainSeq}; 
social network behavior analysis \cite{ahmed2016mining} 
etc., where a benefit of mining weighted frequent sequential patterns is to discover more meaningful hidden knowledge.
As the existing algorithms are not efficient to mine weighted sequential patterns in incremental uncertain databases, finding efficient techniques has become an inevitable research issue.

\subsection{Contributions}
In this work, we propose a new framework to deal with weight constraints in mining sequential patterns in uncertain databases to address the issues mentioned above. In this framework, we introduce a new concept of \textit{weighted expected support} of a sequential pattern. An efficient algorithm \textit{FUWS} is developed based upon this framework to find weighted sequential patterns from any static uncertain databases. Furthermore, we propose two techniques to find the updated set of weighted sequential patterns when the uncertain database is of dynamic nature. To the best of our knowledge, this work is the first to do mining weighted sequential patterns in incremental uncertain databases. 
Our key contributions of this work are as follows:
\begin{enumerate}[]
    \item  An efficient algorithm, \textit{FUWS}, to mine weighted sequential patterns in uncertain databases.
    \item Two new techniques, \textit{uWSInc} and \textit{uWSInc+}, for mining weighted sequential  patterns in incremental database of uncertain sequences.
    \item A new hierarchical index structure, \textit{USeq-Trie}, for maintaining weighted uncertain sequences.
    \item Two upper bound measures, $expSup^{cap}$ and $wgt^{cap}$, are for expected support and weight of a sequence, respectively.
    \item A pruning measure, $wExpSup^{cap}$, to reduce the search space of mining patterns.
    \item An extensive experimental study to validate the efficiency and effectiveness of our approach and its supremacy with respect to the existing methods.
\end{enumerate}


Although there has been a good amount of work in the broad domain of pattern mining, it is clear that mining weighted sequential patterns in uncertain databases is not explored well despite its rising importance. This paper presents solutions to this research issue and provides extensive experimental results to validate our claims.

The rest of the paper is organized in the following sections: background study and discussion of related works in \hyperref[sec:related]{Section 2}, 
our proposed solutions with a proper simulation in \hyperref[sec:soln]{Section 3}, 
analysis of experimental results in \hyperref[sec:results]{Section 4} 
and finally, conclusions in \hyperref[sec:concl]{Section 5}.

\section{Literature Review}  
\label{sec:related}

As with the development of modern technologies, the usages of data are being intensified day by day. A lot of varieties have emerged with a variety of information stored and the knowledge required for different scenarios. Hence, various algorithms have already been developed for mining heterogeneous information. 
A review of the literature has been described below.

\subsection{Sequential Pattern Mining}
A \textit{sequence database} is a list of data tuples where each tuple is an ordered set of itemsets/events.
Unlike itemset mining, it contains order information for events.
Thus, sequential patterns mined from a sequence database can significantly impact many applications~\cite{fournier2017survey}. In recent years, a plethora of research has been conducted on sequential pattern mining~\cite{fournier2017survey,gan2019survey,truong2019survey}. 
The problem of \textit{Sequential Pattern Mining (SPM)} was first discussed in \cite{srikant1996mining_GSP}. Authors proposed a generalized solution named \textit{GSP}~\cite{srikant1996mining_GSP} which is inspired by the itemset mining algorithm \textit{Apriori}~\cite{agrawal1994fast_apriori}.
It has the problem of infrequent pattern generation, and it needs a massive amount of running time.
It also needs extensive memory to store all the $k$-sequences (i.e., sequences of length $k$ to use in the generation of $(k+1)$-sequences.

Afterward, following \textit{FP-Growth}~\cite{han2004mining} for itemset mining, pattern-growth based approach for sequential pattern mining, \textit{PrefixSpan} was introduced in \cite{pei2004mining_prefixSpan}. It uses the divide-and-conquer technique to mine frequent sequences from a precise database. \textit{PrefixSpan} starts mining from frequent sequences of length 1. Then it projects databases into smaller parts by taking the frequent sequences as a prefix. Afterward, it expands longer patterns and further projects into smaller databases recursively.
Later, many improvements have been found in different specific applications by designing efficient pruning techniques in \textit{PrefixSpan} to reduce the search space. A recent work \cite{rizvee2020tree} has proposed a compact and efficient tree-structure, \textit{SP-Tree}, to store the whole database. They have utilized the idea of co-existing item table to facilitate the mining of {sequential patterns} and proposed an algorithm named \textit{Tree-Miner} which holds the \textit{build once mine many} property. Still, \textit{Tree-Miner} algorithm needs a huge amount of memory space to store the whole database which makes it incompetent for incremental mining. 


\subsection{Weighted Sequential Mining}
All frequent items, as well as all sequences, are not equally important.
Such examples have been discussed in Section~\ref{subsec:motivation}.
Different weights are assigned to different items to reflect the significance of various patterns. The weight of an itemset or sequence can be calculated using the item weights.
To mine interesting patterns, the concept of weighted support is introduced
in \cite{yun2007efficient_WIP,yun2008new_WSPAN}. They defined the \textit{weighted support} of a sequence as the resultant value of multiplying its support count with {weight} value. However, this weighted support violates the {anti-monotone} property. 
To apply this property, an upper-bound value of weighted support is used for a sequence.
Based on this upper bound value, sequences are extended from frequent 1-sequences using \textit{PrefixSpan} like approach.
\textit{WIP}~\cite{yun2007efficient_WIP} finds weighted frequent itemsets where
\textit{WSPAN}~\cite{yun2008new_WSPAN} is  popular for mining weighted frequent sequences.
An extra scan of the database is required to find the weighted support of generated patterns and remove the false ones.

\subsection{Mining in Uncertain Databases}
\label{subsec:uncertain_mining}
Handling uncertain databases has become a major concern as their use is increasing in almost every application field.
Several pattern-growth based solutions have been proposed, such as \textit{UFP-Growth}~\cite{leung2008tree} and \textit{PUF-Growth}~\cite{leung2012fast}. Other algorithms for mining patterns in large uncertain datasets~\cite{wang2011efficient}, mining of weighted frequent uncertain itemsets~\cite{ahmed2016mining, lin2016weighted}, mining high-utility uncertain itemsets with both positive and negative utility~\cite{gan2017mining} are proved to be efficient. Yan et al.~\cite{yan2014probabilistic} explored uncertain data in 2D space.

However, in the case of sequential pattern mining and different constraints in it, it still needs researchers' attention. Muzammal et al.~\cite{muzammal2010probabilistic} formulated uncertainty in uncertain sequential pattern mining as source-level uncertainty, i.e., each tuple contains a probability value and element-level uncertainty, i.e., each event in tuples contains a probability value. Two measures of frequentness such as \textit{expected support} and \textit{probabilistic frequentness} are commonly used in frequent itemset and sequential patterns in uncertain databases.
Muzammal et al.~\cite{muzammal2011mining_uSeq1st} explored source-level uncertainty in probabilistic databases and proposed a dynamic programming algorithm to calculate the expected support and also breadth-first and depth-first methods based on candidate generation-and-test paradigm to mine patterns. \textit{U-PrefixSpan}~\cite{ zhao2013mining_uncertainSeq} follows the \textit{PrefixSpan} algorithm~\cite{pei2004mining_prefixSpan} to mine classic sequential patterns in uncertain databases under the probabilistic frequentness measure.  

In contrast, \textit{uWSequence}~\cite{rahman2019mining_uWSeq} mines sequential patterns based on expected support. It uses a weighting filter separately to mine interesting patterns. To the best of our knowledge, \textit{uWSequence}~\cite{rahman2019mining_uWSeq} is the only work to mine weighted sequential patterns from uncertain databases. It proposed
an upper-bound measure of expected support of a sequence called $expSupport^{top}$, which is used in the core mining process to project the database and grow patterns. 
A research concern is how to determine this upper bound measure to reduce the search space of the pattern-growth approach. 

Another line of research focuses on high utility-based sequential pattern mining in uncertain databases. Projection-based \textit{PMiner} algorithm~\cite{lin2019project} takes into account both average utility and uncertainty factors to efficiently mine high average-utility sequential patterns in the uncertain databases.~\cite{lin2020high} proposed a projection-based \textit{PHAUP} algorithm with three novel pruning strategies under an innovative high average-utility sequential pattern mining framework that is superior to mine high average-utility sequential patterns than \textit{PMiner}~\cite{lin2019project}. 
UHUOPM~\cite{DBLP:journals/isci/ChenCGQD21} introduced the concept of  Potential High Utility Occupancy Patterns
(PHUOPs) to incorporate three factors: support, probability, and utility occupancy while maintaining the downward closure property in the mining process.  Ahmed et al.~\cite{ahmed2020evolutionary} proposed an evolutionary model called MOEA-HEUPM to find the non-dominated high expected-utility patterns from uncertain databases without prior knowledge by utilizing a multiobjective evolutionary algorithm based on decomposition (MOEA/D). Authors~\cite{DBLP:journals/tits/SrivastavaLJLD21} presented an efficient algorithm HEUSPM to discover high expected utility sequential patterns in IoCV environments.

\subsection{Incremental Mining Algorithms}
For incremental {frequent itemset} mining, \textit{FUP}~\cite{cheung1996maintenance} is one of the early and well-known contributions which needs to rescan the database when an item is frequent in the incremented portion but was absent in the result set before this increment.
\textit{CanTree}~\cite{leung2005cantree} proposed a tree structure where nodes are ordered canonically instead of frequency. It captures the whole transaction database and does not require any rescan of the whole database or any reconstruction of the tree for any increment. \textit{CP-tree}~\cite{tanbeer2009efficient} periodically restructures the incremental tree structure according to the frequency descending order of items. Thus, it achieves not only a highly compressed tree structure but also a remarkable gain in mining time compared to the corresponding \textit{CanTree}.
Two efficient tree structures~\cite{ahmed2012single} have been proposed to mine weighted frequent patterns from an incremental database with only one scan.
Incremental weighted erasable pattern mining from incremental databases has also been explored in~\cite{nam2020efficient} which proposed efficient list structures. Lin et al.~\cite{lin2017efficiently} proposed RWFI-Mine and its further improvement named as RWFI-EMine to discover recent weighted frequent itemsets efficiently in a temporal database while considering  both weight and the recency constraints of patterns. Gan et al.~\cite{gan2017extracting} introduced the concept of Recent High Expected Weighted Itemset (RHEWI) to take the weight, uncertainty and recency constraints of patterns into account.  Consequently, it proposed two projection-based algorithms RHEWI-P and RHEWI-PS to mine RHEWIs from uncertain temporal databases. ILUNA~\cite{DBLP:journals/isci/Davashi21} proposed a single-pass incremental mining of uncertain frequent patterns without false positive patterns.

Recently, a number of incremental methods have been developed to mine high utility patterns in ~\cite{gan2018survey,lin2020incrementally,wu2020incrementally}.
Wang et al.~\cite{wang2011efficient} proposed two incremental algorithms to mine \textit{Probabilistic Frequent Itemsets (PFI)} from large uncertain databases: (a) \textit{uFUP}, which is a \textit{candidate generation and test} based algorithm inspired by \textit{FUP}\cite{cheung1996maintenance} that mines exact \textit{PFI} and (b) {$uFUP_{app}$}, which shows more efficiency in time and memory than \textit{uFUP}, but mines approximate \textit{PFI}. 
Techniques based on maintaining the whole database in a compact tree enabled efficient mining of the updated set of patterns after each increment.

We naturally need a massive memory space to store the whole database in a tree structure for sequential pattern mining because the number of nodes required to store all data sequences is exponentially high compared to an itemset database.
For example, for an itemset $\{a,b,c\}$, there are a large number of possible sequential instances because of different event formation and their order of appearance, such as $<(a)(b)(c)>$,$<(b)(a)(c)>$,$<(a)(b,c)>$, $<(a,b,c)>$,$<(a,b),(b), (a,c)>$, and so on.
All of the algorithms for sequential pattern mining discussed in the previous sections have a common limitation: they are designed to be performed once on the static database.
If any update in the database occurs, users have to run these algorithms from scratch each time.
This is very inefficient in terms of time and memory, especially if the increments are frequent and small.
In incremental mining, the challenge is to find the updated set of frequent sequences in the shortest possible time. 
Another major concern is that the algorithm cannot be expensive in memory usage.
\textit{IncSpan} \cite{cheng2004incspan_INCSPAN} algorithm proposed a solution for incremental mining of sequential algorithms with the concept of \textit{Semi-Frequent Sequences} (SFS). 
Later, \textit{IncSpan+} \cite{nguyen2005improvements_INCSPAN+} identified its limitation and proposed a corrected version which claimed to give complete results but includes the problem of full database scanning. 
However, \textit{PBIncSpan} \cite{chen2007incremental} proved that \textit{IncSpan+} also fails to find complete results. Using a prefix-based tree structure for pattern maintenance, \cite{chen2007incremental} shows that finding complete results is very challenging when the number of nodes becomes huge, which is obvious in incremental sequence databases.

\textit{PreFUSP-TREE-INS} is proposed in ~\cite{lin2015incrementally} to reduce the number of rescans and it incorporates the
\textit{pre-large concept}. Note that, \textit{pre-large sequences} are the same as \textit{semi-frequent sequences}.
This paper's main limitation is that it performs better only when the increment size is small, such as 0.05\%, 0.1\%, and 0.2\% of the original database.
Besides, there is a safety bound of \textit{pre-large concept}, i.e., the patterns stored in the \textit{pre-large} set can help up to a specific limit of database update.
Results in ~\cite{lin2015incrementally} suggest that this safety bound is very small as it showed efficient performance if the total increment ratio is below 0.2\%.
These results motivated us to design an efficient incremental solution for multiple large increments, i.e., to support both larger {increment size} and larger {total increment ratio}.

Moreover, authors in~\cite{wang2018incremental} proposed incremental algorithms named \textit{IncUSP-Miner, IncUSP-Miner+} to mine high-utility sequential patterns. They proposed a compact tree structure named \textit{Candidate-pattern tree} to maintain the patterns, a tighter upper bound of utility-based measures to prune the tree nodes better, and few strategies to reduce the {tree-node updates} and {database rescans}.
The major limitation is that they perform better only when the 
ratio between the {incremented size} and {original database size} is small.  

Nevertheless, to the best of our knowledge,
\textit{WIncSpan} \cite{ishita2018efficient_WINCSPAN} is the only incremental solution for weighted sequential pattern mining.
Similar to \textit{IncSpan} and \textit{IncSpan+}, it also uses extra buffers to store \textit{semi-frequent sequences (SFS)} from initial database. 
It also has the same limitation of \textit{IncSpan+} that any new pattern arriving later or any pattern that was initially not in the semi-frequent set cannot be found even if it becomes frequent after future increments.
However, it does not need to rescan the database after any increment. Results showed that a reasonable amount of buffering \textit{SFS} becomes enough to find the almost completed set of 
the updated result within a very short time.

\begin{table}[bth]
\centering
\begin{tabular}{|c|c|c|c|c|} \hline 
& \textbf{General} & \textbf{Weighted} & \textbf{High Utility}& \textbf{Uncertain}\\ \hline
Sequential & ~\cite{pei2004mining_prefixSpan},  &~\cite{rahman2019mining_uWSeq} & ~\cite{DBLP:journals/isci/ChenCGQD21},~\cite{lin2020high} &\cite{lin2020high},~\cite{lin2019project},~\cite{muzammal2010probabilistic} \\ 
Pattern Mining &\cite{rizvee2020tree},~\cite{srikant1996mining_GSP} &  ~\cite{yun2008new_WSPAN}&~\cite{DBLP:journals/isci/TruongDLF20},~\cite{DBLP:journals/isci/GanLZCFY20},~\cite{truong2019survey}  &\cite{muzammal2011mining_uSeq1st},~\cite{rahman2019mining_uWSeq},~\cite{zhao2013mining_uncertainSeq} \\ \hline
Incremental &~\cite{cheung1996maintenance},~\cite{leung2005cantree},~\cite{tanbeer2009efficient}  &~\cite{ahmed2012single},~\cite{lee2018single},~\cite{nam2020efficient}  &~\cite{lin2020incrementally},  &~\cite{DBLP:journals/isci/Davashi21},~\cite{wang2011efficient} \\ 
Itemset Mining &  &  &~\cite{wu2020incrementally} & \\ \hline
Incremental  &~\cite{cheng2004incspan_INCSPAN},~\cite{lin2015incrementally},  &~\cite{ishita2018efficient_WINCSPAN}  & ~\cite{wang2018incremental} & \\
Sequential Mining &  &  &  &\\ \hline
\end{tabular}
\caption{\textnormal{ A brief summary of survey algorithms}}
\label{tab:survey_algorithms}
 \end{table}

As shown in Table~\ref{tab:survey_algorithms}, the current literature suggests re-running \textit{uWSequence}~\cite{rahman2019mining_uWSeq} from scratch every time after an increment occurs to uncertain sequential databases. It is very costly in terms of time and space when the database grows larger and larger.
Therefore, we have explored this problem and propose two efficient techniques for incremental mining, \textit{uWSInc} and \textit{uWSInc+}, under a novel framework for mining both unweighted and weighted uncertain sequential patterns, where multiple theoretical tightened pruning upper bound measures and an efficient hierarchical index structure to maintain patterns, \textit{USeq-Trie}, have been developed. Consequently, this framework leads to an efficient design of an algorithm \textit{FUWS} for static uncertain databases. In the following discussions, we will be consistent with weighted uncertain pattern mining. Nonetheless, it is to be noted that we can easily adapt our weighted framework for mining unweighted patterns by setting the weights of all items as 1.0.


\section{A Framework for mining Weighted Uncertain Frequent Sequences} 
\label{sec:soln}

\newdefinition{definition}{Definition}
\newdefinition{example}{Example}
\newdefinition{proposition}{Proposition}
\newdefinition{axiom}{Axiom}
\newtheorem{thm}{Theorem}
\newtheorem{lemma}[thm]{Lemma}
\newproof{pf}{Proof}
\newproof{pot}{Proof of Theorem \ref{thm2}}

Before diving into the detailed description of our proposed framework, we will discuss some preliminaries which will be referred throughout this paper.

\subsection{Preliminaries}

\textbf{\textit{Sequences} and \textit{Uncertain Sequence Database}.}
Let I = \{ i$_1$, i$_2$,..., i$_n$\} be the set of all items in a database. 
An itemset or event e$_i$  = (i$_1$, i$_2$,...,i$_k$) is a subset of I. 
A sequence S = $\textless{}$e$_1$, e$_2$,..., e$_{m}\textgreater{}$ is an ordered set of itemsets \cite{srikant1996mining_GSP}.
For example, S$_1$ = $\textless{}$(i$_2$ ), (i$_1$, i$_5$), (i$_1$)$\textgreater{}$ consists of 3 consecutive itemsets. In case of uncertain sequences, items in each itemset are assigned with their existential probabilities such as $S_{1}$ =$<$(i$_2$: p$_{i_2}$), (i$_1$: p$_{i_1}$, i$_5$: p$_{i_5}$), (i$_1$: p$_{i_1}$)$>$ \cite{muzammal2011mining_uSeq1st,rahman2019mining_uWSeq,zhao2013mining_uncertainSeq}. An \textit{uncertain sequence database} \textit{(DB)} is a collection of uncertain sequences. An example database containing six uncertain sequences is shown in Table \ref{tab:initDB}.

\begin{table}[bh]
    \centering
    \begin{minipage}[t]{0.45\linewidth}
        \begin{tabular}{ll}
        \hline
          \textbf{Id} & \textbf{Uncertain Sequence} \\
        \hline
        1 & (a:0.9, c:0.6)(a:0.7)(b:0.3)(d:0.7) \\
        2 & (a:0.6, c:0.4)(a:0.5)(a:0.4, b:0.3)  \\ 
        3 & (a:0.3)(a:0.2, b:0.2)(a:0.4, b:0.3, g:0.5)  \\ 
        4 & (a:0.1, c:0.1)(a:0.3, b:0.1, c:0.4)   \\ 
        5 & (d:0.1)(a:0.4)(d:0.1)(a:0.5, c:0.6) \\ 
        6 & (b:0.3)(b:0.4)(a:0.1)(a:0.1, b:0.2)   \\ \hline
        \end{tabular}%
        \caption{Uncertain Sequential Database, $DB$}
        \label{tab:initDB}
    	\end{minipage}
	\hspace{4mm}
		\centering
		\begin{minipage}[t]{0.4\linewidth}
		\centering
            \begin{tabular}{cc}
            \hline
              \textbf{Item} & \textbf{Weight} \\
            \hline
            a & 0.8 \\
            b & 1.0  \\ 
            c & 0.9  \\ 
            d & 0.9   \\ 
            e & 0.7 \\ 
            f & 0.9   \\ 
            g & 0.8  \\\hline
            \end{tabular}%
            \caption{Weight of different items}
            \label{tab:wgts}
	\end{minipage}
\end{table}

{\textbf{\textit{Support} and \textit{Expected Support}.}}
\textit{Support} of a sequence in a database is the number of data tuples that contains $S$ as a subsequence. For example, $<(a)(b)>$ has a support of 5 in Table \ref{tab:initDB}. In uncertain databases, \textit{expected support} count makes more sense than this general support count. Different authors defined \textit{expected support} in several ways for an itemset or a sequence in literature. Here, we adopted the definition of \textit{expected support} for a sequence from uWSequence \cite{rahman2019mining_uWSeq} where all items in a sequence are considered independent of each other. The \textit{expected support} is defined as the sum of the maximum probabilities of that sequence in each data tuple. The probability of a sequential pattern is defined simply by multiplying the probability values of that pattern's items in a data tuple.
In Table~\ref{tab:initDB}, for the sequence, $<(a)(b)>$, where \textit{b} occurs in a different event after the occurrence of \textit{a},
the expected support, $expSup(\textless{}(a)(b)\textgreater{})
= max(0.9\times0.3, 0.7\times0.3) 
+ max(0.6\times0.3, 0.5\times0.3) 
+ max(0.3\times0.2, 0.3\times0.3, 0.2\times0.3) 
+ 0.1\times0.1
+ 0 
+ 0.1\times0.2 = 0.57$.
To explain the maximum probability of the pattern $<(a)(b)>$ in 1st sequence, it is the maximum of $0.9\times0.3$ (a in 1st event, b in 3rd event) and $0.7\times0.3$ (a in 2nd event, b in 3rd event). 

\textbf{\textit{Sequence Size} and \textit{Length}.}
\textit{Size} of a sequence $\alpha$ is the number of total itemsets/events in it and is represented by $|\alpha|$. 
\textit{Length} of a sequence is the total count of items present in all events of the sequence.
For example, size of $<(a,c)(a,b,c)>$ is 2 but its length is 5.
Sequence of length $m$ is also called a $m$-sequence.

\textbf{Extension of a Sequence.}
A sequence $\alpha$ can be extended with an item $i$ in two ways, i.e., i-extension and s-extension. 
If item $i$ is added to the last event of $\alpha$ then the resultant sequence, say $\beta$, is called i-extension of $\alpha$. 
For example, $<(a)(b,c)>$ is an i-extension of $<(a)(b)>$ with item $c$. In i-extension, size of a sequence does not increase but its length increases.
Similarly, if item $i$ is appended to $\alpha$ as a new event then the new sequence $\beta$ is called $\alpha$'s s-extension with \textit{i}.
For example, $<(a)(b)(c)>$ is a s-extension of $<(a)(b)>$ with item $c$.

\textbf{\textit{Weight} of a sequence}. Similar to \textit{expected support}, there are several definitions of a \textit{sequence weight}. We adopted the definition of the \textit{weight} of a sequence denoted as \textit{sWeight} in \cite{yun2008new_WSPAN,ishita2018efficient_WINCSPAN} where \textit{sWeight} is the sum of its each individual item’s \textit{weight} divided by the length of the sequence. Weights of the items are belong to the range between 0 to 1 as shown in Table \ref{tab:wgts}. 
For an example,  $sWeight(<(a)(ac)>)=(0.8+0.8+0.9)/3 = 0.833$. 

\textbf{\textit{Frequent} and \textit{Semi-frequent Sequences}.}
The set of sequences that suffice a given {minimum support threshold} (or {expected support threshold} for uncertain database) are called \textit{frequent sequences}.
Similarly, the sequences that meet a given minimum threshold of \textit{weighted expected support}, are called \textit{weighted frequent uncertain sequences} or \textit{weighted uncertain sequential patterns}. A \textit{buffer ratio, $\mu$}, which is of positive value less than 1.0, is chosen to lower the minimum support threshold to find \textit{semi-frequent sequences} that are not frequent but their values are very closed to the minimum support threshold.
As discussed earlier,
\textit{semi-frequent sequences} are helpful to find the updated set of result sequences when the database is incremental. \\

In the following subsections, we propose a new framework for mining weighted frequent sequential patterns in uncertain databases where a new concept of weighted expected support is introduced to incorporate the weight constraint in the mining process. Based on this framework, our proposed algorithms will be discussed, followed by an example simulation. Before diving into the detailed description, 
additional required definitions, proposed measures, and lemmas with proof
are presented in the following discussion.

\begin{definition}
$maxPr$ is the maximum possible probability of the considering sequential pattern in the projected database.
For a pattern $\alpha = <(i_{1})...(i_{m})>$,  \textit{maxPr} used in \textit{uWSequence}  \cite{rahman2019mining_uWSeq} can be defined as
\begin{equation}
    {maxPr(\alpha)} = \prod^{|\alpha|}_{k = 1} \widehat{P}_{DB\mid\alpha_{k-1}}(i_{k}) \ where\ \alpha_{k-1} = (i_{1})...(i_{k-1})
\end{equation}
Here, $\widehat{P}_{DB\mid \alpha}(i)$ = maximum possible probability of item $i$ in the database  $DB\mid \alpha$,  which is projected with $\alpha$ as current prefix. Rahman et. al.~\cite{rahman2019mining_uWSeq} shows that the \textit{maxPr} measure holds the anti-monotone property. 

\begin{example}
 The \textit{maxPr} value for $<(c)(a)>$ = $0.6 \times 0.7$ = 0.42 in Table   \ref{tab:initDB} where maximum possible probability of item $c$ in $DB$ (as shown in Table~\ref{tab:initDB}) is $\widehat{P}_{DB}(<(c)>) = 0.6$ and again,\newline  $\widehat{P}_{DB\mid <(c)>}(<(a)>) = 0.7$ because $0.7$ is maximum possible probability of item $a$ in the $<(c)>$-projected database, $DB\mid<(c)>$.\newline As an another example, we can find that $maxPr(<(a)(c)>)$  = 0.54.
\end{example}
\end{definition}

\begin{definition}
The $maxPr_{S}(\alpha)$ is defined as the maximum probability of a sequential pattern $\alpha$ in a single data sequence $S$. It can be formulated as: 
\begin{equation}
    maxPr_{S}(\alpha) = \max_{\mathcal{E} \in \mathcal{Q}} \bigg( \prod_{k \in \mathcal{E}} p_{k} \bigg)
\end{equation}
where $\mathcal{Q}$ denotes a set of all the sets of the sequential positions for each occurrence of the pattern $\alpha$ in $S$. In addition, $\mathcal{E}$ is a single set of the sequential positions for a particular occurrence of the pattern $\alpha$ in $S$. Moreover, $p_{k}$ is the existential probability of the corresponding item at $k$-th position in $S$.

\begin{example}
Suppose, we need to find the $masPr_S(<(a)(b)>)$ in the 1st sequence $S = S^{<1>}$ in Table \ref{tab:initDB}.
Here, $\mathcal{Q}=\Big\{\{1, 4\}, \{3, 4\} \Big\}$.\newline  
Therefore, $maxPr_{S}(<(a)(b)>) = max(0.9 \times 0.3, 0.7 \times 0.3) = 0.27$.
\end{example}
\end{definition}

\begin{definition}
The \textit{expected support} for a sequential pattern $\alpha$ can be calculated from a database $DB$ using the equation as follows,

\begin{equation}
    expSup(\alpha) = \sum_{i=1}^{|DB|} maxPr_{S^{<i>}}({\alpha})
\end{equation}
Where, $S^{<i>}$ denotes the $i$-th Sequence in the database $DB$ and $|DB|$ denotes the size of the database $DB$.

\begin{example}
To calculate the $expSup(<(ac)>)$ in Table~\ref{tab:initDB}, from the definition of $maxPr_S(\alpha)$, we get\\ 
$maxPr_{S^{<1>}}(<(ac)>) = 0.54$, $maxPr_{S^{<2>}}(<(ac)>) = 0.24$, $maxPr_{S^{<3>}}(<(ac)>) = 0$, \\
$maxPr_{S^{<4>}}(<(ac)>) = 0.12$, $maxPr_{S^{<5>}}(<(ac)>) = 0.30$ and $maxPr_{S^{<6>}}(<(ac)>) = 0$.\\ 
Thus, $expSup(<(ac)>) = 1.20$.
\end{example}
\end{definition}

\begin{definition}
\label{def:escap}
 ${expSup^{cap}(\alpha)}$ is an upper bound of expected support of a pattern $\alpha$ which is defined as
 \begin{equation}
     expSup^{cap}(\alpha)= maxPr(\alpha_{m-1})\times\sum_{\forall S\in (DB|{\alpha_{m-1}})} maxPr_{S}(i_{m})
 \end{equation}

\begin{example}
To compute the $expSup^{cap}(<(ac)(b)>)$ in Table~\ref{tab:initDB}, as per the definition, \\ 
$expSup^{cap}(<(ac)(b)>) =  maxPr(<(ac)> \times \sum_{\forall S \in DB|(<(ac)>)} maxPr_S(<(b)>)$ \\
where, $maxPr(<(ac)>) = 0.54$ and $\sum_{\forall S \in DB|(<(ac)>)} maxPr_S(<(b)>) = 0.7$: because $DB|(<(ac)>)$ has three non-empty sequences i.e., $<(a:0.7)(b:0.3)(d:0.7)>$, $<(a:0.5)(a:0.4,b:0.3)>$, and $<(a:0.3, b:0.1, c:0.4)>$. The values of $maxPr_S(<(b)>)$ for these three sequences are $0.3$, $0.3$, and $0.1$. \\
Consequently, $expSup^{cap}(<(ac)(b)>) = 0.54 \times 0.7 = 0.378$.\\
Similarly, the $expSup^{cap}$ value of another sequence $<(ac)>$, $expSup^{cap}(<(ac)>) = 1.8$.
\label{exp:expsup_cap}
 \end{example}
\end{definition}

\begin{lemma}
The \textbf{$expSup^{cap}$} of a sequential pattern is always greater than or equal to the actual expected support of that pattern.
\label{lem:es01}
\end{lemma}

\begin{pf}
To keep the proof less complicated, we consider a sequential pattern $\alpha = <(i_{0})(i_{1})....(i_{m})>$ where each event/itemset consists of a single item, $i_{k}$. \\
According to the definitions,  $ \forall i_{k} \in \alpha : maxPr(i_{k}) \geq maxPr_{S}(i_{k})$.\\
$\Rightarrow maxPr({i_{0}}) \times \sum_{\forall S \in (DB|{i_{0}})} maxPr_{S}(i_{1}) \geq  \sum_{\forall S \in DB} maxPr_{S}({\textless{}(i_{0})(i_{1})}\textgreater{})$\\
$\Rightarrow maxPr({\alpha_{m-1}}) \times \sum_{\forall S \in (DB|\alpha_{m-1})} maxPr_{S}(i_{m}) \geq  \sum_{\forall S \in DB} maxPr_{S}(\alpha)$\\
$\Rightarrow expSup^{cap}(\alpha) \geq expSup(\alpha)$ \\
$\therefore$ The equality holds only when each item has same existential probability for its all positions in whole database. Otherwise, $expSup^{cap}(\alpha) > expSup(\alpha)$ will always be true. \hfill \boxed{}
\end{pf}

\begin{lemma}
For any sequence $\alpha$, the value of $expSup^{cap}(\alpha)$ is always less than or equal to the $expSupport^{top}(\alpha)$ which is used as an upper bound of expected support in uWSequence \cite{rahman2019mining_uWSeq} and can be equivalently defined as $expSupport^{top}(\alpha) = maxPr(\alpha_{m-1})\times maxPr(i_{m}) \times sup_{i_{m}}$ where $sup_{i_{m}}$ denotes the support count of $i_{m}$.
\label{lem:03_capLessTop}
\end{lemma}

\begin{pf} According to definitions,  $\forall S,  \forall i_{k} \in \alpha : maxPr_{S}(i_{k}) \leq maxPr(i_{k})$ \\
$\Rightarrow \sum maxPr_{S}(i_{m}) \leq maxPr(i_{m}) \times sup_{i_{m}}$  \\
$\Rightarrow maxPr(\alpha_{m-1}) \times \sum maxPr_{S}(i_{m}) \leq maxPr(\alpha_{m-1}) \times maxPr(i_{m}) \times sup_{i_{m}}$  \\
$\Rightarrow expSup^{cap}(\alpha) \leq expSupport^{top}(\alpha)$
\hfill \boxed{}
\end{pf}

Thus, being a tighter upper bound of expected support, $expSup^{cap}$ can reduce the search space more in pattern-growth based mining process and hence it generates less false positive patterns than $expSupport^{top}(\alpha)$.

\begin{definition}
$WES(\alpha)$ is the \textit{weighted expected support} of a sequential pattern $\alpha$ defined as
\begin{equation}
    WES(\alpha) = expSup(\alpha) \times sWeight(\alpha)
\end{equation}
It is inspired by the widely used concept of \textit{weighted support} for precise databases as described in Section~\ref{sec:related}.
\end{definition}
\begin{example}
 According to Table \ref{tab:initDB} and Table \ref{tab:wgts},  weighted expected support of  $<(a)(ac)>$,\newline 
 $WES(<(a)(ac)>) =  (0.1\times0.3\times0.4 + 0.4\times0.5\times0.6) \times (0.8+0.8+0.9)/3 = 0.11$.
\end{example}

\begin{definition}
 Weighted Frequent Sequential Pattern: a sequence $\alpha$ is called weighted frequent sequential pattern if $WES(\alpha)$ meets a minimum weighted expected support threshold named as \textit{minWES}. This minimum threshold is defined to be,
\begin{equation}
    minWES = min\_sup\times database\ size\times WAM \times wgt\_fct
\end{equation}
\end{definition}

Here, $min\_sup$ is user given value in range [0,1] related to a sequence's expected support,
$WAM$ is weighted arithmetic mean of all item-weights present in the database defined as
\begin{equation}
        WAM = \frac{\sum_{i\in I}^{} f_{i} \times w_{i}}{\sum_{i\in I}^{} f_{i}}  
\end{equation}
where $w_{i}$ and $f_{i}$ are the weight and frequency of item $i$ in current updated database. 
The value of \textit{WAM} changes after each increment in the database.
$wgt\_fct$ is user given positive value chosen for tuning the mining of weighted sequential patterns.
Choice of $min\_sup$ and $wgt\_fct$ depends on aspects of application.

\begin{example}
Let us assume that $min\_sup = 0.2$ and $wgt\_fct = 0.75$.\newline 
From Table~\ref{tab:initDB} and Table~\ref{tab:wgts}, 
$WAM = \dfrac{(14\times0.8)+(8\times1.0)+(5\times0.9)+(3\times0.9)+(1\times0.8)}{14+8+5+3+1}$ $= 0.88$; database size = 6;
Therefore, $minWES = 0.792$.
\end{example}

However, the measure $WES$ does not hold anti-monotone property as any item with higher weight can be appended to a weighted-infrequent sequence, and the resulting super-sequence may become weighted-frequent. So, to employ anti-monotone property in mining weighted frequent patterns, we propose two other upper bound measures, $wgt^{cap}$  and $wExpSup^{cap}$, which are used as upper bound of weight and weighted expected support, respectively. 

\begin{definition}
We define an upper bound of weight for a pattern $\alpha$ of length $m$, $wgt^{cap}(\alpha)$ as follows, 
 \begin{equation}
    wgt^{cap}(\alpha) = \max (mxW_{DB}(DB|\alpha_{m-1}),  mxW_{s}(\alpha))
\end{equation}
where $mxW_{DB}(DB|\alpha_{m-1})$ is the weight of the item with maximum weight value in the projected database and $mxW_{s}(\alpha)$ is the weight of the item with maximum weight value in the sequential pattern $\alpha$.
\label{def:mxw}
\end{definition}

\begin{example}
 Suppose, we need to calculate the $wgt^{cap}(\alpha)$, where, $\alpha=<(ac)>$.\newline 
 From Table~\ref{tab:wgts},
 $mxW_s(<(ac)>) = \max(0.8, 0.9) = 0.9$.\newline  
 Again, items of ($DB|<(a)>$) are $a$, $b$, $c$, $d$, and $g$ in Table~\ref{tab:initDB}.\newline  
 So, $mxW_{DB}(DB|(<(a)>) = \max(0.8, 1.0, 0.9, 0.9, 0.8)$ $ = 1.0$.\newline 
 Therefore, $wgt^{cap}(<(ac)>) = \max(1.0, 0.9) = 1.0$.
 \label{exp:wgt_cap}
 \end{example}

To handle the downward property of weighted frequent patterns in precise databases, authors in \cite{yun2008new_WSPAN, ishita2018efficient_WINCSPAN} attempted to use the maximal weight of all items in the whole database as the upper bound of the weight of a sequence, noticing that this upper bound may generate much more false-positive patterns. To narrow the search space and keep the number of false candidate patterns as small as possible, we use $wgt^{cap}$ as an upper bound in mining weighted patterns. Intuitively, using $wgt^{cap}$ instead of the maximal weight of all items is more beneficial. Moreover, there may be many patterns for which the value of $wgt^{cap}$ is less than the maximal weight of all items. Hence, the patterns may meet the minimum threshold because of the higher value of maximal weight but may not satisfy the threshold due to the lower value of $wgt^{cap}$, resulting in a narrower search space of mining patterns and fewer candidate patterns. 

\begin{lemma}
For any sequential pattern $\alpha$ of length m, the value of $wgt^{cap}(\alpha)$ is always greater than or equal to the \textit{sWeight} value of its all super patterns.
\label{lem:mw}
\end{lemma}

\begin{pf}
Let us assume that $\alpha  \subset \alpha^{'}$ for some sequential pattern $\alpha^{'}$ of length $m^{'}$ where $m^{'}> m$.

According to Definition \ref{def:mxw}, $mxW_{S}(\alpha) \geq sWeight(\alpha)$.
The equality holds when the weights of all item in $\alpha$ are equal. Similarly, $mxW_{S}(\alpha^{'}) \geq sWeight(\alpha^{'})$.

Now, if the weights of all items in database are not equal, then $mxW_{DB}(DB|\alpha_{m-1}) \geq mxW_{DB}(DB|\alpha^{'}_{m^{'}-1})$ must hold since $DB|\alpha_{m-1}$ contains all frequent items of $DB|\alpha'_{m^{'}-1}$. 

Moreover, it is straightforward that $mxW_S(\alpha^{'})$ is always greater than or equal to $mxW_S(\alpha)$. Nevertheless, when $mxW_S(\alpha^{'}) > mxW_S(\alpha)$, the item with maximum weight in $\alpha^{'}$ must come from the projected database, $DB|\alpha_{m-1}$ and thus $mxW_S(\alpha^{'})\leq mxW_{DB}(DB|\alpha_{m-1})$.\\
$\therefore \max( mxW_S(\alpha^{'}),  mxW_{DB}(DB|\alpha^{'}_{m^{'}-1})) \leq \max( mxW_S(\alpha),  mxW_{DB}(DB|\alpha_{m-1}))$\\
$\Rightarrow$ $ wgt^{cap}(\alpha^{'}) \leq wgt^{cap}(\alpha)$ \\
$\Rightarrow  sWeight(\alpha^{'}) \leq wgt^{cap}(\alpha)$ \\
Again,  if the weights of all items are equal,  then
$wgt^{cap}(\alpha^{'}) = sWeight(\alpha^{'}) = sWeight(\alpha) = wgt^{cap}(\alpha)$ \\
Therefore,  we can conclude that the value of $wgt^{cap}(\alpha)$ is always greater or equal to the value of $sWeight(\alpha)$ or $sWeight(\alpha^{'})$ which is true for all cases.
\hfill \boxed{}
\end{pf}

\begin{definition}
As mentioned before, the upper bound of weighted expected support is $wExpSup^{cap}(\alpha)$, defined as:
\begin{equation}
    wExpSup^{cap}(\alpha) = expSup^{cap}(\alpha) \times wgt^{cap}({\alpha})
    \label{eq_wExpSup_cap}
\end{equation}
\end{definition}

\begin{example}
  Considering the pattern $\alpha=<(ac)>$, the value of $wExpSup^{cap}(<(ac)>) = 1.8 \times 1.0 = 1.8$ where $expSup^{cap}(\alpha) = 1.8$ (as computed in Example~\ref{exp:expsup_cap}) and $wgt^{cap}(<(ac)>) = 1.0$ (see in Example~\ref{exp:wgt_cap}).
 \end{example}

\begin{lemma}
\label{lem:wes01}
The value of $wExpSup^{cap}(\alpha)$ is always greater than or equal to weighted expected support of a sequential pattern $\alpha$,  $WES(\alpha)$. Hence,  using the $wExpSup^{cap}$ value of any pattern as the upper bound of weighted expected support in mining patterns,  it may generate some false positive frequent patterns.
\end{lemma}

\begin{pf}
Lemma  \ref{lem:es01} and  \ref{lem:mw} has showed that for a sequential pattern $\alpha$, \linebreak
$expSup^{cap}(\alpha) \geq expSup(\alpha)$ and $wgt^{cap}(\alpha) \geq sWeight(\alpha)$ \\
$ \Rightarrow  expSup^{cap}(\alpha) \times wgt^{cap}(\alpha) \geq expSup(\alpha) \times sWeight(\alpha) $ \\
$ \Rightarrow  wExpSup^{cap}(\alpha) \geq WES(\alpha) $ \\
$\therefore$ The value of $wExpSup^{cap}(\alpha)$ is always greater than or equal to the value of $WES(\alpha)$ for any sequence $\alpha$. As a result,  some patterns might be introduced as frequent patterns due to its higher value of $wExpSup^{cap}$ being not actually weighted frequent.
\hfill \boxed{}
\end{pf}

\begin{lemma}
\label{lem:wescap}
 If the value of $wExpSup^{cap}$ for a sequential pattern, $\alpha$, is below the minimum weighted expected support threshold \textit{minWES}, the value of \textit{WES} for that pattern and its all super patterns must not satisfy the threshold. In other words, the pattern $\alpha$ and its all super patterns must not be frequent if the value of $wExpSup^{cap}(\alpha)$ does not satisfy the threshold. Thus, the anti-monotone property holds while mining patterns.
\end{lemma}

\begin{pf}
Assume that $\alpha  \subseteq \alpha^{'}$ for some patterns $\alpha^{'}$.
Recall that, expSup($\alpha$) = $\sum_{\forall S\in DB} maxPr_{S}(\alpha)$ \cite{rahman2019mining_uWSeq}.
By definition, $expSup(\alpha)\geq expSup(\alpha')$.\\
Again,  $expSup^{cap}(\alpha) \geq  expSup(\alpha) \Rightarrow$  $expSup^{cap}(\alpha) \geq  expSup(\alpha')$.\\
Moreover, $wgt^{cap}(\alpha)\geq sWeight(\alpha').$\\
Now, $wExpSup^{cap}(\alpha) = expSup^{cap}(\alpha)\times wgt^{cap}(\alpha) \geq expSup(\alpha')\times sWeight(\alpha') = WES(\alpha')$\\
So, if $wExpSup^{cap}(\alpha) < minWES$ holds, then $WES(\alpha')< minWES$ must hold for any $\alpha'\supseteq\alpha$.\\
Therefore, upper bound $wExpSup^{cap}$ could be able to find out the complete set of frequent patterns.
\hfill \boxed{}
\end{pf}

\textbf{Pruning Condition.} In pattern-growth based mining algorithms, we define a pruning condition to reduce the search space. 
According to Lemma \ref{lem:wescap}, we can safely define our pruning condition which is to be used in mining algorithm as follows:
\begin{equation}
    wExpSup^{cap}(\alpha) \geq minWES
\label{equ:prucon}
\end{equation}
The $wExpSup^{cap}(\alpha)$ determines whether the current pattern $\alpha$ will be used to generate its super patterns or not. To prune weighted infrequent patterns earlier during the mining process but maintain the anti-monotone property,  the $wExpSup^{cap}$ is calculated instead of \textit{WES} to mine potential candidate patterns. If, for any $k$-sequence $\alpha$, $wExpSup^{cap}(\alpha)<minWES$, then any possible extension of $\alpha$ to a $(k+1)$-sequence can be safely pruned, i.e, mining of longer patterns with prefix $\alpha$ can be ignored without missing any actual frequent patterns. 
However, to get actual frequent patterns, we must prune out false positive patterns from the set of candidate patterns because $wExpSup^{cap}$ is an approximate value.


\subsection{USeq-Trie: Maintenance of Patterns}
In our proposed algorithms,  
we have used a hierarchical data structure, \textit{USeq-Trie}, to store patterns compactly and to update their weighted expected support efficiently. 
Each node in \textit{USeq-Trie} will be created as either an s-extension or i-extension from its parent node. Each edge is labeled by an item. In an s-extension, the edge label is added as a different event. In an i-extension, it is added in the same event as its parent. The sequence of the edge labels in a path to a node from the root node denotes a pattern.

\begin{figure}[tbhp] 
    \centering
    \includegraphics[width=0.3\textwidth]{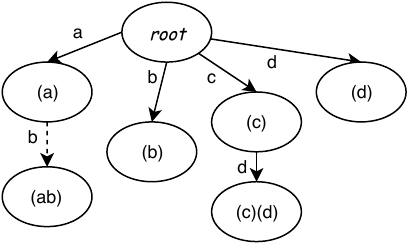}
    \caption{Storing frequent sequences in a \textit{USeq-Trie}}
    \label{fig:init_trie}
\end{figure}
For example,  $\textless{}(a)\textgreater{}$,  $\textless{}(ab)\textgreater{}$, $\textless{}(b)\textgreater{}$, $\textless{}(c)\textgreater{}$, $\textless{}(c)(d)\textgreater{}$,  and $\textless{}(d)\textgreater{}$ are frequent sequential patterns which are stored into \textit{USeq-Trie} shown in Figure  \ref{fig:init_trie}. In this figure, the s-extensions are denoted by solid lines and i-extensions by dashed lines. By traversing the \textit{USeq-Trie} in depth-first order, we will get all patterns stored in it. Each node represents the pattern up to that node from the root node and stores its weighted expected support which is ignored in Figure  \ref{fig:init_trie} for simplicity.

\begin{figure}[tb]
    \begin{minipage}[t]{0.48\linewidth}
    \centering
    		\includegraphics[width=.6\textwidth,  height=.4\textwidth]{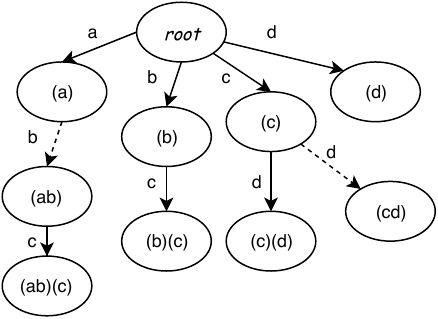}
    			\caption{After insertion of patterns}
    			\label{fig:insertion_1}
    	\end{minipage}
		\begin{minipage}[t]{0.48\linewidth}
		\centering
		\includegraphics[width=.6\textwidth,  height=.4\textwidth]{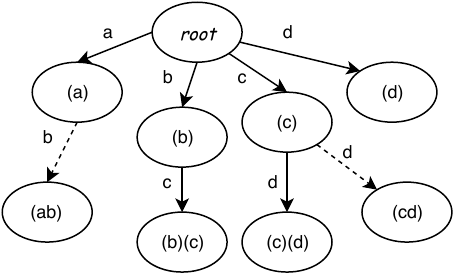}
			\caption{After deleting $\textless{} (ab)(c) \textgreater{}$ pattern}
			\label{fig:deletion_1}
	\end{minipage}

\end{figure}

\textbf{Insertion and Deletion}. To insert patterns into the \textit{USeq-Trie},  we start to traverse it from the root node and take the first item in a pattern as the current item. If there is a child node with an edge labeled by the current item, we go to that node. Otherwise, we create a child node with an edge labeled by the current item. Then we move to the child node and set the next item in the pattern as the current item. Recursively,  we follow the same process up to the last item in the pattern.
For example,  now we insert $\textless{} (ab)(c) \textgreater{}$ into the \textit{USeq-Trie} shown in Figure  \ref{fig:init_trie}. The root node has a child node with an edge labeled by \textit{a}. So,  we go to the child node labeled by (a) and check whether it has an edge labeled by \textit{b}. The extension has to be an i-extension as \textit{a} and \textit{b} are in the same event. As we can see, such a node already exists.
So, we go to the node which is labeled by \textit{(ab)}. Afterward, we have to check if there is an s-extension by an edge labeled $c$. But there is no such child node. Therefore, we create a node by s-extension and set \textit{c} as the edge label and \textit{(ab)(c)} as the node label. Similarly, we insert $\textless{} (b)(c) \textgreater{}$ and $\textless{} (cd) \textgreater{}$ into the \textit{USeq-Trie}. The updated \textit{USeq-Trie} is shown in Figure  \ref{fig:insertion_1}. To delete a pattern from \textit{USeq-Trie}, we traverse the corresponding path which represents the pattern in bottom-up order. The node in the path having no child nodes in the \textit{USeq-Trie} will be removed while traversing the corresponding path from the leaf node to the root node.
The resulting \textit{USeq-Trie} after deleting $\textless{}(ab)(c)\textgreater{}$  is shown in Figure \ref{fig:deletion_1}.\\

\begin{figure}[tbhp] 
    \centering
    \includegraphics[width=.8\linewidth, height=.4\linewidth]{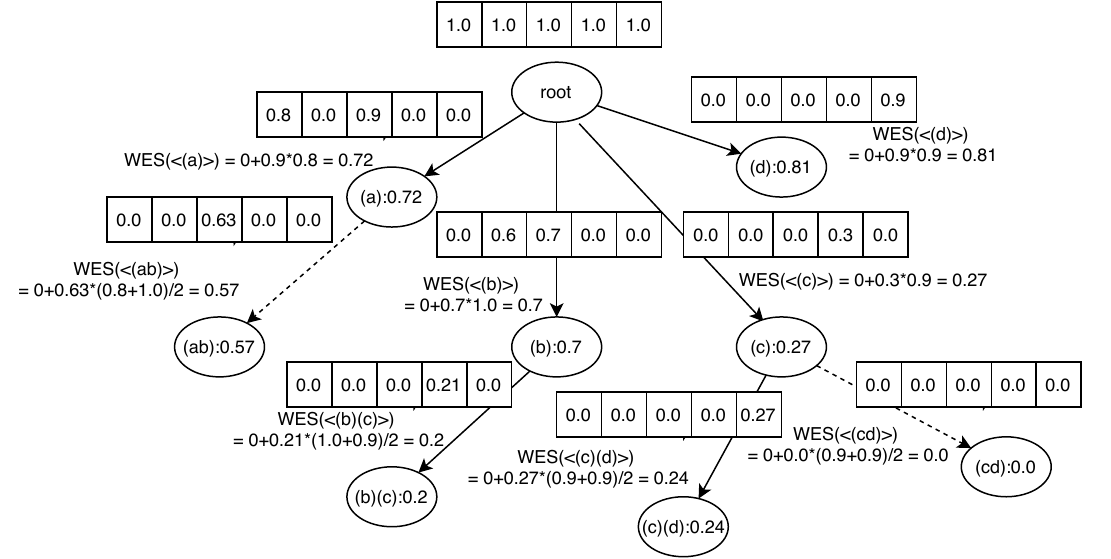}
			\caption{Pattern Maintenance and WES calculation using \textit{USeq-Trie}}
			\label{fig:useq-trie}
\end{figure}

\textbf{Support Calculation}. 
We propose an efficient method denoted as \textit{SupCalc} by using \textit{USeq-Trie} to calculate weighted expected support of patterns. In this method, it reads sequences from the dataset one by one and updates the support of all candidate patterns stored into the \textit{USeq-Trie} against this one. For a sequence $S = \textless{}e_{1}e_{2}..e_{n}\textgreater{}$ (where $e_{i}$ is an event or itemset), the steps are defined as follows:
\begin{itemize}
    \item At each node, define an array the $size\ of\ S$, which is \textit{n} in this case. At the root node, all values of the array will always be 1.0.
    \item Traverse the \textit{Useq-Trie} in depth-first order. After following an edge, let the current pattern from root to a particular node be $\alpha$ which ends with the item \textit{$i_{k}$}. The maximum \textit{weighted expected support} of  
    the pattern $\alpha$
    is stored at proper indices of the following node's array. These proper indices are the ending positions of 
    $\alpha$
    as a sub-sequence in $S$. Set values to zero at other indices.
    \item While traversing the \textit{USeq-Trie}, iterate all events in $S$. (i)
    For \textit{s-extension} with an item \textit{$i_{k}$}, we calculate the support of the current pattern $\alpha$ (ends with $i_{k}$ in a new event) by multiplying the probability of item \textit{$i_{k}$} in current event, $e_{m}$ with the maximum probability in the parent node's array up to the event $e_{m-1}$. The resulting support is stored at position m in the following node's array. 
    (ii) For i-extension, the support will be calculated by multiplying the probability of the item \textit{$i_{k}$} in $e_{m}$ with the value at position m in the parent node's array and stored at position m in the following child node's array.
    After that, the maximum value in the resulting array multiplied with its weight will be added to the weighted expected support of the current patterns at the corresponding node.
    \item Use the resultant array to calculate the support of all super patterns while traversing the next nodes.
\end{itemize} 

\makeatletter
\renewcommand{\ALG@beginalgorithmic}{\small}
\makeatother
\algrenewcommand{\alglinenumber}{\normalsize}

\renewcommand{\algorithmicrequire}{\textbf{Input:}}
\renewcommand{\algorithmicensure}{\textbf{Output:}}
\makeatletter
\renewcommand{\ALG@beginalgorithmic}{\small }
\makeatother
\algrenewcommand{\alglinenumber}{\small}

\begin{algorithm}[tbh]
\caption{Procedure of SupCalc}
\begin{algorithmic}[1]
\Statex
\Require \textit{DB:} initial database,  \textit{candidateTrie}: stores candidate patterns
\Ensure Calculated weighted expected supports for all patterns 
\Statex
\Procedure{SupCalc}{$DB$,  \textit{candidateTrie}}
\ForAll{${S =  < e1, e2, e3,...,en > } \in {DB}$} \Comment{$e_{k}$ is an itemset/event}
\State $ar \gets\ $the array of size equals to the number of events in $|S| $ which is initialized as 1
\State $wgt\_sum,  itm\_cnt \gets$ 0 and 0 \Comment{$wgt\_sum$ - sum of items' weights and $itm\_cnt$ - their counts in a pattern}
\State TrieTraverse($S$, \textit{null}, \textit{candidateTrie.root},  \textit{ar},  \textit{wgt\_sum},  \textit{itm\_cnt})
\EndFor
\EndProcedure                                                
\Statex 

\Procedure{TrieTraverse}{$S$, \textit{cur\_itmset}, \textit{cur\_node},  \textit{ar},  \textit{wgt\_sum},  \textit{itm\_cnt}}
\ForAll{$ node \in cur\_node.descendents$}
\State $cur\_edge \gets$ edge label between current child \textit{node} and \textit{cur\_node}
\State $cur\_ar \gets$ the array of size \textit{ar} initialized as 0 
\ForAll{ $e_{k} \in S$}
\If{S-Extension Is TRUE}
\State $cur\_itmset \gets cur\_edge$
\If{ $cur\_itmset \in e_{k}$ }
\State mxSup $\gets \max^{k-1}_{i = 1} ar_{i}$
\State $cur\_ar_{k}$ $\gets mxSup \times p_{cur\_edge}$ \Comment{$p_{cur\_edge}$ denotes the existential probability of $cur\_edge$ in $e_{k}$} 
\EndIf
\If{ $cur\_itmset \notin e_{k}$ }
\State $cur\_ar_{k}$ $\gets 0$
\EndIf
\EndIf
\If{I-Extension Is TRUE}
\State $ cur\_itmset \gets (cur\_itmset \cup cur\_edge)$ \Comment{$cur\_itmset$ is extended with $cur\_edge$}
\If{ $cur\_itmset \in e_{k}$ }
\State $cur\_ar_{k}$ $\gets ar_{k} \times p_{cur\_edge}$ \Comment{$p_{cur\_edge}$ denotes the existential probability of $cur\_edge$ in $e_{k}$}  
\EndIf
\If{ $cur\_itmset \notin e_{k}$ }
\State $cur\_ar_{k}$ $\gets 0$
\EndIf
\EndIf

\EndFor
\State $mxSup \gets \max^{|S|}_{i=1} cur\_ar_{i}$
\State $cur\_wgt\_sum \gets wgt\_sum + wgt_{cur\_edge}$ \Comment{$wgt_{cur\_edge}$ denotes the weight of $cur\_edge$}
\State $cur\_itm\_cnt \gets itm\_cnt + 1$
\State $node.WES \gets node.WES + mxSup \times \frac{cur\_wgt\_sum}{cur\_itm\_cnt}$
\State TrieTraverse($S$, \textit{cur\_itmset}, \textit{node},  \textit{cur\_ar},  \textit{cur\_wgt\_sum},  \textit{cur\_itm\_cnt})
\EndFor
\EndProcedure
\Statex
\end{algorithmic}
\label{algo:supcal}
\end{algorithm}

For example,  we consider a data sequence,  $S = \textless{} (a:0.8)(b:0.6)(a:0.9,  b:0.7)(c:0.3)(d:0.9)\textgreater{}$. To calculate weighted expected support, the size of array is equal to 5 where each value is set initially as 1. For the edge label \textit{a},  the values at indices 1 and 3 of the array will be 0.8 and 0.9, respectively, and at other indices will be 0. The maximum value in the array which is 0.9 multiplied with its weight 0.8 will be added to \textit{WES} of the current pattern at the child node which is labeled by \textit{(a)}. \textit{(ab)} is an i-extension of \textit{(a)} with item \textit{b}. Item \textit{b} is only in the $e_{3}$ itemset in $S$. The value of \textit{(ab)} at position 3 in the array of node labeled by \textit{(ab)} will be the probability of item \textit{b} in $e_{3}$ itemset in $S$ multiplied with the value at position 3 in the array of node labeled by \textit{(a)} which is $0.7 \times 0.9 = 0.63$. The values at other positions in the array of node \textit{(ab)} will be 0.0. The maximum value in the resulting array (0.63) multiplied with its \textit{sWeight} ($(0.8+1.0)/2=0.9$) will be added to the \textit{WES} of node \textit{(ab)}. Afterward, by traversing the \textit{USeq-Trie} in depth-first order, let us consider the next branch from the root node of the \textit{USeq-Trie}. The array for the edge label $b$ contains values of 0.6 and 0.7 at indices 2 and 3, respectively. The \textit{WES} of the node labeled by $(b)$ equals to the value of max(0.6,0.7) multiplied with the weight of $b$. Next, $(b)(c)$ is a s-extension of $(b)$ with item \textit{c}. Item \textit{c} is only in the $e_{4}$ event in $S$. The value of $(b)(c)$ at position 4 in the array of node labeled by ${(b)(c)}$ will be the probability of item \textit{c} in $e_{4}$ event in $S$ multiplied with the maximum value up to the index 3 in the array of node labeled by $(b)$, which is $0.3 \times 0.7 = 0.21$. The values at other positions in the array of node $(b)(c)$ will be 0.0. The maximum value in the resulting array (0.21) multiplied with its \textit{sWeight} ($(1.0+0.9)/2=0.95$) will be added to the $WES$ of node $(b)(c)$. Similarly, the support of all other patterns at corresponding nodes will be calculated. The results are shown in Figure \ref{fig:useq-trie}.


The pseudo-code has been given in Algorithm  \ref{algo:supcal}.
It takes $O(N\times |S|)$ to update \textit{N} number of nodes against the sequence $S$. Therefore,  the total time complexity of actual support calculation is $O(|DB|\times N\times k)$ where \textit{k} is the maximum sequence length in the dataset. It outperforms the procedure used in \textit{uWSequence}  \cite{rahman2019mining_uWSeq}
which needs $O(|DB|\times N\times k^{2})$.
Moreover,  we can remove false-positive patterns and find frequent ones with \textit{O(N)} complexity.
Thus,  the use of \textit{USeq-Trie} leads our solution to become more efficient.

\subsection{FUWS : Faster Mining of Uncertain Weighted Frequent Sequences}
In order to reduce the number of false-positive patterns by introducing a sophisticated upper bound measure, $wExpSup^{cap}$, and to make mining patterns more efficient, we develop an algorithm named as \textit{FUWS} inspired by \textit{PrefixSpan} \cite{pei2004mining_prefixSpan}, to mine weighted sequential patterns in an uncertain database. The sketch of \textit{FUWS} algorithm  is as follows,
\begin{itemize}
    \item Process the database such that the existential probability of an item in a sequence is replaced with the maximum probability of its all  next occurrences in this sequence. This idea is similar to the preprocess function of \textit{uWSequence} \cite{rahman2019mining_uWSeq}. In addition, sort the items in an event/itemset in lexicographical order. This preprocessed database will be used to run the \textit{PrefixSpan} like mining approach to find the candidates for frequent sequences. 
    \item Calculate \textit{WAM} of all items present in the current database and set the threshold of weighted expected support,  \textit{minWES}.
    \item Find 1-length frequent items and, for each item,  project the preprocessed database into smaller parts and expand longer patterns recursively. Store the candidates into a \textit{USeq-Trie}.
    \item While growing longer patterns,  extend current prefix $\alpha$ to $\alpha'$ with an item $\beta$ as  s-extension or i-extension according to the pruning condition defined in Equation  \ref{equ:prucon}.
    \item Use of $wExpSup^{cap}$ value instead of actual support generates few false-positive candidates. Scan the whole actual database,  update weighted expected supports and prune false-positive candidates based on their weighted expected support.
\end{itemize}


\makeatletter
\renewcommand{\ALG@beginalgorithmic}{\small}
\makeatother
\algrenewcommand{\alglinenumber}{\normalsize}

\renewcommand{\algorithmicrequire}{\textbf{Input:}}
\renewcommand{\algorithmicensure}{\textbf{Output:}}
\makeatletter
\renewcommand{\ALG@beginalgorithmic}{\small }
\makeatother
\algrenewcommand{\alglinenumber}{\small}

\begin{algorithm}[tbh]
\caption{Procedure of \textit{FUWS}}
\begin{algorithmic}[1]
\Statex
\Require \textit{DB:} initial database,  \textit{min\_sup:} support threshold,  \textit{wgtFct:} weight factor
\Ensure \textit{FS:} set of weighted frequent patterns 
\Statex
\Procedure{FUWS}{$DB$,  \textit{min\_sup},  \textit{wgtFct}}
\State $pDB,  WAM \gets preProcess($DB$)$ 
\State $minWES \gets min\_sup \times|pDB| \times WAM \times wgt\_fct$
\State $extItms,  maxPrs,  iWgts \gets Determine(\textit{pDB})$ to find the set of potential s-extendable items.
\ForAll{${\beta } \in {extItms}$}
\State $expSup^{cap}(\beta) \gets \times \sum_{\forall S \in (pDB)}maxPr_{S}({\beta})$
\State $wgt^{cap}(\beta) \gets \max(iWgts)$ \Comment{\textit{iWgts} - List of all items' weights in $pDB$}
\If{$wExpSup^{cap}{(\beta)} = expSup^{cap}(\beta) \times wgt^{cap}(\beta) \geq minWES$}
\State $\textit{candidateTrie} \gets FUWSP(pDB|{\beta }, \beta,  maxPrs_{\beta }, iWgts_{\beta},  iWgts_{\beta},  1)$ 
\EndIf
\EndFor
\State \textbf{Call} SupCalc(\textit{DB},  \textit{candidateTrie})

\State FS $\gets$ Remove false positives and find frequent patterns from \textit{candidateTrie}
\EndProcedure                                                
\Statex 


\Procedure{FUWSP}{$DB, \ \alpha, \ maxPr_{\alpha },\ mxW_{\alpha}, \ sWgt_{\alpha}, \ |\alpha|$}
\State $extItms, \ mxPrs, \ iWgts \gets Determine(DB)$ to find set of potential i or s-extendable items for prefix $\alpha$.
\ForAll{$ \beta \in extItms$}
\State $expSup^{cap}(\alpha \cup \beta) \gets maxPr_{\alpha } \times \sum_{\forall S \in (DB|\alpha)}maxPr_{S}({\beta})$
\State $wgt^{cap}(\alpha \cup \beta) \gets \max(mxW_{\alpha}, iWgts)$ \Comment{\textit{iWgts} - List of all items' weights in $DB$}
\If{$wExpSup^{cap}{(\alpha \cup \beta)} = expSup^{cap}(\alpha \cup \beta) \times wgt^{cap}(\alpha \cup \beta) \geq minWES$}
\State $maxPr_{\alpha \cup \beta} \gets maxPr_{\alpha} \times mxPrs_{\beta}$
\State $sWgt_{\alpha \cup \beta} \gets sWgt_{\alpha} + sWgts_{\beta}$
\State $FUWSP(DB|{\beta}, (\alpha \cup \beta),  maxPr_{\alpha \cup \beta}, \max(mxW_{\alpha}, sWgts_{\beta}),  sWgt_{\alpha \cup \beta},  |\alpha \cup \beta|)$
\EndIf
\EndFor
\EndProcedure
\Statex
\end{algorithmic}
\label{algo:fuws}
\end{algorithm}

Pseudo-code for \textit{FUWS} is shown in Algorithm  \ref{algo:fuws}. The function \textit{preProcess} in Line 2 prepares the input database for pattern-growth approach \textit{FUWSP} in \textit{FUWS}.
It computes \textit{WAM} to incorporate our weight constraint. In Algorithm  \ref{algo:fuws}, the function, \textit{Determine}, finds the set of all the potential extendable items,  the maximum probabilities of items in the processed database, and their weights from the weight vector.
In our mining process \textit{FUWSP},  we have used $wExpSup^{cap}$, the upper bound for weighted expected support of a sequence,  to find out the set of i-extendable and s-extendable items and their probabilities and weights in Line 13. In Lines 14-20,  when the value of $wExpSup^{cap}(\alpha \cup \beta )$ satisfies the threshold \textit{minWES},  \textit{FUWSP} is called recursively for the pattern $(\alpha \cup \beta)$ to generate its super sequences. Otherwise,  the algorithm prunes the current pattern $(\alpha \cup \beta)$ and its super patterns. As a result,  \textit{FUWSP} generates all potential candidate sequences recursively.
In addition to that,  we use \textit{candidateTrie} which is a \textit{USeq-Trie} to store candidates and update their weighted expected support efficiently. However,  at Lines 10-11,  \textit{FUWS} calls the \textit{SupCalc} function in Algorithm \ref{algo:supcal} to calculate weighted expected support for all candidate sequences stored into the \textit{candidateTrie} and remove false ones. An extensive simulation of \textit{FUWS} and its experimental results are discussed in Section \ref{sec:simulation} and \ref{results} respectively. 

\subsection{Two Approaches for Incremental Database}
As we have mentioned, mining the complete set of frequent weighted sequential patterns is very expensive with respect to time and space when the database grows dynamically, so to find out the almost complete set of weighted frequent sequences (FS), we propose two techniques,  \textit{uWSInc} and \textit{uWSInc+}. In both approaches,  we lower the minimum threshold \textit{minWES} to $minWES^{'}=minWES\times \mu$ where $0<\mu<1$ is an user-chosen buffer ratio. 
Sequences with weighted expected support less than \textit{minWES} but at least equal to $minWES'$ are stored as weighted semi-frequent sequences, which are named as \textit{SFS}.  

\subsubsection{uWSInc : Faster Incremental Mining of Uncertain Weighted Frequent Sequences}
Instead of running an algorithm from scratch after each increment, \textit{uWSInc} algorithm works only on the appended part of the database. At first, it runs \textit{FUWS} once to find \textit{FS} and \textit{SFS} from the initial dataset and uses \textit{USeq-Trie} to store frequent and semi-frequent sequences. 
After each increment $\Delta DB$,  the algorithm follows the steps as listed below:
\begin{enumerate}
    \item Update database size and \textit{WAM} value. Calculate the new threshold of weighted expected support.
    \item For each sequence $\alpha$ in \textit{FS} and \textit{SFS},  update its weighted expected support,  \textit{WES$_{\alpha}$},  by using our proposed faster support calculation method, \textit{SupCalc}, described in Algorithm \ref{algo:supcal}.
    \item Update the \textit{FS} and \textit{SFS} by comparing updated \textit{WES$_{\alpha}$} values with the new \textit{minWES} and $minWES^{'}$ respectively. A sequence may go to one of the updated \textit{FS'} or \textit{SFS'}, or vice versa or it may become infrequent. Once a pattern becomes infrequent, it will be removed and  its information will get lost.
    \item Use \textit{FS'} and \textit{SFS'} as \textit{FS} and \textit{SFS} for the next increment.
\end{enumerate}

\makeatletter
\renewcommand{\ALG@beginalgorithmic}{\small}
\makeatother
\algrenewcommand{\alglinenumber}{\normalsize}
\renewcommand{\algorithmicrequire}{\textbf{Input:}}
\renewcommand{\algorithmicensure}{\textbf{Output:}}
\makeatletter
\renewcommand{\ALG@beginalgorithmic}{\small}
\makeatother
\algrenewcommand{\alglinenumber}{\small}

\begin{algorithm}[tbh]
\caption{Procedure of \textit{uWSInc}}
\begin{algorithmic}[1]
\Statex
\Require \textit{DB:} initial database,  $\Delta {DB}$ : new increments,  \textit{min\_sup:} support threshold,  $\mu$: buffer ratio,  \textit{wgt\_fct:} weight factor 
\Ensure \textit{FS:} set of weighted frequent patterns 
\Statex
\Procedure{InitialMining}{$DB$,  $\Delta{DB}_{i}$,  \textit{min\_sup},  $\mu$,  \textit{wgt\_fct}}
\State $seqTrie \gets FUWS(DB,  min\_sup \times \mu, \textit{wgt\_fct})$

\ForAll{$\Delta{DB}_{i}$}
\State $DBSize \gets DBSize + \Delta{DB}_{i}Size $
\State $seqTrie \gets$ uWSInc($\Delta{DB}_{i}$,  $min\_sup$,  $\mu$,  \textit{wgt\_fct}, $seqTrie$)
\EndFor
\EndProcedure                                                
\Statex 

\Procedure{uWSInc}{$\Delta{DB}_{i}$,  $min\_sup$,  $\mu$,  \textit{wgt\_fct}, $seqTrie$}
\State \textbf{Call} SupCalc($\Delta{DB}_{i}$,  \textit{seqTrie})
\State $minWES \gets min\_sup \times DBsize \times WAM \times wgt\_fct$
\ForAll{$\beta \in (FS \lor SFS )\ stored\ into\ seqTrie$}
\If{$wExpSup(\beta) < (minWES \times \mu)$}
\State Remove pattern $\beta$ from \textit{seqTrie}
\EndIf
\EndFor
\State $FS \gets seqTrie.find\_frequent\_patterns(\textit{minWES})$
\EndProcedure
\Statex
\end{algorithmic}
\label{algo:uWSInc}
\end{algorithm}

Pseudocode for \textit{uWSInc} is given in Algorithm  \ref{algo:uWSInc}. The \textit{uWSInc} algorithm runs \textit{FUWS} once on the initial database,  \textit{DB}. In Line 2, it stores both \textit{FS} and \textit{SFS} into a single \textit{seqTrie} which brings out more space compactness. For each increment $\Delta DB_{i}$,  it calls \textit{SupCalc} function in Algorithm \ref{algo:supcal} to update the weighted expected support (WES) values of all \textit{FS} and \textit{SFS} accordingly. In Line 9-11,  it traverses the \textit{seqTrie} and removes patterns whose \textit{WES} is less than $minWES^{'}$. Finally,  it traverses the \textit{seqTrie} to find frequent patterns whose \textit{WES} is greater or equal to the threshold \textit{minWES} that is \textit{FS}.
The simulation of \textit{uWSInc} by using an example has been explained in Section  \ref{sec:simulation} and the details of result analysis are shown in Section \ref{sec:results}. 


\subsubsection{uWSInc+ : Incremental Mining of Uncertain Weighted Frequent Sequences for Better Completeness}
Let us consider some cases: (a)  an increment to the database may introduce a new sequence which was initially absent in both \textit{FS} and \textit{SFS} but appeared frequently in later increments;  (b) a sequence had become infrequent after an increment but could have become semi-frequent or even frequent again after next few increments. There are many real-life datasets where new frequent patterns might appear in future increments due to their seasonal behavior, different characteristics, or concept drift.
The \textit{uWSInc} algorithm does not handle these cases. To address these cases, we maintain another set of sequences denoted as promising frequent sequences (\textit{PFS}) after each increment $\Delta DB$ introduced into \textit{DB}. Promising frequent sequences are neither globally frequent nor semi-frequent, but their weighted expected supports satisfy a user-defined support threshold named as \textit{LWES} that is used to find locally frequent patterns in $\Delta DB$  at a particular point. Here, whether a pattern is globally frequent (or semi-frequent) or not is determined by its frequency in the entire database, and whether a pattern is locally frequent or not is determined by its frequency in an increment. Intuitively, it can be assumed that locally frequent patterns may become globally frequent or semi-frequent after the next few increments. The patterns whose \textit{WES} values do not meet the local threshold, \textit{LWES}, are very unlikely to become globally frequent or semi-frequents. Thus maintaining \textit{PFS} may significantly increase the performance of an algorithm in finding the almost complete set of frequent patterns after each increment. To incorporate the concept of promising frequent sequences (\textit{PFS}) in mining patterns,  we propose another approach, called \textit{uWSInc+}, shown in Algorithm  \ref{algo:uWSInc+}.  

\begin{figure}[!tbh] 
    \centering
    \includegraphics[width=0.9\textwidth, height=.45\textwidth]{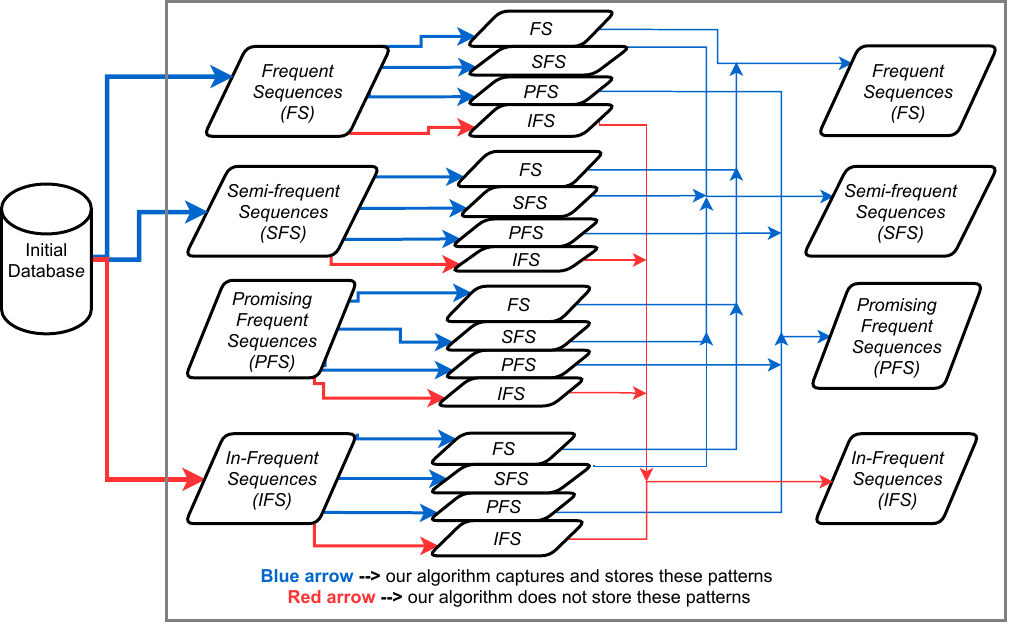}
			\caption{Determination of sequences in our proposed $uWSInc+$ architecture.}
			\label{fig:allcases}
\end{figure}

In Figure~\ref{fig:allcases}, the determination of sequences in our proposed \textit{uWSInc+} architecture has been presented. Each frequent sequence will be either frequent, semi-frequent, promising frequent, or infrequent after each increment. Similarly, one of the four cases will occur to each semi-frequent or promising frequent sequence. $uWSInc+$ stores FS, SFS and PFS into \textit{USeq-Trie} and maintains them for next increments. Nevertheless, when any sequence becomes an infrequent sequence, it will not be stored further. As a result, $uWSInc+$ loses that information. Again, any infrequent sequence can be \textit{PFS} or $SFS$ or $FS$ after several increments. Intuitively, an infrequent sequence will become $PFS$ before being $SFS$ or $FS$ as the size of an increment is usually smaller than that of the whole database. Since $uWSInc+$ stores $PFS$, consequently, it would be able to capture them. Therefore, the maintenance of $PFS$ makes $uWSInc+$ more robust to any concept drifts in incremental databases.

\renewcommand{\algorithmicrequire}{\textbf{Input:}}
\renewcommand{\algorithmicensure}{\textbf{Output:}}
\makeatletter
\renewcommand{\ALG@beginalgorithmic}{\small}
\makeatother
\algrenewcommand{\alglinenumber}{\small}
\makeatletter
\renewcommand{\ALG@beginalgorithmic}{\small}
\makeatother
\algrenewcommand{\alglinenumber}{\small}
\begin{algorithm}[htb] 
\caption{Procedure of \textit{uWSInc+}}
\begin{algorithmic}[1] 
\Statex
\Require \textit{DB:} initial database,  $\Delta {DB}$ : new increments,  \textit{min\_sup:} minimum support threshold,  $\mu$: buffer ratio,  \textit{wgt\_fct:} weight factor
\Ensure \textit{FS:} set of frequent patterns 
\Statex
\Procedure{InitialMining}{$DB$,  $\Delta{DB}_{i}$,  \textit{min\_sup},  $\mu$, \textit{wgt\_fct}}
\State $seqTrie \gets FUWS(DB,  min\_sup \times \mu, \textit{wgt\_fct})$
\State $pfsTrie \gets\ Trie\ to\ store\ PFS\ which\ is\ initialized\ as\ empty$
\ForAll{$\Delta{DB}_{i}$}
\State $DBSize\gets DBSize + \Delta{DB}_{i}Size $
\State $seqTrie$, $pfsTrie$ $\gets$ uWSInc+($\Delta{DB}_{i}$,  $min\_sup$,  $\mu$, \textit{wgt\_fct}, $seqTrie$, $pfsTrie$)
\EndFor
\EndProcedure                                                
\Statex                                                 

\Procedure{uWSInc+}{$\Delta{DB}_{i}$,  $min\_sup$,  $\mu$, \textit{wgt\_fct}, $seqTrie$, $pfsTrie$}
\State $LWES = 2 \times min\_Sup \times \mu \times \Delta{DB}_{i}Size \times WAM^{'} \times wgt\_fct$ \Comment{choice of \textit{LWES} may vary}
\State $lfsTrie \gets FUWS(\Delta{DB}_{i},  2 \times min\_Sup \times \mu, \textit{wgt\_fct})$
\State \textbf{Call} SupCalc($\Delta {DB}_{i}$,  \textit{seqTrie})
\State \textbf{Call} SupCalc($\Delta {DB}_{i}$,  \textit{pfsTrie})
\State $minWES \gets min\_sup \times DBsize \times WAM \times wgt\_fct$

\ForAll{$\alpha \in (FS \lor SFS)$ stored into \textit{seqTrie}}
\If{$wExpSup(\alpha) < (minWES \times \mu $)}
\State Delete pattern $\alpha$ from \textit{seqTrie}
\If{$wExpSup(\alpha) \geq LWES $}
\State Insert pattern $\alpha$ into \textit{pfsTrie}
\EndIf
\EndIf
\EndFor
\ForAll{$\beta \in PFS$ stored into \textit{pfsTrie}}
\If{$wExpSup(\beta) \geq (minWES \times \mu)$}
\State Delete pattern $\beta$ from \textit{pfsTrie}
\State Insert pattern $\beta$ into \textit{seqTrie}
\ElsIf{$wExpSup(\beta) < LWES$}
\State Delete pattern $\beta$ from \textit{pfsTrie}
\EndIf
\EndFor
\ForAll{$\gamma \in LFS$ stored into \textit{lfsTrie}}
\If{$wExpSup(\gamma) \geq (minWES \times \mu)$ }
\State Insert $\gamma$ into \textit{seqTrie}
\ElsIf{$wExpSup(\gamma) \geq LWES)$ }
\State Insert $\gamma$ into \textit{pfsTrie}
\EndIf
\EndFor
\State $FS \gets seqTrie.find\_frequent\_patterns(\textit{minWES})$
\EndProcedure
\end{algorithmic}
\label{algo:uWSInc+}
\end{algorithm}

Similar to \textit{uWSInc},  frequent and semi-frequent sequences generated by running \textit{FUWS} on the initial database are stored as \textit{FS} and \textit{SFS}. For space efficiency,  we store \textit{FS} and \textit{SFS} together into a single \textit{USeq-Trie} instead of using two different \textit{USeq-Trie} structures. In addition,  a different \textit{USeq-Trie}, which is initially empty,  is used to store promising frequent sequences (\textit{PFS}).

After each increment $\Delta DB$,  the steps of the algorithm are as follows:
\begin{enumerate}
    \item Update database size,  \textit{WAM}, \textit{minWES}, and \textit{$minWES^{'}$}.
    \item Run \textit{FUWS} only in $\Delta DB$ to find locally frequent sequences against a local threshold,  \textit{LWES} and store them into a \textit{USeq-Trie}, named as \textit{LFS}. Users can choose \textit{LWES} based on the aspects of application. 
    \item For all $\alpha$ in \textit{FS},  \textit{SFS} and \textit{PFS},  update \textit{WES$_{\alpha}$} by using \textit{SupCalc} method in Algorithm \ref{algo:supcal}.
    \begin{itemize}
        \item if $WES_{\alpha} <LWES$,  delete $\alpha$'s information.
        \item else if $WES_{\alpha} <minWES'$,  move $\alpha$ to \textit{PFS'}. 
        \item else if $WES_{\alpha} < minWES$, move $\alpha$ to \textit{SFS'}.
        \item else move $\alpha$ to \textit{FS'}.
    \end{itemize}
    \item Move each pattern $\alpha$ from \textit{LFS} to \textit{PFS'} or \textit{SFS'} or \textit{FS'} based on \textit{WES$_{\alpha}$}.
    \item Use \textit{FS'},  \textit{SFS'},  and \textit{PFS'} as \textit{FS},  \textit{SFS},  and \textit{PFS} respectively for the next increment.
\end{enumerate}

In Algorithm  \ref{algo:uWSInc+},  \textit{uWSInc+} stores \textit{FS} and \textit{SFS} into \textit{seqTrie} which are generated from running \textit{FUWS} once on the initial databae \textit{DB} in Line 2. It initializes \textit{pfsTrie} as an empty \textit{USeq-Trie} in Line 3. Then for each increment $\Delta {DB}_{i}$,  it runs \textit{FUWS} only in $\Delta{DB}_{i}$ to find locally weighted frequent sequences considering \textit{LWES} and stores them into another \textit{USeq-Trie} named as \textit{lfsTrie} in Line 9. After that,  it calls \textit{SupCalc} function to update the weighted expected support for all patterns ($FS \lor SFS$) stored into \textit{seqTrie} and \textit{PFS} stored into \textit{pfsTrie}. In Line 13-28,  it updates the \textit{seqTrie} and \textit{pfsTrie} according to the updated \textit{WES} of all patterns. Finally,  it finds the frequent sequences \textit{FS} by traversing the \textit{seqTrie}. 

The proposed algorithms $uWSInc$ and $uWSInc+$ never generate any false positive patterns which can be verified from Lemma \ref{lem:6}.
\begin{lemma}
\label{lem:6}
  uWSInc and uWSInc+ algorithms never generate any false-positive sequences. In other words, any generated sequential pattern $\alpha$, that is generated by one of uWSInc algorithm or uWSInc+ algorithm, must be a true-positive pattern.
 \end{lemma}
 \begin{pf}
Assume that a mined weighted frequent sequential pattern, $\alpha$, is a false positive pattern. Then, there could be two sources of this pattern.
\begin{enumerate}
    \item Case 1: The pattern $\alpha$ is mined by $uWSIn$c algorithm
    \item Case 2: The pattern $\alpha$ is mined by $uWSInc+$ algorithm
\end{enumerate}

 As, every mined sequential pattern, $\alpha$, is mined using either $uWSInc$ algorithm or $uWSInc+$ algorithm in the proposed incremental architectures.
 

 According to the proposed system architecture, both $uWSInc$ and $uWSInc+$ approaches apply $FUWS$ algorithm to mine weighted frequent sequences(\textit{FS})  and semi-frequent sequences(\textit{SFS}) from initial database. 

Now, let us assume that a mined weighted frequent sequence, $\alpha$, is a false positive pattern. Then, 
 \begin{enumerate}
     \item Case 1: The sequence $\alpha$ is mined by $uWSInc$ algorithm: 
     
     From Algorithm~\ref{algo:uWSInc}, we can see that the $uWSInc$ process applies the $FUWS$ algorithm to find all the weighted frequent sequence $FS$ and weighted semi-frequent sequence $SFS$ from the initial database.\newline  
     If $\alpha$ is a sequence mined by $uWSInc$, then $\alpha$ must be a sequence from the set of $FS$ or $SFS$. \hfill(1)\newline
     So, a sequence $\alpha$ mined by $FUWS$ algorithm can be a false positive sequence. \hfill(2)\newline
     But, according to Lemma \ref{lem:wescap}, all the sequences mined by $FUWS$ must be true positive. \hfill(3)\newline
    Thus, analyzing the Statements (1), (2), and (3), we can say that it is clearly a contradiction.\newline 
    $\therefore$ A sequence $\alpha$, which is mined by $uWSInc$ algorithm, must be a true positive sequence. And none of the mined sequences by $uWSInc$ algorithm can be a false-positive sequence. \hfill(4)   
 
     \item Case 2: The sequence $\alpha$ is mined by $uWSInc+$ algorithm: 
     
     From Algorithm~\ref{algo:uWSInc+}, we can see that all the sequences generated by the $uWSInc+$ algorithm mined from one of the two processes below:
     \begin{itemize}
         \item Process 1: using the $FUWS$ algorithm to find all the weighted frequent sequence $FS$ and weighted semi-frequent sequence $SFS$ from the initial database
         \item Process 2: measuring the frequent measure for the incremental databases, it finds two sets of sequences as local frequent sequences $LFS$ and promising frequent sequence $PFS$.
     \end{itemize}  
     For Process 1,\newline  
     If $\alpha$ is a sequence mined by $uWSInc+$, then $\alpha$ must be a sequence from the set of $FS$ or $SFS$. \hfill(5)\newline
     So, a sequence $\alpha$ mined by $FUWS$ algorithm can be a false positive sequence. \hfill(6)\newline
     But, according to Lemma \ref{lem:wescap}, all the sequences mined by $FUWS$ must be true positive. \hfill(7)\newline
    Thus, analyzing the Statements (5), (6) and (7), we can say that it is clearly a contradiction, too.\newline 
    
    For Process 2,\newline 
    It can be said that the local frequent measure lacks the global occurrences\newline 
    Thus, all the local measures must be lower than the global measures. So, there is no chance of getting any false-positive sequences. \newline 
    
    $\therefore$ A sequence $\alpha$ which is mined by $uWSInc+$ algorithm, must be a true positive sequence. And none of the mined sequences by the $uWSInc+$ algorithm can be a false-positive sequence. \hfill(8)\newline  

 \end{enumerate}
 $\therefore$ From the Statements (4) and (8), we can conclude that "$uWSInc$ and $uWSInc+$ algorithms never generate any false-positive sequences. In other words, any generated sequence $\alpha$ which is generated by the $uWSInc$ or $uWSInc+$ algorithms, must be a true positive sequence."
 \hfill \boxed{}
 \end{pf}

 Details of an example simulation and result analysis are shown in Section  \ref{sec:simulation} and Section  \ref{sec:results}, respectively.
\subsection{Example Simulation}
\label{sec:simulation}

Let us consider Table  \ref{tab:initDB} as the initial database \textit{DB} and the increments shown in Table \ref{tab:increments}.  For this simulation, set the support threshold  min\_sup=20\%,  buffer ratio  $\mu$=0.7, and wgt\_fct = 1.0. As a result,  the minimum weighted expected support threshold for frequent sequences,  \textit{minWES} = 1.06 and for semi-frequent sequences,  \textit{$minWES^{'} = 0.74$}.

\begin{figure}[!tbh] 
    \centering
    \includegraphics[width=\linewidth]{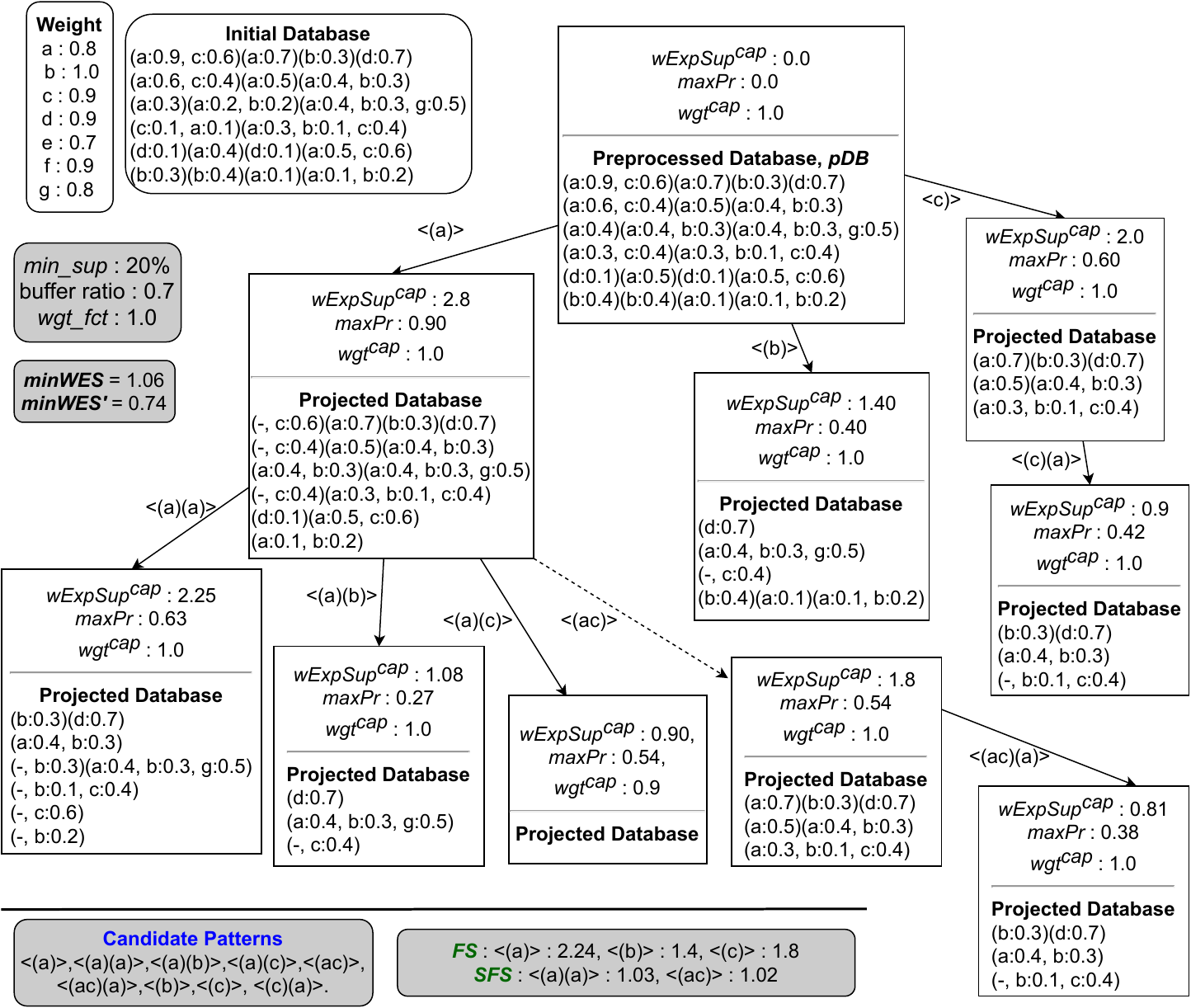}
			\caption{FUWS simulation for Initial Database, DB in Table \ref{tab:initDB}}
			\label{fig:fuws_simulation}
\end{figure}

The detailed simulations of \textit{FUWS} on \textit{DB} in Table  \ref{tab:initDB} are shown in Figure  \ref{fig:fuws_simulation} where we compared the value of $wExpSup^{cap}$ for a sequence with $minWES^{'}$ to find out the candidates of frequent and semi-frequent sequences together. The semi-frequent sequences will be used in our incremental techniques later.

\textit{FUWS} algorithm processes \textit{DB} in the way that has been described in Algorithm \ref{algo:fuws}. The preprocessed database has been shown as \textit{pDB} in Figure \ref{fig:fuws_simulation}. The \textit{FUWS} algorithm uses \textit{pDB} to find the potential candidates. An item is called s-extendable (or i-extendable) when the the value of $wExpSup^{cap}$ for the sequence extension (or itemset extension) of a prefix by that item satisfies the threshold $minWES^{'}$. First of all, it finds extendable items considering the prefix pattern which is empty initially. Thus the resulting super sequences  are $\textless{}(a)\textgreater{} : 2.8, \textless{}(b)\textgreater{} : 1.40, \textless{}(c)\textgreater{} : 2.0$ where the real numbers associated with each item denote the value of $wExpSup^{cap}$ for the respective super patterns. It projects the database recursively while considering the newly generated super pattern as prefix pattern and finds out extendable items following the same process. In Figure \ref{fig:fuws_simulation}, an edge label indicates an extendable item and the edge type indicates the extension type; solid lines are for s-extensions and dashed lines are for i-extensions. 

In this example, \textit{FUWS} algorithm considers $\textless{}(a)\textgreater{}$ as first prefix pattern and it projects \textit{pDB} accordingly. After that,  the algorithm finds i-extendable item $\textless{}(\_c)\textgreater{}$ and s-extendable items for the prefix pattern $\textless{}(a)\textgreater{}$ which are $\textless{}(a)\textgreater{}, \textless{}(b)\textgreater{}$ and $\textless{}(c)\textgreater{}$. Consequently, newly generated super patterns are $\textless{}(a)(a)\textgreater{}:2.25, \textless{}(a)(b)\textgreater{}:1.08, \textless{}(a)(c)\textgreater{}:0.90$ and $\textless{}(ac)\textgreater{}:1.8$. Then it considers each super patterns as prefix pattern individually and tries to find out longer patterns by following the same process recursively.  In this example, the algorithm does not find any extendable items for prefix patterns $\textless{}(a)(a)\textgreater{}, \textless{}(a)(b)\textgreater{}, \textless{}(a)(c)\textgreater{}$, except for $\textless{}(ac)\textgreater{}:1.8$ which has only one super sequence $<(ac)(a)>:0.81$, so then it backtracks to project the database considering $\textless{}(b)\textgreater{}$ as next 1-length prefix pattern. Again, there are no extendable items for $\textless{}(b)\textgreater{}$. It takes $\textless{}(c)\textgreater{}$ as next prefix pattern and finds only one extendable item, which is $\textless{}(a)\textgreater{}$. The algorithm repeats the same process for $\textless{}(c)(a)\textgreater{}$. No extendable items are found for prefix pattern $\textless{}(c)(a)\textgreater{}$. As there are no unexplored patterns, the recursive process terminates. It stores all potential candidate patterns into \textit{USeq-Trie}. 

To remove false positive patterns, the algorithm scans the database \textit{DB} to calculate the actual weighted expected support \textit{WES} for all candidates. Finally,  it finds frequent and semi-frequent sequences by comparing the values of \textit{WES} with \textit{minWES} and \textit{$minWES^{'}$}. The resultant \textit{FS}: $\textless{}(a)\textgreater{}:2.24,\textless{}(b)\textgreater{}:1.4,\textless{}(c)\textgreater{}:1.8$ and \textit{SFS}: $\textless{}(a)(a)\textgreater{}:1.03,  \textless{}(ac)\textgreater{}:1.02$ where the real values correspond to $WES$.

\begin{table}[h]
    \centering
    \begin{tabular}{|c|c|l|}
        \hline
        \textbf{Increment} & \textbf{Id} &      \textbf{Sequence} \\ \hline \hline

        \multirow{4}{*}{$\Delta DB_{1}$}
        &
          7 &  (c:0.6, a:0.7)(a: 0.8)(f:0.9, a:0.6)            \\ \cline{2-3}
        & 8 & (c:0.6, a:0.4)(c:0.8)(a:0.6)(f:0.5)(g:0.4, c:0.7) \\ \cline{2-3}
        & 9 & (f:0.8)(a:0.3)(c:0.9)(d:0.9)(f:0.5, a:0.7, d:0.4)  \\ \cline{2-3}
        & 10 & (c:0.7)(a:0.1)(a:0.8, c:0.6, d:0.8) \\ \hline \hline 
        \multirow{3}{*}{ $\Delta DB_{2}$ }
        &
          11 & (f:0.1)(f:0.3, c:0.7)(a:0.9)(d:0.9)(f:0.2, g:0.1) \\ \cline{2-3}
        & 12 & (a:0.2, c:0.1)(b:0.8)(f:0.4, e:0.4)(g:0.1)(e:0.5, g:0.2)    \\ \cline{2-3}
        & 13 & (c:0.6)(a:0.9)(d:0.6)(e:0.6)(a:0.5, e:0.4, c:0.1)\\ \hline
    \end{tabular}
    \caption{Increments,  $\Delta DB_{i}$}
    \label{tab:increments}
\end{table}

\begin{table}[h]
    \centering
    \begin{tabular}{|c|c|l|c|l|}
    \hline
    \textbf{Inc.} & \multicolumn{2}{|c|}{\textbf{uWSInc}} & \multicolumn{2}{|c|}{\textbf{uWSInc+}} \\ \hline     
    \multirow{7}{*}{ $\Delta DB_{1}$ } & \multirow{5}{*}{FS:}  & $<(a)> : 4.56$ & \multirow{3}{*}{LFS:} & $<(a)>:2.32, <(c)>:2.7, <(d)>:1.53, <(f)>:1.98,$\\ 
    & & $<(a)(a)> : 1.90$ & & $<(a)(f)>:0.99, <(ac)>:0.97, <(c)(a)>:1.83,$ \\          
    & & $<(ac)> : 1.99$  & & $ <(c)(f)>:1.23, <(f)(c)>:0.96$ \\ \cline{4-5}    
    & & $<(c)> : 4.50$ & \multirow{2}{*}{FS:} & $<(a)> : 4.56, <(a)(a)> : 1.9, <(ac)> : 1.99,$\\    
    & & & & $<(c)> : 4.50, <(c)(a)> : 1.83, <(f)> : 1.98$ \\ \cline{2-5}    
    & \multirow{2}{*}{SFS:} & \multirow{2}{*}{$<(b)>:1.4$} & SFS: & $<(c)(d)>:1.23,<(b)>:1.4,<(c)(f)>:1.25,<(d)>:1.53$  \\ \cline{4-5}    
    & & & PFS: & $<(a)(f)>:0.99, <(f)(c)>:0.96$ \\ \hline \hline             
    
    \multirow{8}{*}{ $\Delta DB_{2}$ } & \multirow{5}{*}{FS:}  &   & \multirow{3}{*}{LFS:} &  $<(a)>:1.6, <(a)(d)>:1.15, <(b)>:0.8, <(c)>:1.26,$ \\ 
    & & $<(a)> : 6.16$ & & $<(d)>:1.35, <(c)(d)>:0.89, <(c)(a)>:1.99, $ \\            
     & & $<(a)(a)> : 2.26$ & & $<(c)(a)(d)>:0.77,<(e)>:0.77$ \\ \cline{4-5}    
    & & $<(c)> : 5.76$ & \multirow{2}{*}{FS:} & $<(a)>:6.16, <(a)(a)>:2.26, <(c)>:5.76,$ \\    
    & &  &  & $<(c)(a)>:2.82, <(d)>:2.88, <(f)>:2.61$ \\ \cline{2-5}    
    & \multirow{3}{*}{SFS:} & $<(ac)> : 2.05$ & SFS: &  $<(ac)>:2.05, <(b)>:2.2, <(c)(d)>:2.12$ \\ \cline{4-5}    
    &  & $<(b)> : 2.20$ & \multirow{2}{*}{PFS:}  & $<(a)(d)>:1.15, <(c)(f)>:1.41, <(a)(f)>:1.22,$\\     
    & & &  & $<(e)>:0.77, <(f)(c)>:1.03, <(c)(a)(d)>:0.77$ \\ \hline 

    \end{tabular}%
    \caption{Simulation for Increments,  $\Delta DB_{i}$}
    \label{tab:increments_simulation}
\end{table}

After the first increment,  $\Delta DB_{1}$,  the updated value of \textit{minWES}
is 1.74. The \textit{uWSInc} algorithm scans the $\Delta DB_{1}$ and updates the weighted expected support values for patterns in \textit{FS} and \textit{SFS} which are found in the initial \textit{DB}. As a result,  \textit{FS} and \textit{SFS} are updated as shown in Table  \ref{tab:increments_simulation}.
The second approach,  \textit{uWSInc+} runs \textit{FUWS} and finds the locally frequent set, \textit{LFS},  for
$\Delta DB_{1}$ using 0.96 as \textit{LWES}. Users can set different local threshold \textit{LWES} based on the size and nature of increments,  distribution of items,  etc. In this simulation, let us assume that $LWES = 2\times min\_sup \times |\Delta DB_{i}| \times WAM \times \mu \times wgt\_fct $ for $\Delta DB_{i}$. By scanning $\Delta DB_{1}$,  it updates the WES for \textit{FS},  and \textit{SFS} which are found in initial database \textit{DB}. After that \textit{FS},  \textit{SFS} and \textit{PFS} have been updated according to the updated \textit{minWES} and \textit{LWES}. The results are shown in Table  \ref{tab:increments_simulation}. From Table  \ref{tab:increments_simulation},  we can see that new pattern \textless{}(c)(a)\textgreater{} and \textless{}(f)\textgreater{} appear in \textit{FS} and $<(b)>$, \textless{}(d)\textgreater{}, \textless{}(c)(d)\textgreater{} and \textless{}(c)(f)\textgreater{} appear in \textit{SFS} of \textit{uWSInc+} but not in \textit{uWSInc}. These patterns \textless{}(d)\textgreater{}, \textless{}(c)(d)\textgreater{} and \textless{}(c)(f)\textgreater{} might be frequent later after few increments. The \textit{uWSInc+} might be able to find them which \textit{uWSInc} could never do.

Similarly for the second increment $\Delta DB_{2}$,  the \textit{uWSInc} and \textit{uWSInc+} algorithms use \textit{FS},  \textit{SFS},  and \textit{PFS} which are updated after first increment and follow the same process to generate updated \textit{FS},  \textit{SFS},  and \textit{PFS}.  The results are shown in Table  \ref{tab:increments_simulation}. Finally, we can see that three patterns  \textless{}(c)(a)\textgreater{}, \textless{}(d)\textgreater{} and \textless{}(f)\textgreater{} have become frequent after this increment  which are found by \textit{uWSInc+} but not \textit{uWSInc}. This makes the difference between our two approaches clear as \textit{uWSInc+} can find them but \textit{uWSInc} cannot.

\subsection{Analysis of Time and Space Complexity}
Before going to the experimental performance evaluation in Section \ref{results}, it is necessary to discuss the runtime and memory complexity of our proposed algorithms. We use Big-O-notation to denote the upper bound of complexity while considering that each computer operation takes approximately constant time. Here, throughout this section, M indicates the total number of sequences in the given dataset. Similarly, N - the number of nodes in \textit{USeq-Trie}, L - the maximum length of a data sequence, D - the maximum depth of \textit{USeq-Trie} (aka maximum length of candidate patterns), and S - the maximum size of an extendable itemset.  
\begin{itemize}

    \item SupCalc (in Algorithm  \ref{algo:supcal}) - This function scans the given dataset sequence by sequence and calls \textit{TrieTraverse} function for each data sequence to calculate the support of all candidate patterns where \textit{TrieTraverse} function takes $O(N \times L)$. Therefore,  overall runtime complexity will be $O(M \times N \times L)$. Moreover, in the worst possible cases,  memory complexity will be $O(D \times L)$ due to the depth-first traversal on \textit{USeq-Trie}. 

    \item FUWS (in Algorithm  \ref{algo:fuws}) - To process the whole data-set by using \textit{preProcess} function, it may require $O(M \times L) $ units of time and $O(M \times L)$ units of memory space. Then \textit{Determine} function may take $O(M \times L)$ time along with $O(M \times S)$ unit memory. Moreover, the execution tree of the recursive process \textit{FUWSP} could expand exhaustively, though any branch could be pruned out. Thus in the worst possible case, the runtime will be $O(S^{D} \times M \times L) $ to find potential candidate patterns from an uncertain database.
    \item uWSInc (in Algorithm  \ref{algo:uWSInc}) - This algorithm will take $O(N \times M \times L)$ to find updated set of \textit{FS} and \textit{SFS} after updating their support using \textit{SupCalc} function for each increment into the dataset.
    \item uWSInc+ (in Algorithm  \ref{algo:uWSInc+}) - Since it runs \textit{FUWS} on each increment $\Delta DB_{i}$ to find \textit{LFS}, the first part of its complexity will be same as the runtime complexity of \textit{FUWS} algorithm. Its complexity mostly depends on the complexity of \textit{FUWS}. Besides that, the complexity for the rest of this algorithm depends on the complexity of \textit{SupCalc} algorithm to update the weighted support of \textit{FS},  \textit{SFS}, and \textit{PFS}, which are generated from previous increments along with to update the \textit{FS},  \textit{SFS}, and \textit{PFS} after each increment. Approximately, it will be $O(N \times M \times L)$.
\end{itemize} 

The above complexity analysis shows that our proposed algorithms can find weighted frequent sequences from both static and incremental datasets efficiently with respect to time and memory. Extensive experimental performance analysis on different real-life datasets is provided in the following section to validate this claim. 

\section{Performance Evaluation} 
\label{sec:results}

\label{results}

To evaluate our algorithms and show the effect of weights and uncertainty in data mining, we needed real-life standard datasets with noise and probability. 
Unfortunately none of the datasets given in data mining repositories such as SPMF\footnote{\label{spfmlink}\url{http://www.philippe-fournier-viger.com/spmf/index.php?link=datasets.php}}{\fontsize{7pt}{6pt}\selectfont} and FIMI\footnote{\url{http://fimi.uantwerpen.be/data/}}{\fontsize{7pt}{6pt}\selectfont} are uncertain.
Hence, we have used the general sequence datasets after assigning weights and probabilities by using different distributions to demonstrate and analyze the performance of our proposed solutions. 

\textbf{Datasets.}
Among the datasets that we have used, \textit{Retail}, \textit{Foodmart}, \textit{OnlineRetail}, and \textit{Chainstore} are market-basket data; 
\textit{Kosarak}, \textit{FIFA}, and \textit{MSNBC} are click stream data;
\textit{Sign} contains sign language utterance;
\textit{Accident} contains traffic accident data 
and \textit{Leviathan} is a dataset of word sequences converted from a famous novel, Leviathan.  
However, the \textit{Retail}, \textit{OnlineRetail}, \textit{Foodmart}, and \textit{Chainstore} datasets are given in itemset format. So, we have converted them into \textit{SPMF sequential format} where each \textit{transaction} is a single \textit{sequence} and its each individual \textit{item} is considered to be a single \textit{event}.
A short description of all the tested datasets is given in Table \ref{tab:datasets}.

\begin{table}[bth]
\centering
\begin{tabular}{|p{.17\textwidth}|p{.135\textwidth}|p{.11\textwidth}|p{.11\textwidth}|p{.3\textwidth}|}
  \hline 
  
\rule{0pt}{2ex} \textbf{Dataset} & \textbf{Total \newline Sequences} & \textbf{Average\newline Lenth} & \textbf{Distinct Items}& \textbf{Remarks}\\
\hline

\rule{0pt}{2ex} Retail
& 88,163 & 11.306 & 16,471  & Customar transcations data of 5 consecutive months\\ \hline
 
\rule{0pt}{2ex} Kosarak
& 990,000  & 8.1 & 41,270  & Click-stream Data from a news portal\\ \hline
   
\rule{0pt}{2ex} Accidents
& 340,184 & 34.808 & 469  & Traffic-accident data for the period 1991-2000  \\ \hline

\rule{0pt}{2ex} Chainstore
& 1,112,949 & 7.2 & 46,086  & Customer transaction data where incremental mining can be greatly effective \\ \hline

\rule{0pt}{2ex} Foodmart
& 4,141 & 4.424 & 1,559  & Market-basket data with huge variety of items \\ \hline

\rule{0pt}{2ex} OnlineRetail
& 541,909 & 4.37 & 2,603  & A sparse market-basket dataset\\ \hline

\rule{0pt}{2ex} Leviathan
& 5,834 & 26.34 & 9,025  & Conversion of the novel \textit{Leviathan} into word sequences \\ \hline


\rule{0pt}{2ex} FIFA
& 26,198 & 34.74 & 2,990  & click stream data from \textit{FIFA World Cup 98} website\\ \hline

\rule{0pt}{2ex} Sign
& 730 & 51.997 & 267  & Sign language utterance \\ \hline

\rule{0pt}{2ex} MSNBC
& 31,790  & 13.33 & 17  & Click-stream data from news website \\ \hline

\end{tabular}
\caption{\textnormal{ Dataset Description}}
\label{tab:datasets}
 \end{table}

\begin{figure}[!bth]
 \centering
    \includegraphics[width=.6\linewidth,height=.35\linewidth]{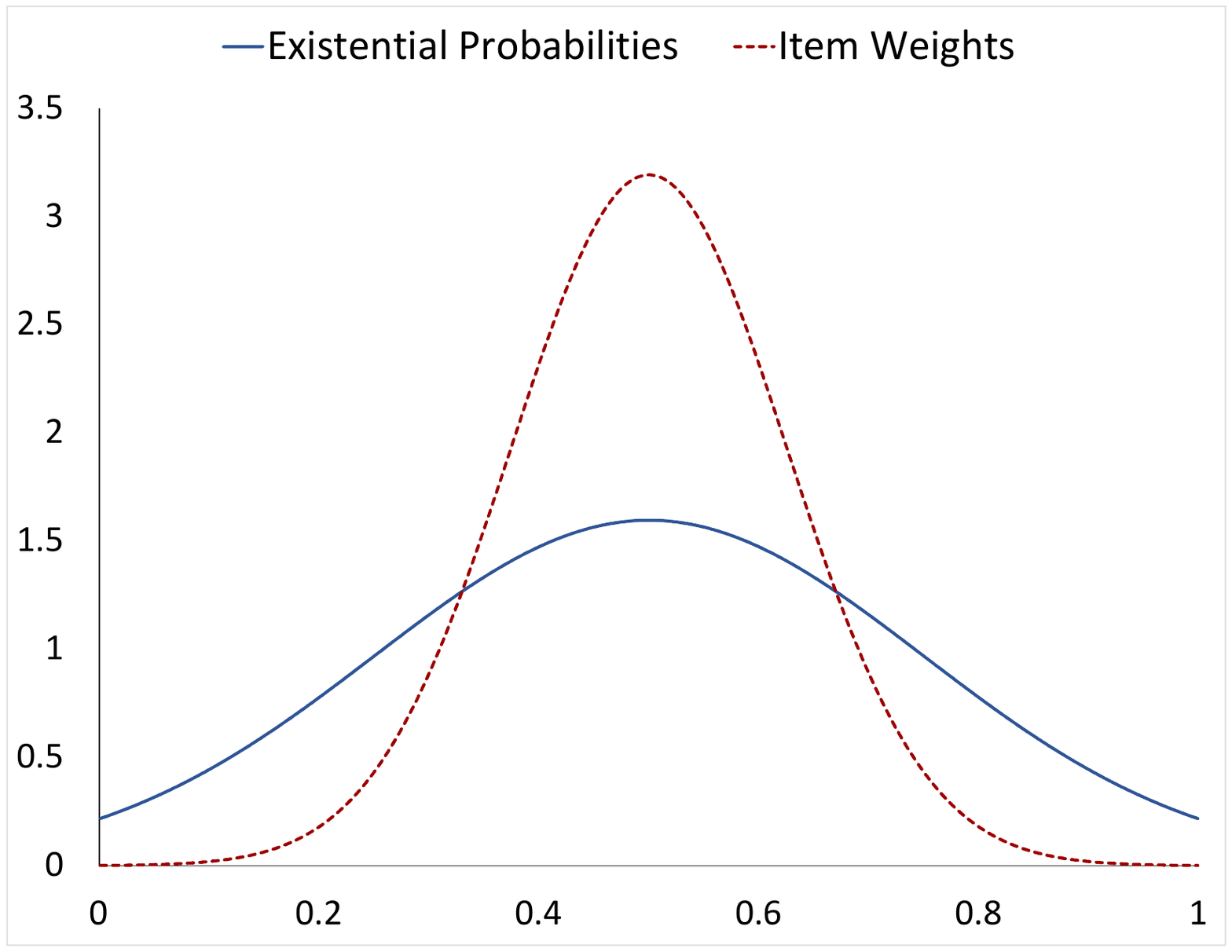}
    \caption{Distribution of Probability Values and Item Weights}
    \label{fig:dist_curve}
\end{figure}

\textbf{Assignment of Probability and Weight. }
We have used a normal distribution to assign probability in the existing popular real-life datasets to reflect the nature of uncertainty. 
The normal distribution is more prominent in statistics and is widely used in the field of data mining as it fits many natural phenomena. 
We have assigned the existential probability to each item using a gaussian distribution with mean $\mu$ = 0.5 and standard deviation $\sigma$ = 0.25 for our general experimental purpose. Moreover, the weight for an item has been assigned using a gaussian distribution with $\mu$ = 0.5 and $\sigma$ = 0.125 as a general-purpose. Figure \ref{fig:dist_curve} shows the distributions for probability and weight values in the general setting of our experiments.

Besides this general processing, we have tested the performance of our proposed algorithm with respect to different distributions of probability and weight values. We have also analyzed the results by changing different parameters to verify the correctness and efficiency of our algorithms. All of these analyses are discussed in the following sections.
 Section~\ref{initial_phase} demonstrates the performance of our static algorithm, \textit{FUWS}, for mining \textit{weighted/unweighted sequential patterns} from uncertain databases. Section~\ref{incremental_phase} shows the efficiency of our proposed techniques, \textit{uWSInc} and \textit{uWSInc+}, for the incremental mining.
We have implemented our algorithms using \textit{Python} programming language and a machine that has Core™ i5-9600U 2.90GHz CPU with 8GB RAM.  

\subsection{Performance of  Uncertain Sequential Pattern Miner, \textit{FUWS}}
\label{initial_phase}
Here we provide the experimental results that show the performance of \textit{FUWS}. We have compared the performance with the existing algorithm, \textit{uWSequence} \cite{rahman2019mining_uWSeq}, which is discussed in Section~\ref{subsec:uncertain_mining}. \textit{uWSequence} proposed a framework where the definition of \textit{weighted sequential pattern} is different from our proposed definition that is inspired by the widely accepted concept of \textit{weighted support} in the literature of weighted pattern mining from precise datasets.
While \textit{uWSequence} is the current best algorithm of mining \textit{weighted sequential patterns} from uncertain databases, 
the authors also showed that it outperforms the existing methods for mining \textit{sequential patterns} in {uncertain databases} without weight constraint. 
Hence, it is sufficient to compare the performance with \textit{uWSequence} to show the efficiency of our proposed \textit{FUWS}.
We have set the weights of all items to 1.0, which brings both algorithms under a unifying framework as the definition of the \textit{weighted sequential pattern} differs. 

\textbf{Evaluation Criteria. }
We have considered two main criteria to evaluate the performance of \textit{uWSequence} and \textit{FUWS} for the same support threshold and same assignment of probability values in a dataset.
\begin{enumerate}[label=(\alph*)]
    \item \textbf{Total number of candidates generated. } Recall that both \textit{FUWS} and \textit{uWSequence}, like \textit{PrefixSpan}, use different upper bounds for actual expected support value. For this reason, both of them generate some \textit{false positive candidates} which get removed when their actual expected support values are measured through an extra scan of the database. The performance of a mining algorithm should be called superior if it generates fewer candidates than other state-of-art algorithms and finds an equal number of actual patterns.
    \item \textbf{Total running time required. } Runtime is an established criterion to evaluate the efficiency of pattern mining algorithms. The less the total required time, the more efficient the algorithm is. We conduct experiments on several real-life datasets to demonstrate the performance of \textit{FUWS} compared to the current best state-of-art algorithm \textit{uWSequence}.
\end{enumerate}

Experiments on this section are conducted by varying the following parameters:
\begin{enumerate}[label=(\alph*)]
    \item \textbf{Minimum support threshold. } The minimum support threshold, \textit{min\_sup\%}, is a value between 0.0 to 1.0, which is used to calculate the minimum (weighted) expected support threshold for determining the (weighted) sequential patterns. Experiments are conducted for different \textit{min\_sup\%} for each dataset to show the effectiveness of \textit{FUWS} at any threshold based on the appl ication requirements and discussed elaborately in Sections~\ref{subsub:min_sup} and~\ref{subsub:min_sup_rt}.
    \item \textbf{Probability distribution. } As the item existential probability values are assigned by ourselves, we have run both the algorithms several times with different probability distributions and discussed the experimental results in Section~\ref{subsub:prob_dis}.
    \item \textbf{Choice of weight factor. } The parameter \textit{wgt\_fct} has been introduced to tune the mining of weighted patterns. The concept is to get patterns with more weighted expected support by setting this factor to a higher value. We have run our algorithm on several datasets with the same support threshold, same probability, and weight assignment but different weight factors to demonstrate this feature. The results are shown in Section~\ref{subsub:wgt_fct}.
\end{enumerate}

\begin{figure}[!htb]
    \centering 
    \begin{subfigure}{0.45\linewidth}
      \includegraphics[width=\linewidth, height = .6\linewidth]{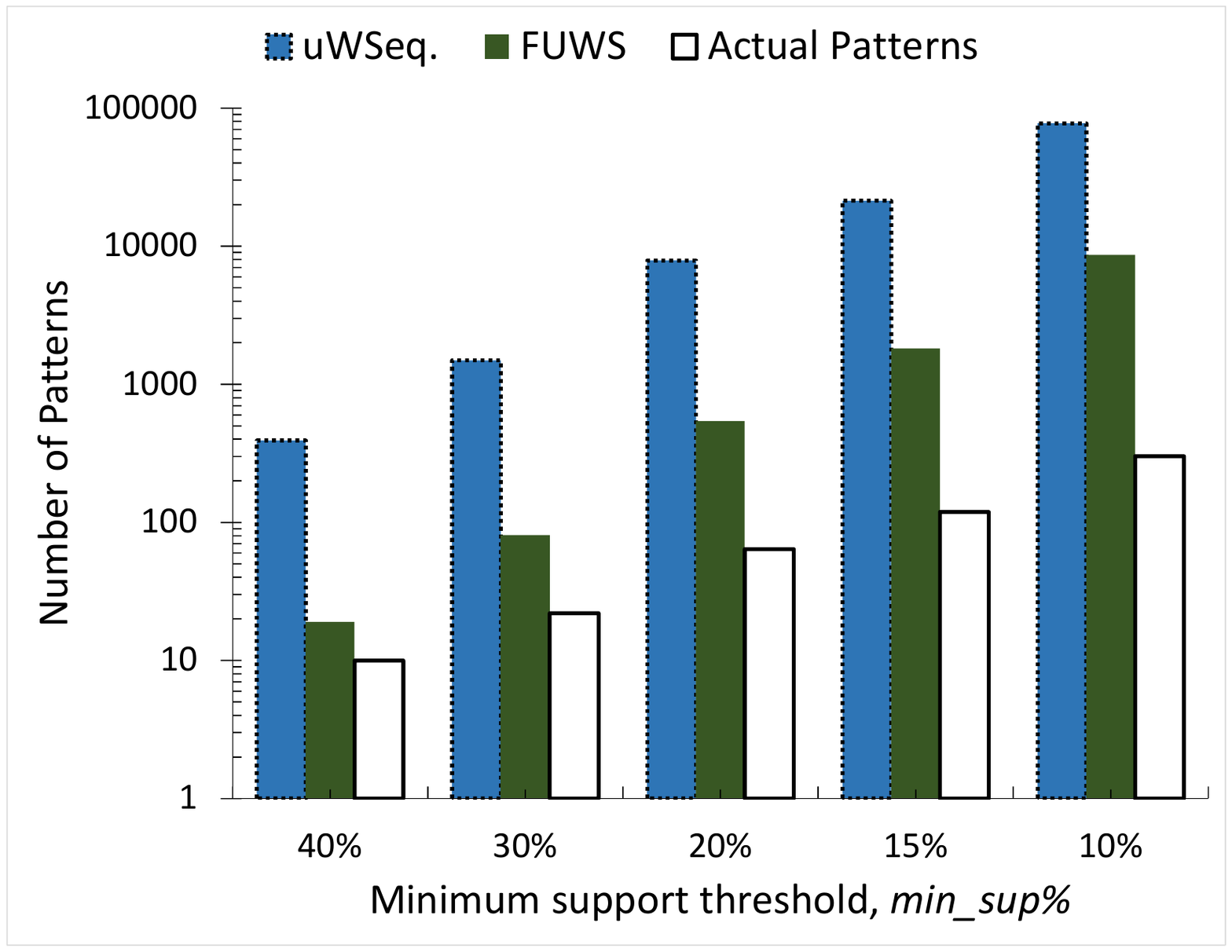}
      \caption{Number of candidate patterns in \textit{Sign} dataset}
      \label{sign_cand}
    \end{subfigure}\hfil 
    \begin{subfigure}{0.45\linewidth}
      \includegraphics[width=\linewidth, height = .6\linewidth]{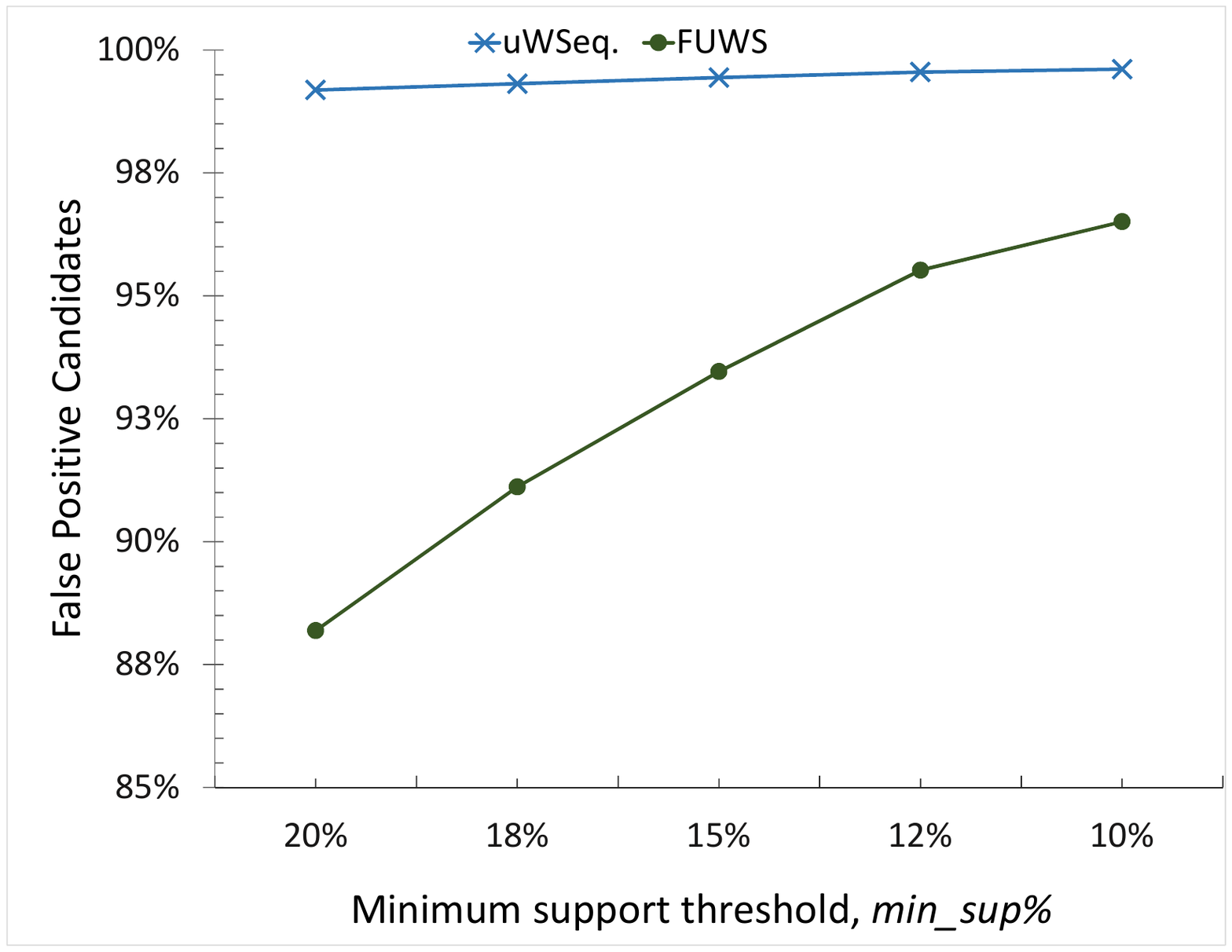}
      \caption{\% of false positive candidates in \textit{Sign} dataset}
      \label{sign_false}
    \end{subfigure}
    \medskip
    \begin{subfigure}{0.45\linewidth}
      \includegraphics[width=\linewidth, height = .6\linewidth]{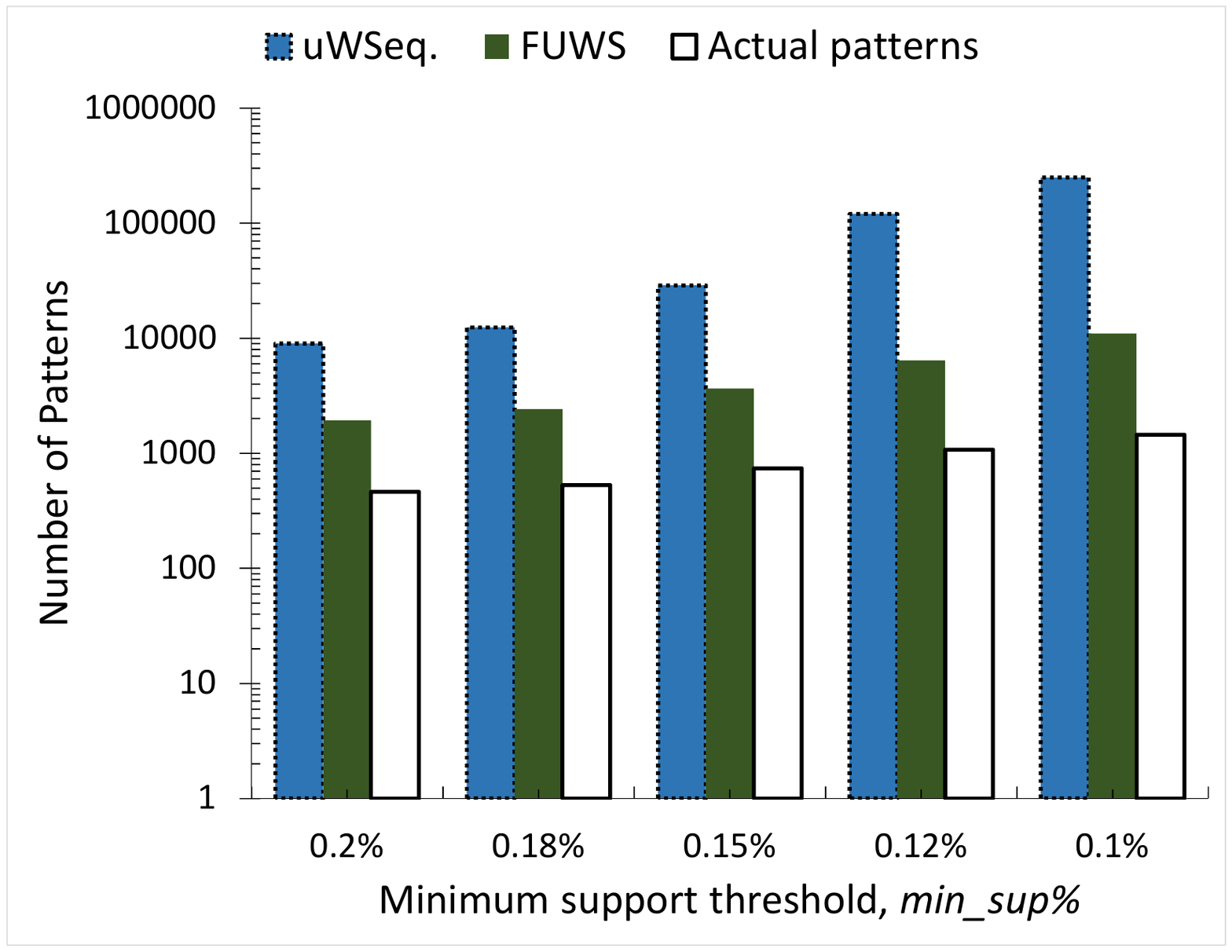}
      \caption{Number of candidate patterns in \textit{Kosarak} dataset}
      \label{kos_cand}
    \end{subfigure}
    \hfil 
    \begin{subfigure}{0.45\textwidth}
      \includegraphics[width=\linewidth, height = .6\linewidth]{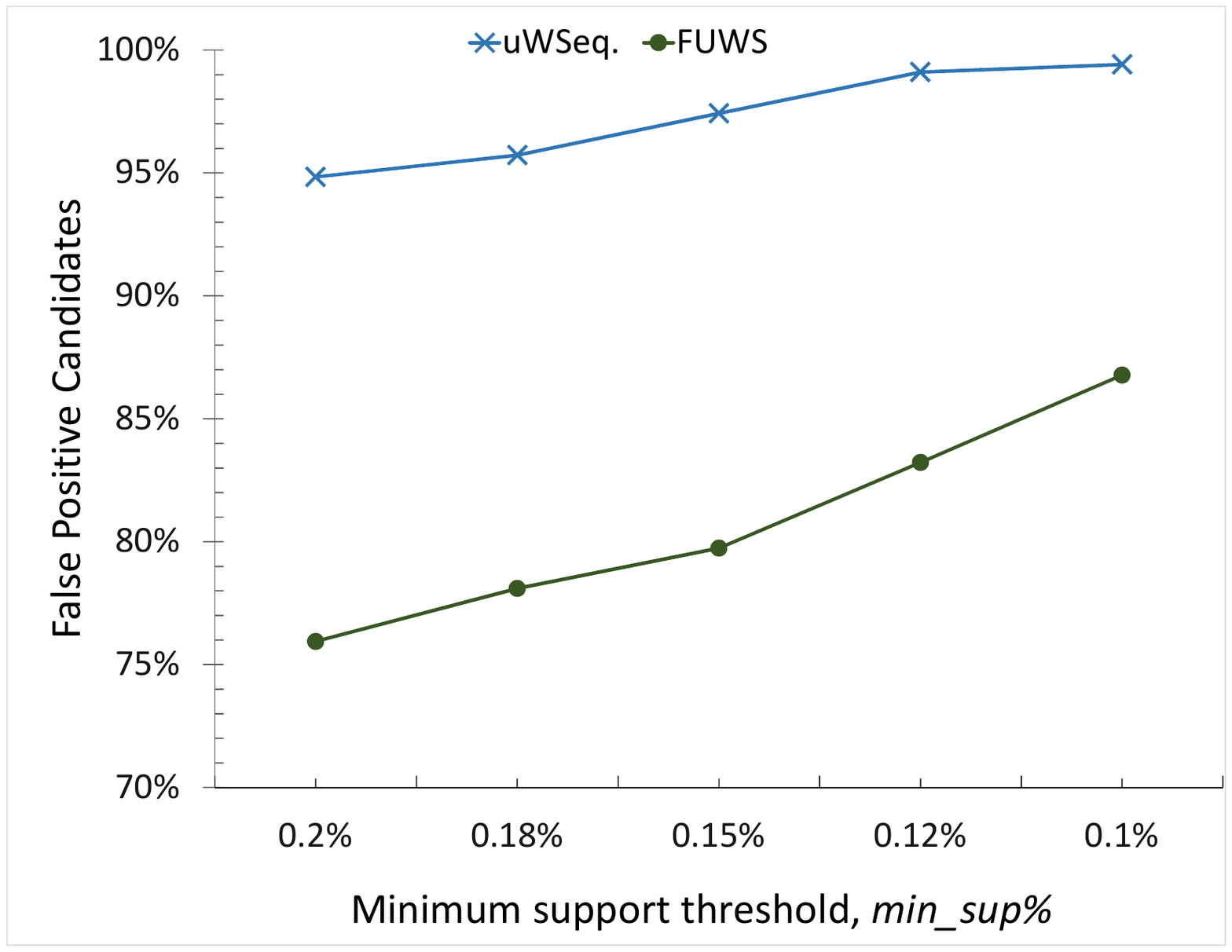}
      \caption{\% of false positive candidates in \textit{Kosarak} dataset}
      \label{kos_false}
    \end{subfigure}

\caption{Comparison of candidate generation between \textit{FUWS} and \textit{uWSequence} in \textit{Sign} and \textit{Kosarak} datasets}
\label{false_comp1}
\end{figure}

\begin{figure}[!bth]
    \centering 
    \begin{subfigure}{0.45\textwidth}
      \includegraphics[width=\linewidth, height = .6\linewidth]{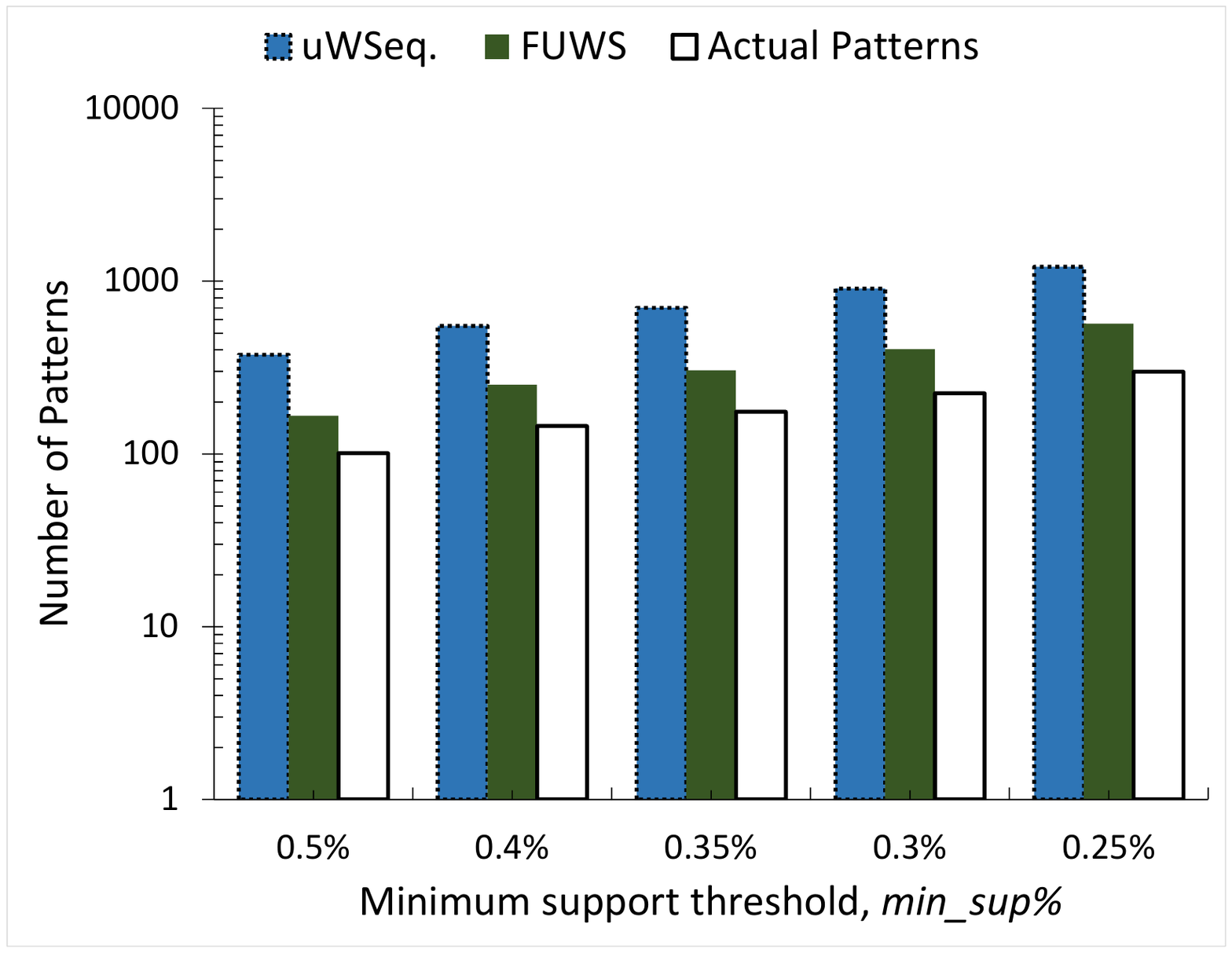}
      \caption{\textit{Retail} dataset}
      \label{retail_cand}
    \end{subfigure}
    \hfil 
    \begin{subfigure}{0.45\textwidth}
      \includegraphics[width=\linewidth, height = .6\linewidth]{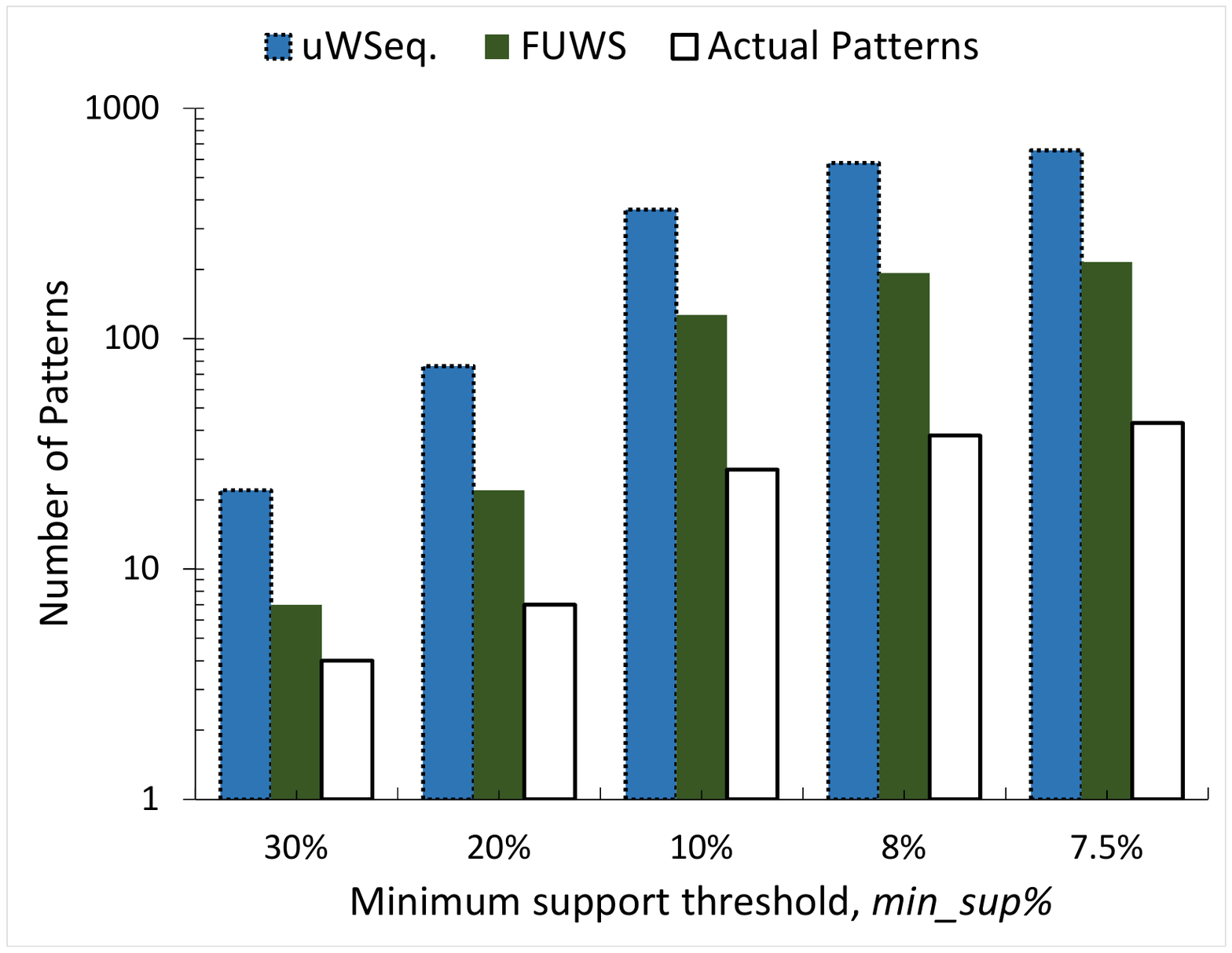}
      \caption{\textit{MSNBC} dataset}
      \label{msnbc_cand}
    \end{subfigure}
    \medskip
    \begin{subfigure}{0.45\textwidth}
      \includegraphics[width=\linewidth, height = .6\linewidth]{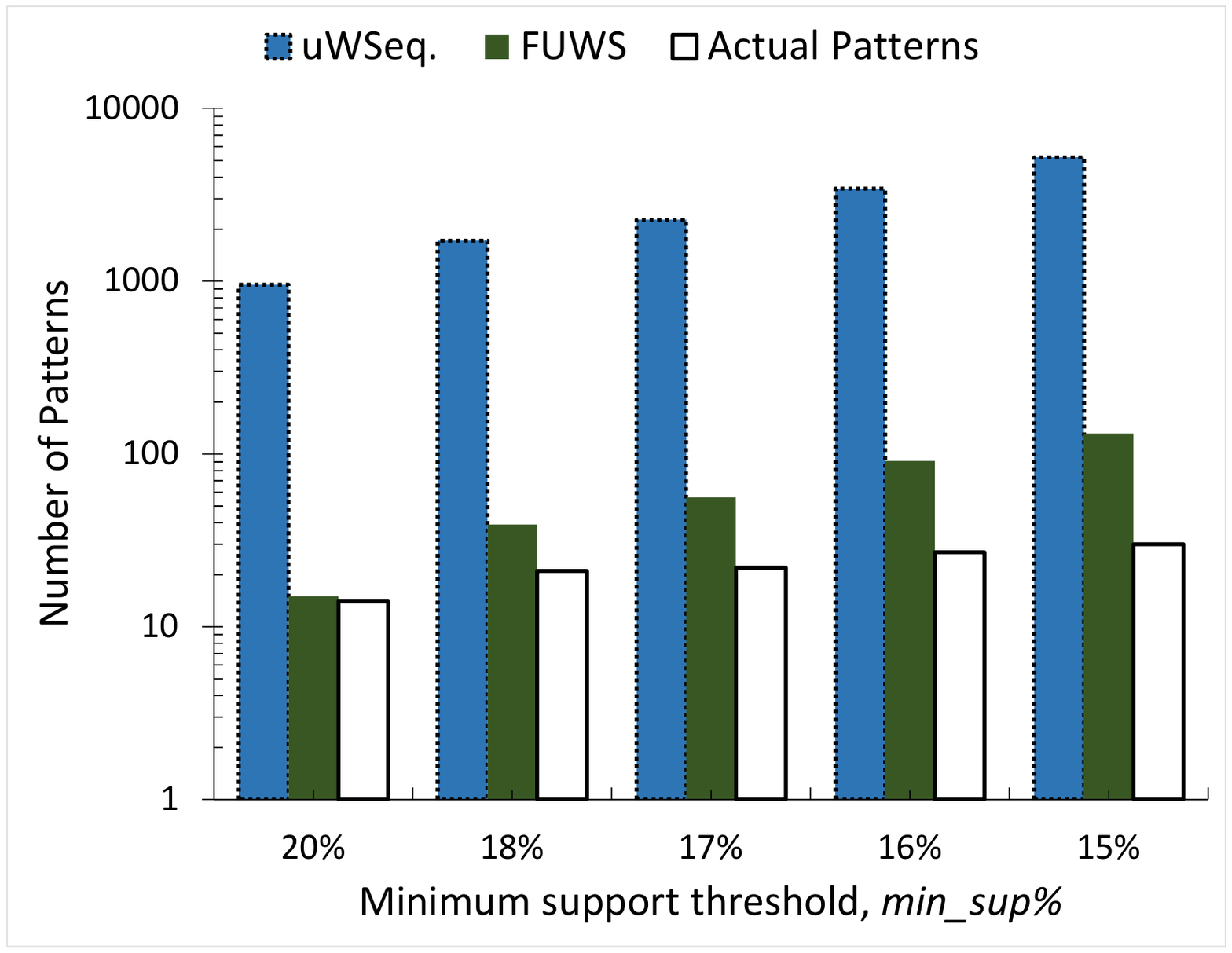}
      \caption{\textit{FIFA} dataset}
      \label{fifa_cand}
    \end{subfigure}\hfil 
    \begin{subfigure}{0.45\textwidth}
      \includegraphics[width=\linewidth, height = .6\linewidth]{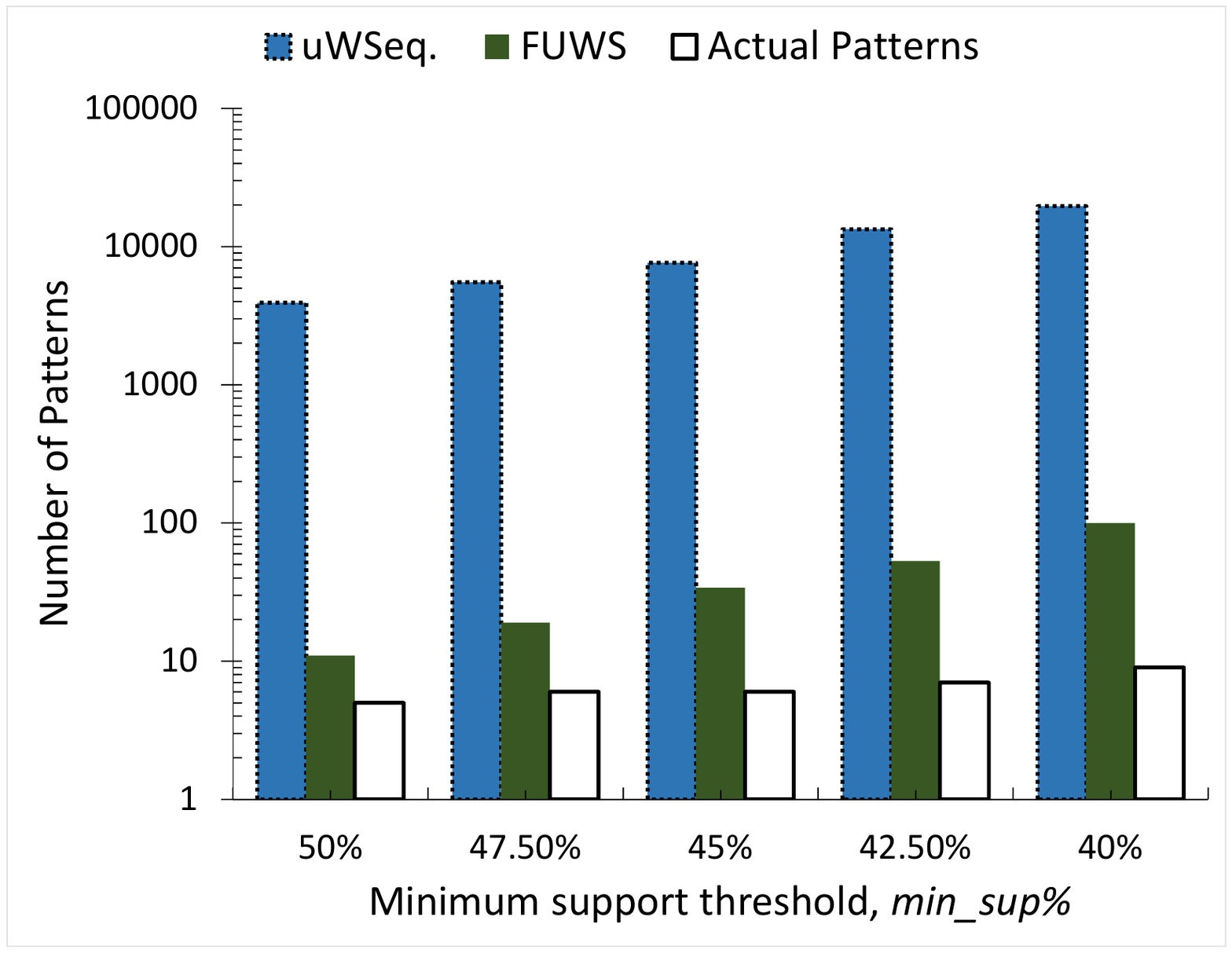}
      \caption{\textit{Accident} dataset}
      \label{acc_cand}
    \end{subfigure}
\caption{Comparison of candidate generation between \textit{FUWS} and \textit{uWSequence} in \textit{Retail, MSNBC, FIFA,} and \textit{Accident}}
\label{cand_comp1}
\end{figure}
\subsubsection{Comparison of Candidate Generation with \textit{uWSequence}}
\label{subsub:min_sup}

\textit{FUWS} uses $expSup^{cap}$ as the upper bound measure which is theoretically tighter than the $expSupport^{top}$ used in \textit{uWSequence}.
As a result, it generates fewer false positive candidates (Lemma \ref{lem:03_capLessTop}).
Figure  \ref{sign_cand} shows the comparison of candidate generation between \textit{FUWS} and \textit{uWSequence} in \textit{Sign} dataset for different support thresholds.
We have plotted the count of sequences in the y-axis on a logarithmic scale for a better graphical representation.
As we can see, \textit{FUWS} generates 542 candidates, where \textit{uWSequence} generates 7874 candidates when the support threshold is 20\%. However, only 64 of them are found to be weighted frequent after calculating their \textit{actual weighted expected support} (using the efficient \textit{SupCalc method}). 

Thus, \textit{FUWS} generates 88.19\% false positive candidates whereas this number is 99.18\% for \textit{uWSequence} which is shown in Figure \ref{sign_false}. 
The difference is about 11 percentage points in this case.
We have observed that this difference gets lower when the support threshold decreases.
Both \textit{FUWS} and \textit{uWSequence} generate more false positive candidates for a smaller support threshold.

Similarly, comparison of candidate generation in other datasets such as \textit{Kosarak}, \textit{Retail}, \textit{MSNBC}, \textit{FIFA} and \textit{Accident} are shown in Figures \ref{kos_cand}, \ref{retail_cand}, \ref{msnbc_cand}, \ref{fifa_cand}, and \ref{acc_cand}.
The difference between \textit{FUWS} and \textit{uWSequence} is huge in the \textit{Accident} dataset. 
We could not run the \textit{uWSequence} algorithm for lower thresholds on this dataset within our 8GB memory capacity. We were able only to find the results with a support threshold not less than 40\%.
From the results in all datasets, we can conclude that a strict upper bound of weighted expected support calculation leads to a significant reduction in candidate generation in the mining process. However, the ratio between the number of candidate patterns and frequent patterns increases with the decrease in the support threshold. Nonetheless, the ratio for \textit{FUWS} is much lower than \textit{uWSequence} for all datasets. 

\begin{figure}[!thb]
    \centering 
    \begin{subfigure}{0.45\linewidth}
      \includegraphics[width=\linewidth, height = .6\linewidth]{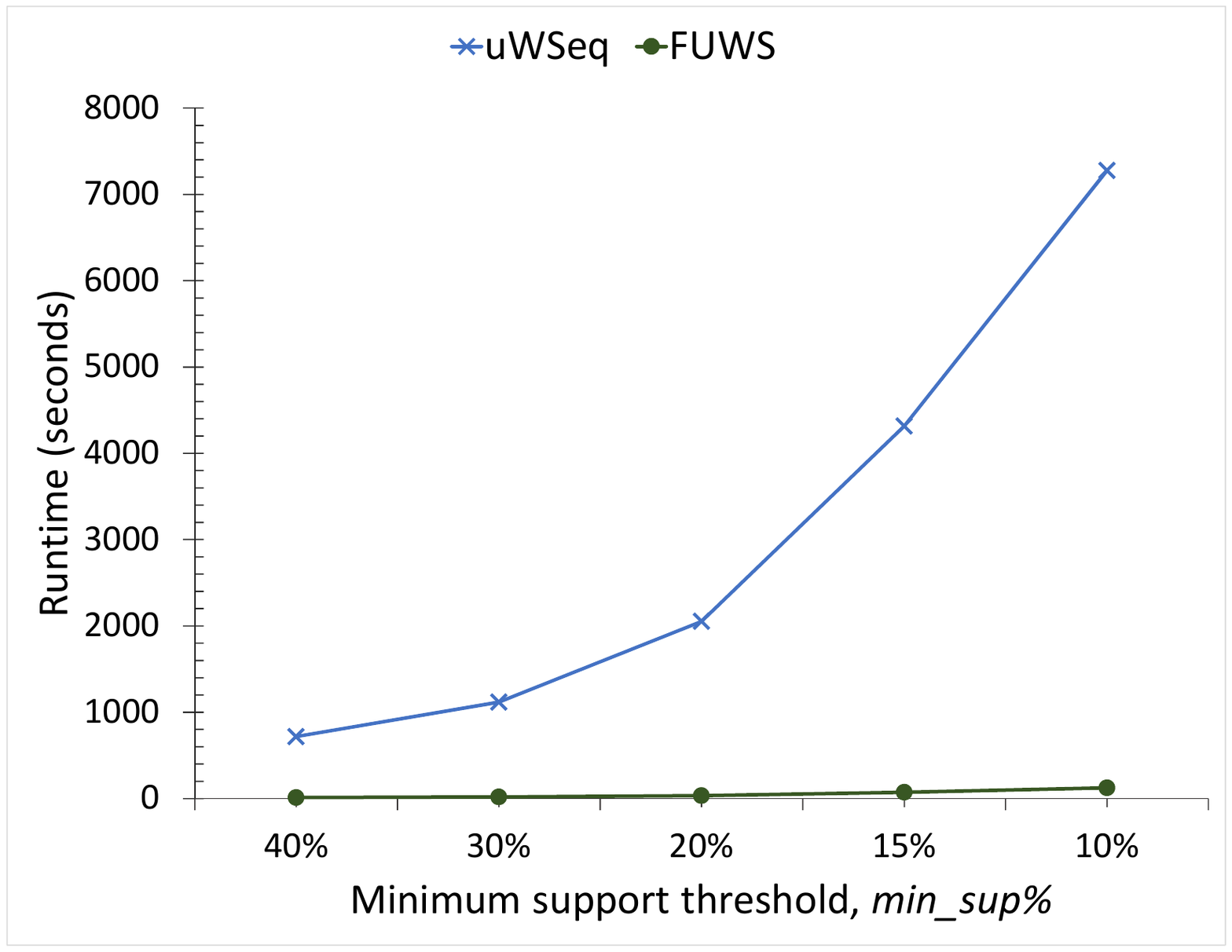}
      \caption{\textit{Sign} dataset}
      \label{sign_rt}
    \end{subfigure}\hfil 
    \begin{subfigure}{0.45\linewidth}
      \includegraphics[width=\linewidth, height = .6\linewidth]{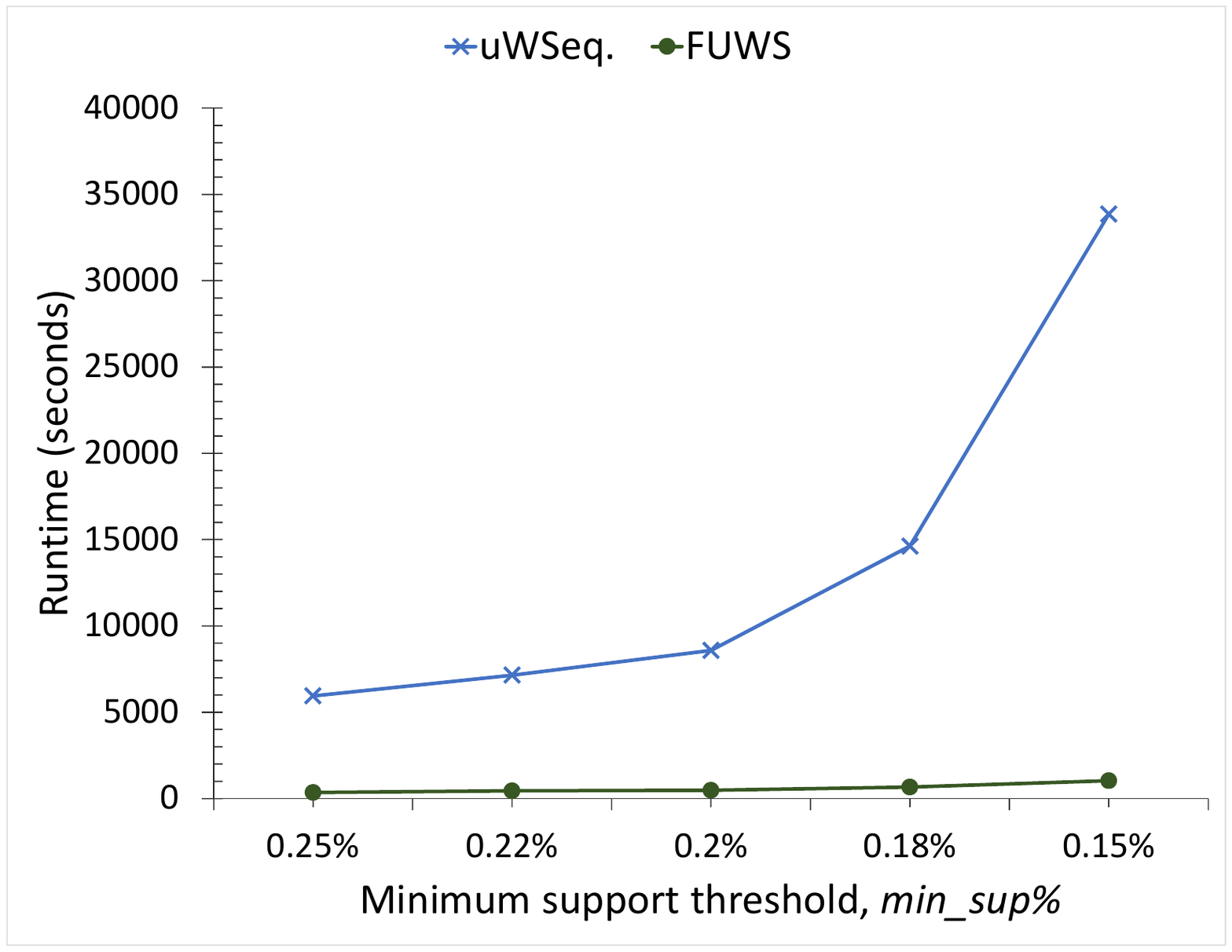}
      \caption{\textit{Kosarak} dataset}
      \label{kos_rt}
    \end{subfigure}
    \medskip
    \begin{subfigure}{0.45\linewidth}
      \includegraphics[width=\linewidth, height = .6\linewidth]{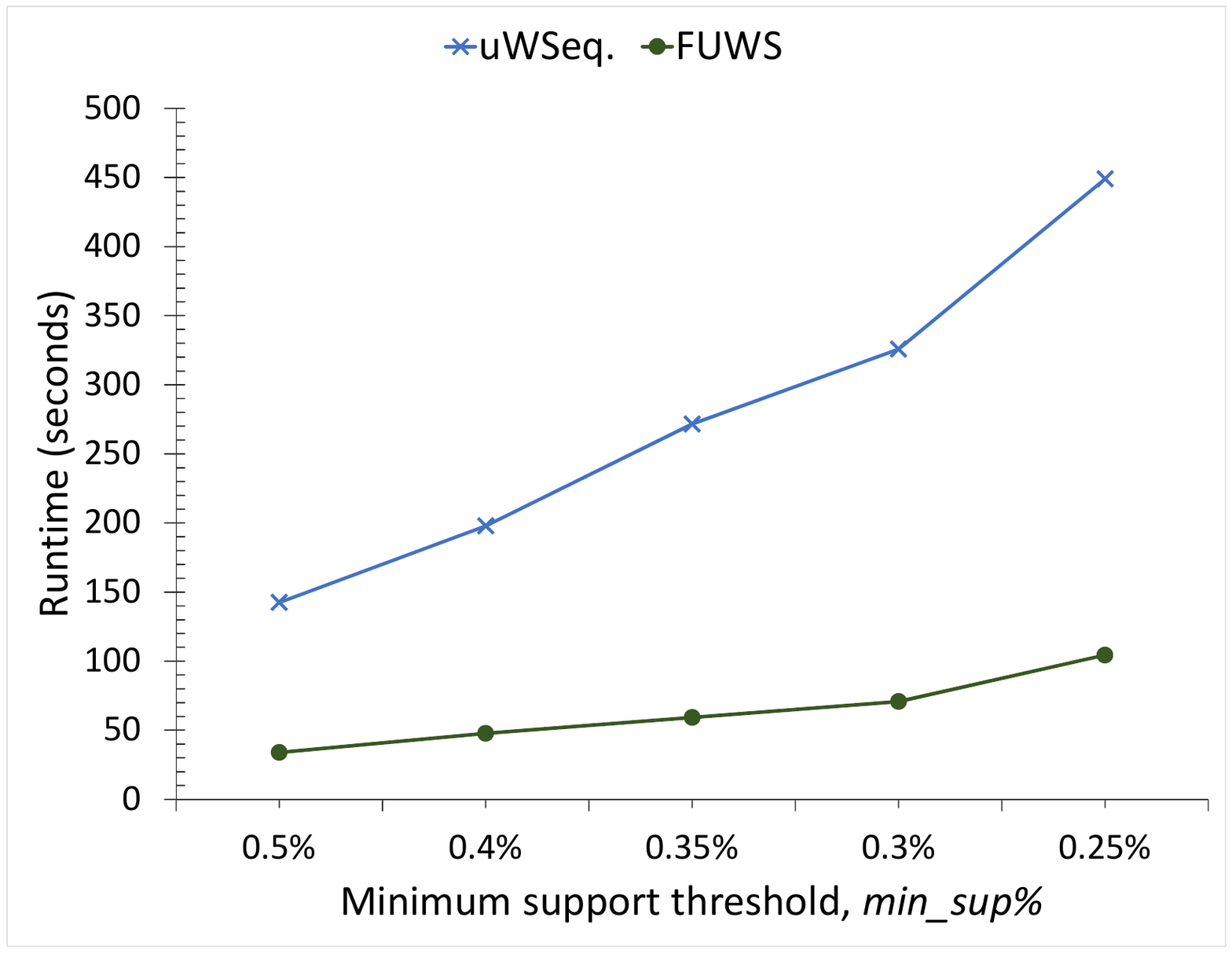}
      \caption{\textit{Retail} dataset}
      \label{ret_rt}
    \end{subfigure}
    \hfil 
    \begin{subfigure}{0.45\textwidth}
      \includegraphics[width=\linewidth, height = .6\linewidth]{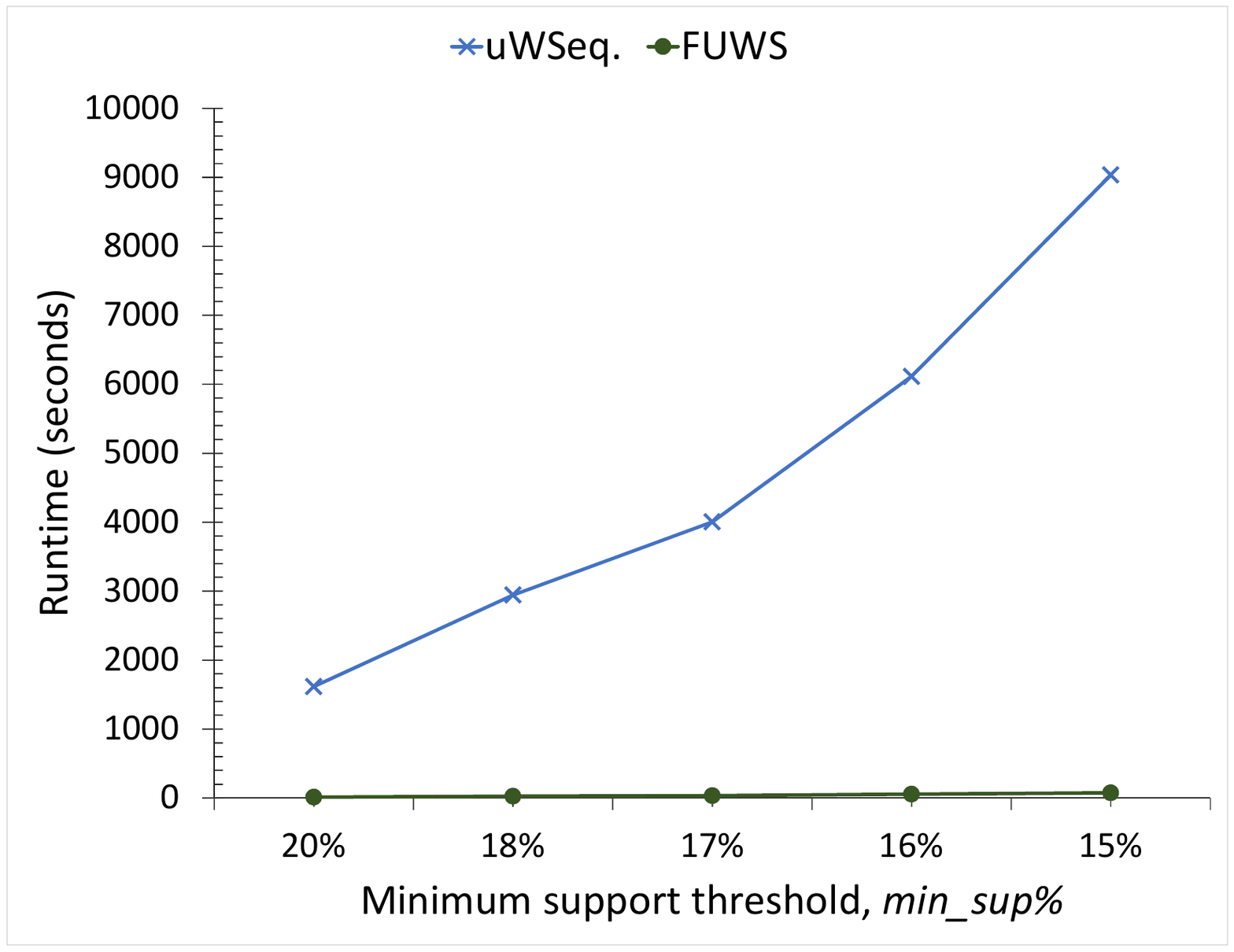}
      \caption{\textit{FIFA} dataset}
      \label{fifa_rt}
    \end{subfigure}
    \medskip
    \begin{subfigure}{0.45\linewidth}
      \includegraphics[width=\linewidth, height = .6\linewidth]{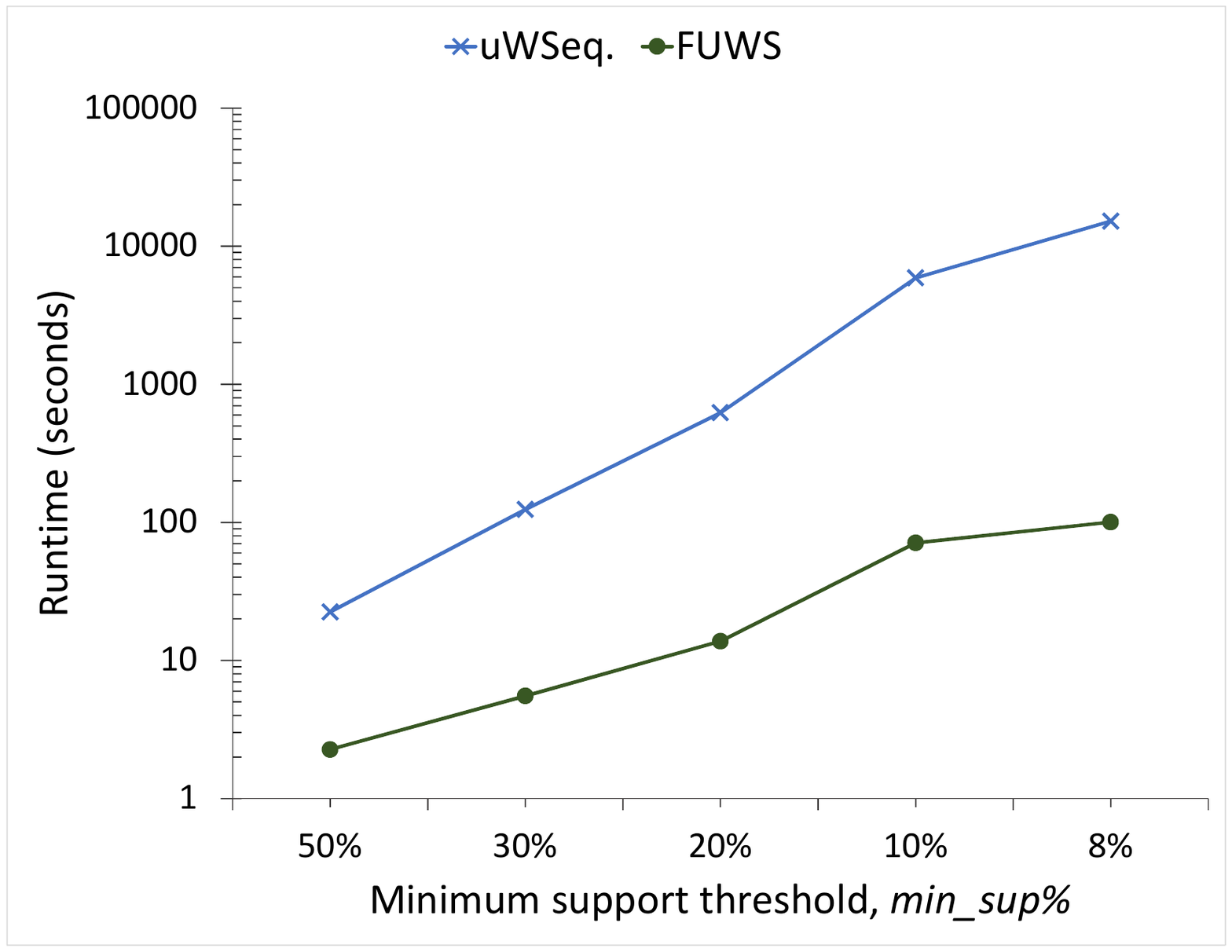}
      \caption{\textit{MSNBC} dataset}
      \label{fuws_msn_rt}
    \end{subfigure}\hfil 
    \begin{subfigure}{0.45\linewidth}
      \includegraphics[width=\linewidth, height = .6\linewidth]{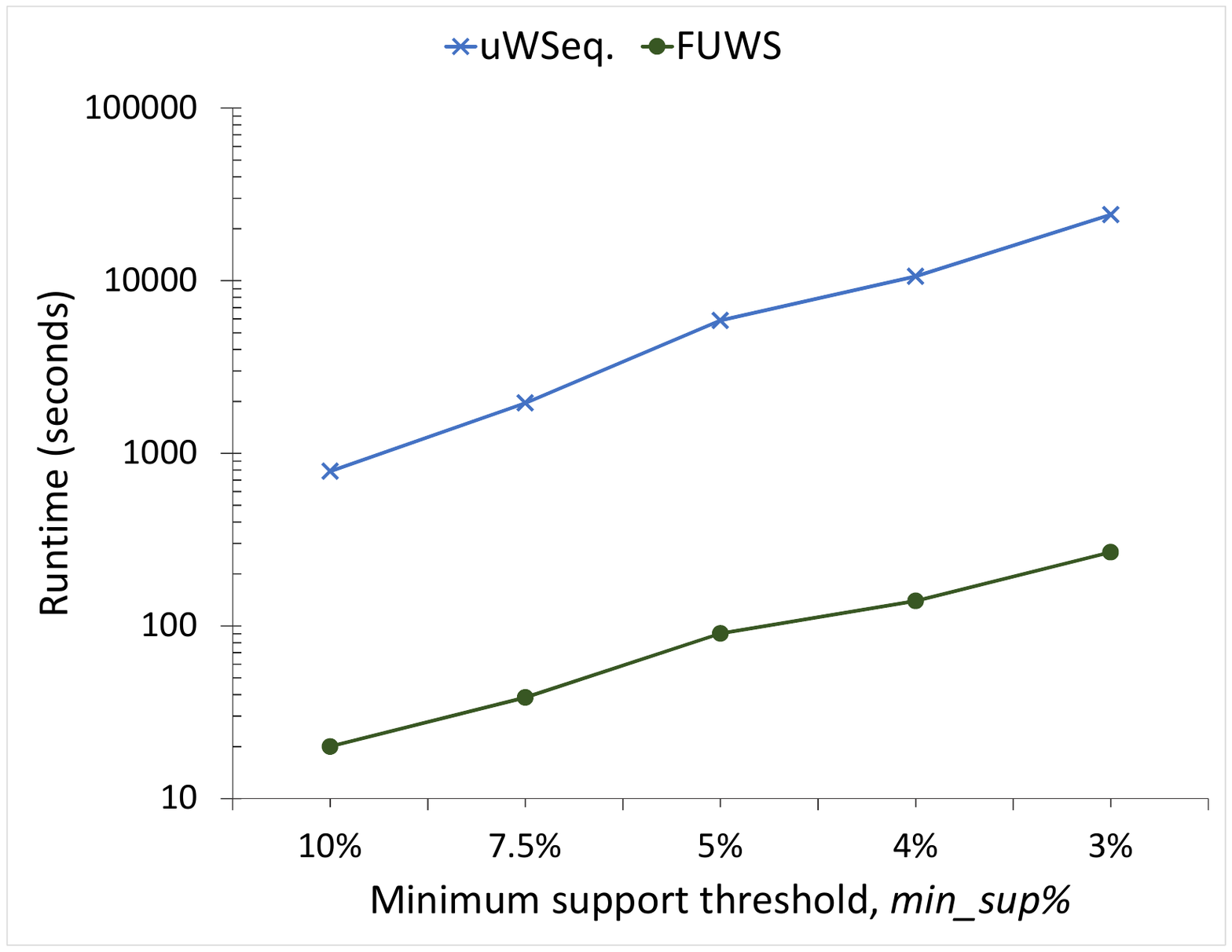}
      \caption{\textit{Leviathan} dataset}
      \label{lev_rt}
    \end{subfigure}

\caption{Comparison of runtime between \textit{FUWS} and \textit{uWSequence}
for various support threshold in different datastets 
}
\label{fuws_rt_1}
\end{figure}

\subsubsection{Comparison of Runtime with \textit{uWSequence}}
\label{subsub:min_sup_rt}

Figure \ref{sign_rt} shows comparison of runtime in \textit{Sign} dataset for different support thresholds. 
As we can see, the difference between the runtime of the two algorithms increases with the decrease in the support threshold. 
The curve representing \textit{FUWS} rises slowly, but the curve of \textit{uWSequence} rises up very fast with a slight decrease in the threshold.
Figure \ref{kos_rt} shows the runtime comparison between \textit{FUWS} and \textit{uWSequence} in \textit{Kosarak} dataset.
It is quite similar to the result in \textit{Sign}. 
The difference in the runtime between \textit{FUWS} and \textit{uWSequence} algorithms increases exponentially with the decreasing values of the support threshold.

The reasons behind increasing difference in runtime are as follows, 
\begin{enumerate}[label=(\alph*)]
    \item \textit{uWSequence} generates more false-positive candidates than \textit{FUWS} for the same support threshold.
    \item even if the number of candidates was the same, \textit{uWSequence} would consume more time as its complexity of calculating actual weighted expected support is worse than that of \textit{FUWS}.
\end{enumerate}
The use of faster support calculation method \textit{SupCalc} based on \textit{USeq-Trie} gives a benefit to \textit{FUWS} in the latter case.
Results in  \textit{Retail}, \textit{FIFA}, \textit{MSNBC}, and \textit{Leivathan} datasets are also shown in Figure \ref{fuws_rt_1}. 
In every dataset, \textit{FUWS} outperforms the existing \textit{uWSequence} at any minimum support threshold.

\subsubsection{Analysis for Different Probability Distributions }
\label{subsub:prob_dis}
\begin{figure}[tbh]
    \centering 
    \begin{subfigure}{0.45\linewidth}
      \includegraphics[width=\linewidth, height = .6\linewidth]{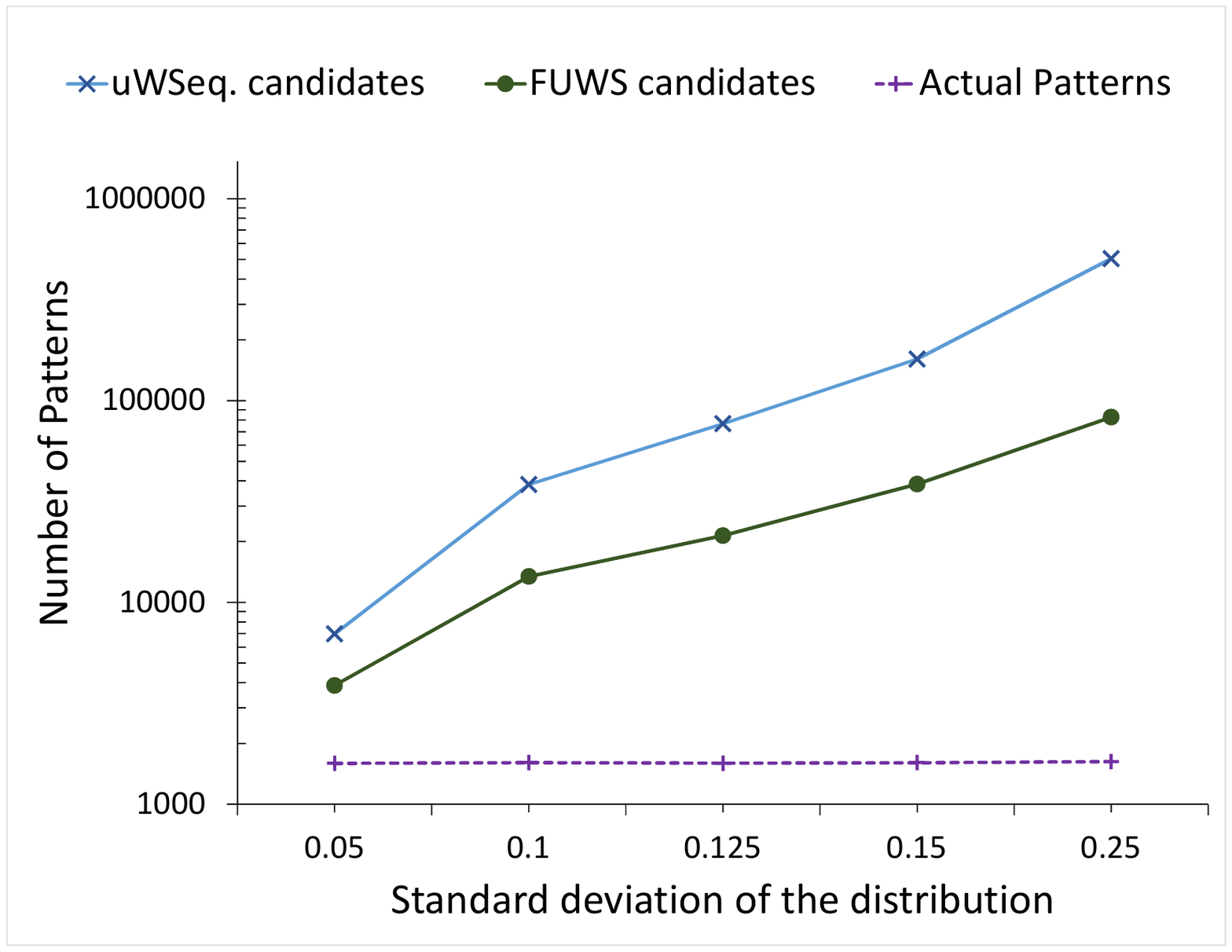}
      \caption{\textit{Sign} dataset}
      \label{sign_prob}
    \end{subfigure}\hfil 
    \begin{subfigure}{0.45\linewidth}
      \includegraphics[width=\linewidth, height = .6\linewidth]{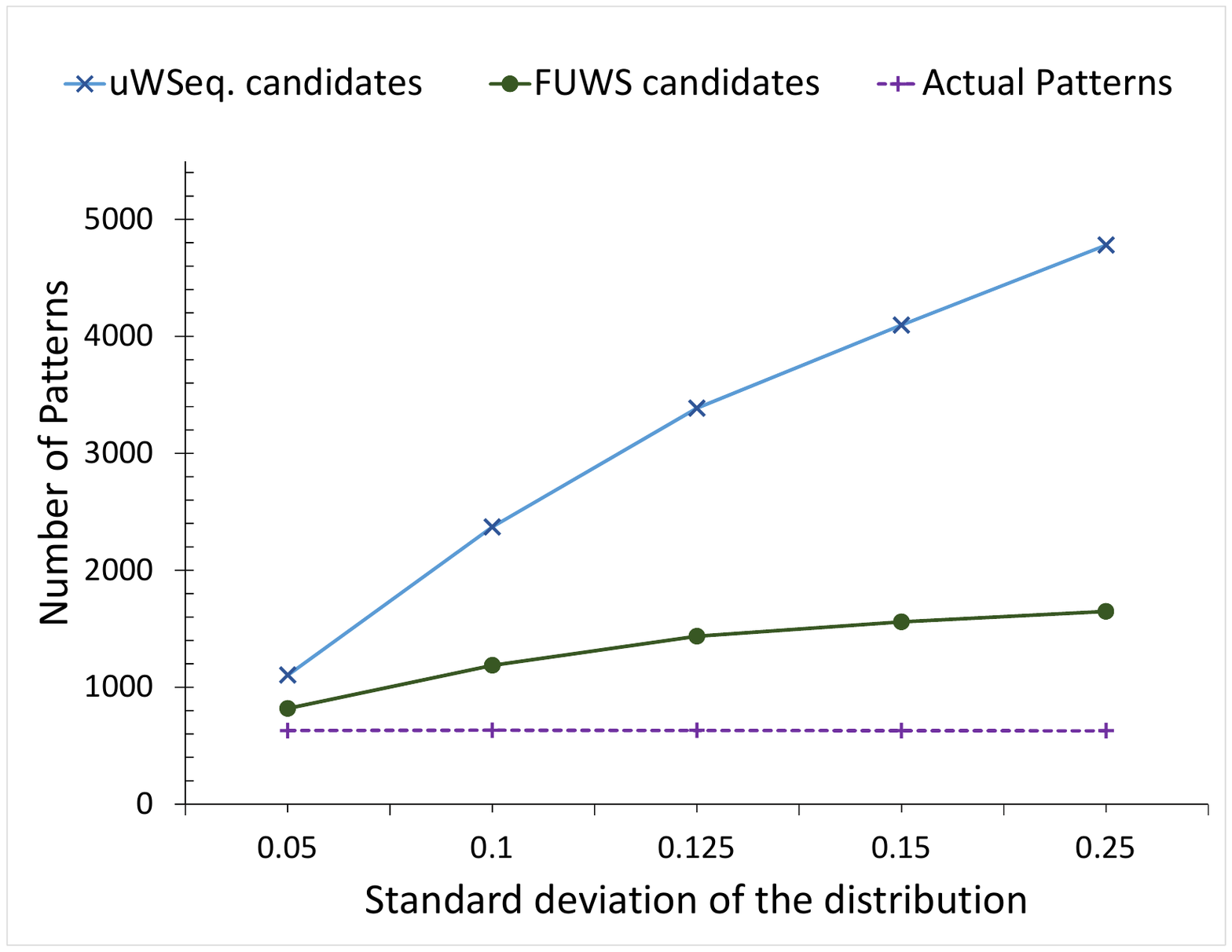}
      \caption{\textit{OnlineRetail} dataset}
      \label{online_prob}
    \end{subfigure}

\caption{Change in candidate generation for various distribution of probabilities in \textit{Sign} and \textit{OnlineRetail} datasets}
\label{prob_dist}
\end{figure}

Figure \ref{sign_prob} shows the analysis for different values of standard deviation in \textit{Sign} dataset with 5\% min\_sup.
When the standard deviation value is large, the difference between an item's minimum and maximum existential probability in data sequences is most likely to become larger. 
This affects the calculation of $expSup^{cap}$ in \textit{FUWS} and $expSupport^{top}$ in \textit{uWSequence} for a prefix sequence.
The larger standard deviation value makes the upper bound of expected support calculation less tight in both algorithms. Consequently, both algorithms generate more false-positive candidates than those with a smaller standard deviation.
Figure \ref{online_prob} shows a similar result in \textit{OnlineRetail} dataset with a 0.1\% support threshold. 
Note that, whatever the distribution is, $expSup^{cap}$ is always less than or equal to $expSupport^{top}$.
Hence, \textit{FUWS}, using $expSup^{cap}$, always generates fewer false-positive candidates than \textit{uWSequence}.
Based on the results, it can be said that
\textit{FUWS} outperforms \textit{uWSequence} in any dataset for any kind of distribution of the uncertainty values.

\subsubsection{Analysis with respect to Different Choice of Weight Factor}
\label{subsub:wgt_fct}
\begin{figure}[bht]
 \centering
    \includegraphics[width=.5\textwidth,height=.3\textwidth]{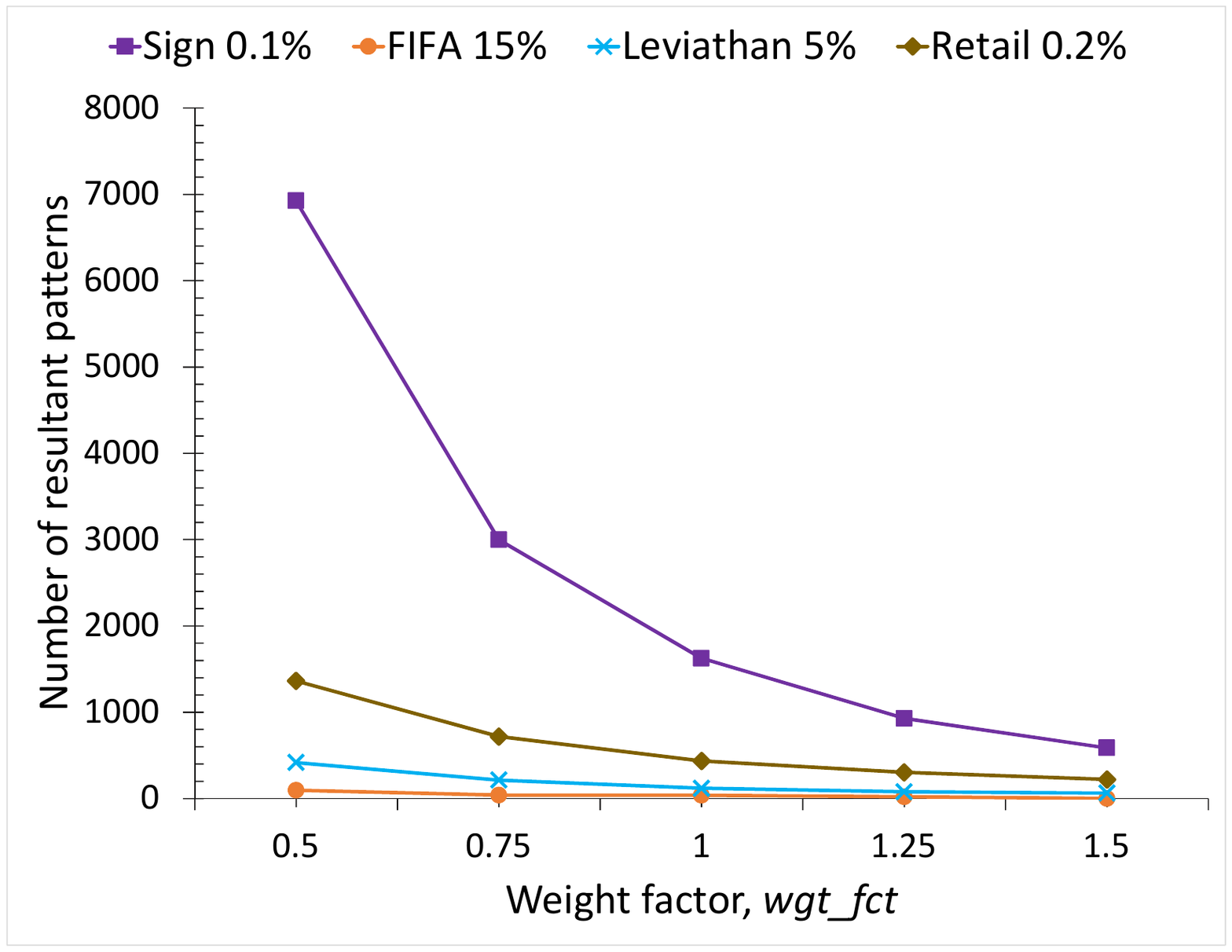}
    \caption{Change in number of patterns for different weight factors in \textit{Sign}, \textit{FIFA}, \textit{Leviathan}, and \textit{Retail} datasets}
    \label{wght_fct}
\end{figure}

The results in \textit{FIFA}, \textit{Leviathan}, and \textit{Retail} datasets are shown in Figure \ref{wght_fct} with support thresholds 15\%, 5\%, and 0.2\% respectively. As we can see, when the weight factor is 0.75, \textit{FUWS} finds 718 patterns to be weighted frequent. When this factor is increased to 1.25 to find patterns with more weight values, \textit{FUWS} gives 303 patterns as weighted frequent sequences. Thus, the choice of weight factor lets the user find interesting sequences according to their necessity.

\subsection{Performance of {the incremental approaches,} \textit{uWSInc} and \textit{uWSInc+}}
\label{incremental_phase}

To the best of our knowledge, there is no algorithm to mine weighted sequential patterns from incremental uncertain databases. The baseline approach is to run the \textit{FUWS} algorithm from scratch after every increment. 
This section compares the two proposed algorithms with the baseline and highlights the differences between them in different datasets.

\textbf{Evaluation Criteria. }
The criteria to evaluate the performance of incremental approaches are as follows,
\begin{enumerate}[label=(\alph*)]
    \item \textbf{Required time to find the updated result. } An incremental approach is efficient if it requires significantly less time than this baseline approach which runs the \textit{FUWS} algorithm in the whole updated database after every increment. The baseline approach is naturally very expensive.
    \item \textbf{Completeness of the result. } The set of weighted frequent sequences found by the baseline approach is the complete set of actual patterns. The completeness of an incremental algorithm is defined to be the percentage of patterns found compared to this complete set.
\end{enumerate}
Note that an incremental approach may require scanning the whole database after each increment in the worst case to ensure a complete result by an incremental solution. On the other hand, a significant improvement in the runtime can be achieved with only a small sacrifice in completeness, as we will see in the following subsections.

To use the datasets as incremental ones, we used 50\% of the dataset to be the initial part. We then introduced five increments (each one randomly chosen to be 20 to 60\% of the initial size). However, in the case of the \textit{Retail} dataset, which is mentioned to be a dataset of five months, we used the first one-fifth of its transactions to be the initial portion. We then introduced four increments which roughly represent the next four months. We have also conducted experiments by varying initial size and increment size and tested the scalability.
Parameters for these experiments are:
\begin{enumerate}[label=(\alph*)]
    \item \textbf{Minimum support threshold. } 
    Experiments are conducted for different \textit{min\_sup\%} for each dataset to validate the efficacy of our incremental approaches in finding almost complete results at any threshold based on the application requirements, which is discussed elaborately in Sections~\ref{subsub:inc_min_sup_rt} and~\ref{subsub:inc_min_sup_com}.
    \item \textbf{Buffer ratio. } Buffer ratio, $\mu$, is used to lower the minimum weighted expected support.  When $\mu=1.0$, it means no buffer to store \textit{semi-frequent sequences} \textit{ (in uWSInc, uWSInc+)}. 
    Lower values of $\mu$ mean larger buffers. With this lowered threshold, more candidates are generated and tested. As a result, this requires more time. However, it can find more patterns using the \textit{semi-frequent sequences}.  Detailed results are shown in Section~\ref{subsub:inc_buffer_ratio}.
    \item \textbf{Increment size. } Many incremental algorithms have the limitation that they do not perform well beyond a certain increment size. This is also called update ratio, i.e., $\frac{(size\ of\ \Delta DB)}{(size\ of\ initial\ DB)}$. We have run our algorithms several times by changing the update ratio when other parameters are fixed and showed the efficiency in Section~\ref{subsub:inc_inc_size}.
    \item \textbf{Dataset size. } Besides the increment size, this is another form of testing the scalability of the incremental approach. We have gradually increased the dataset and plotted the total time needed to find the updated result after each increment starting from an initial dataset. The update ratio is not fixed in this case, rather drawn from a range of 0.2 to 0.6 randomly. Results in Section~\ref{subsub:inc_dataset_size} show how the algorithms performs for such scaled datasets.
    \item \textbf{Intial dataset size. } Existing incremental mining algorithms focus on almost complete results by only buffering \textit{semi-frequent sequences (SFS)} depend on the initial dataset size. Thus, we develop our \textit{uWSInc+} algorithm to overcome this limitation. Hence, we have compared between \textit{uWSInc} (buffers \textit{SFS} only) and \textit{uWSInc+} (along with \textit{SFS}, it buffers extra promising frequent sequences that are mined locally in the increments) in Section~\ref{subsub:inc_initial_size} by setting different initial dataset size when other parameters are fixed.
\end{enumerate}

\subsubsection{Runtime Analysis with respect to Support Threshold}
\label{subsub:inc_min_sup_rt}
\begin{figure}[!htb]
    \centering 
    \begin{subfigure}{0.45\linewidth}
	\includegraphics[width=\textwidth,height=.6\textwidth]{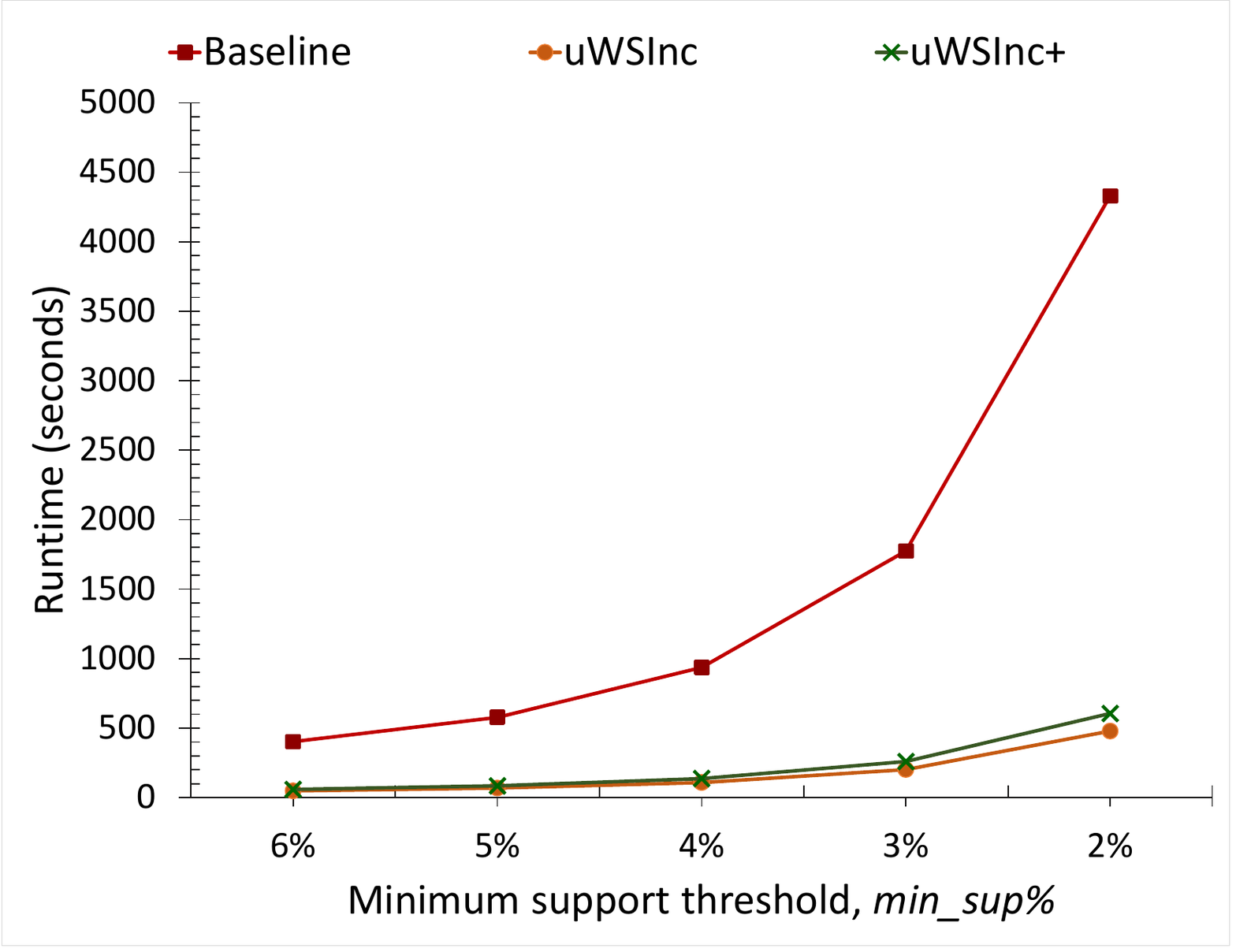}
    \caption{Runtime in \textit{Leviathan}}
    \label{fig:lev_rt_minsup}
    \end{subfigure}\hfil 
    \begin{subfigure}{0.45\linewidth}
    \includegraphics[width=\textwidth,height=.6\textwidth]{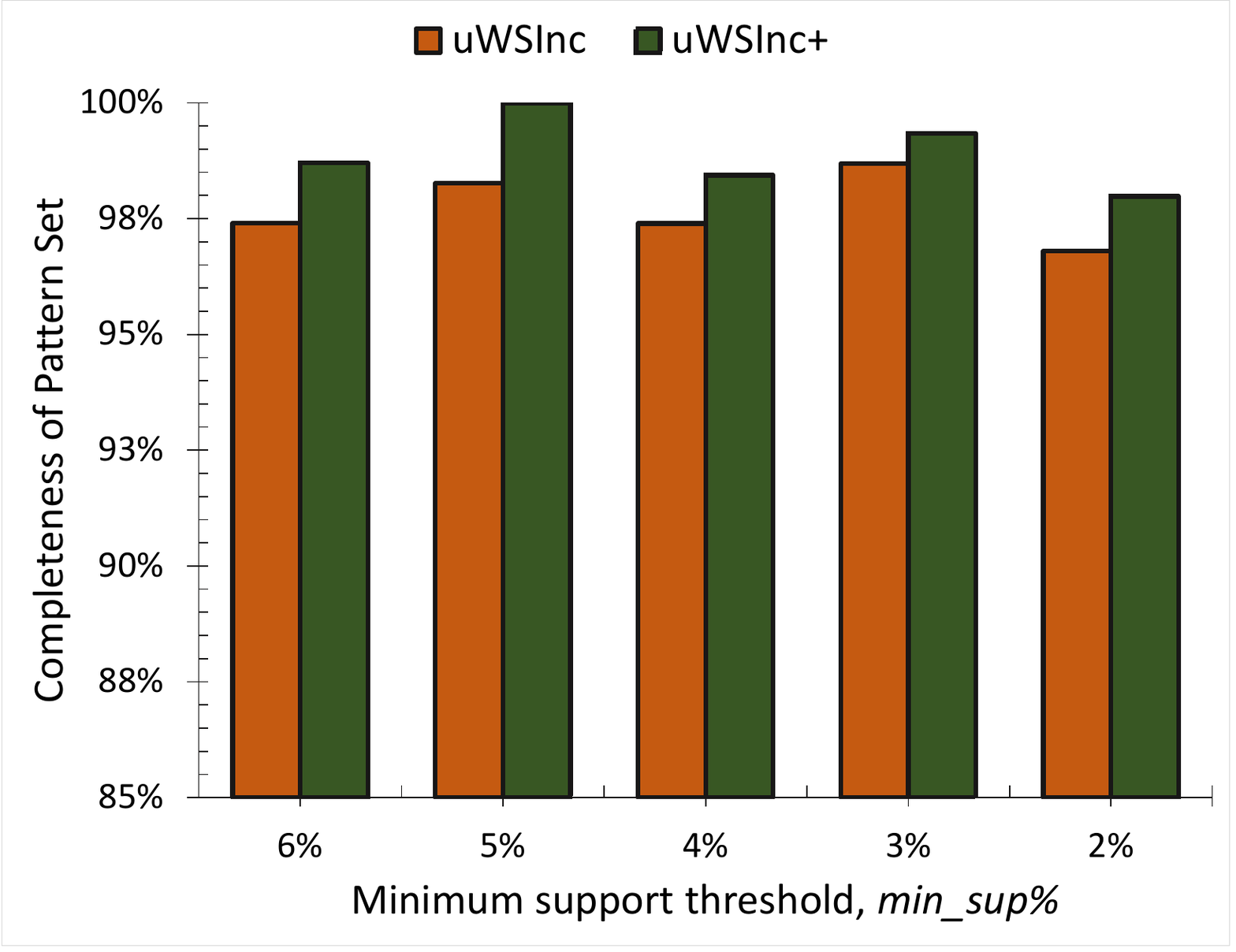}
    \caption{Completeness in \textit{Leviathan}}
    \label{fig:lev_comp_minsup}
    \end{subfigure}
    \medskip
    \begin{subfigure}{0.45\linewidth}
    \includegraphics[width=\textwidth,height=.6\textwidth]{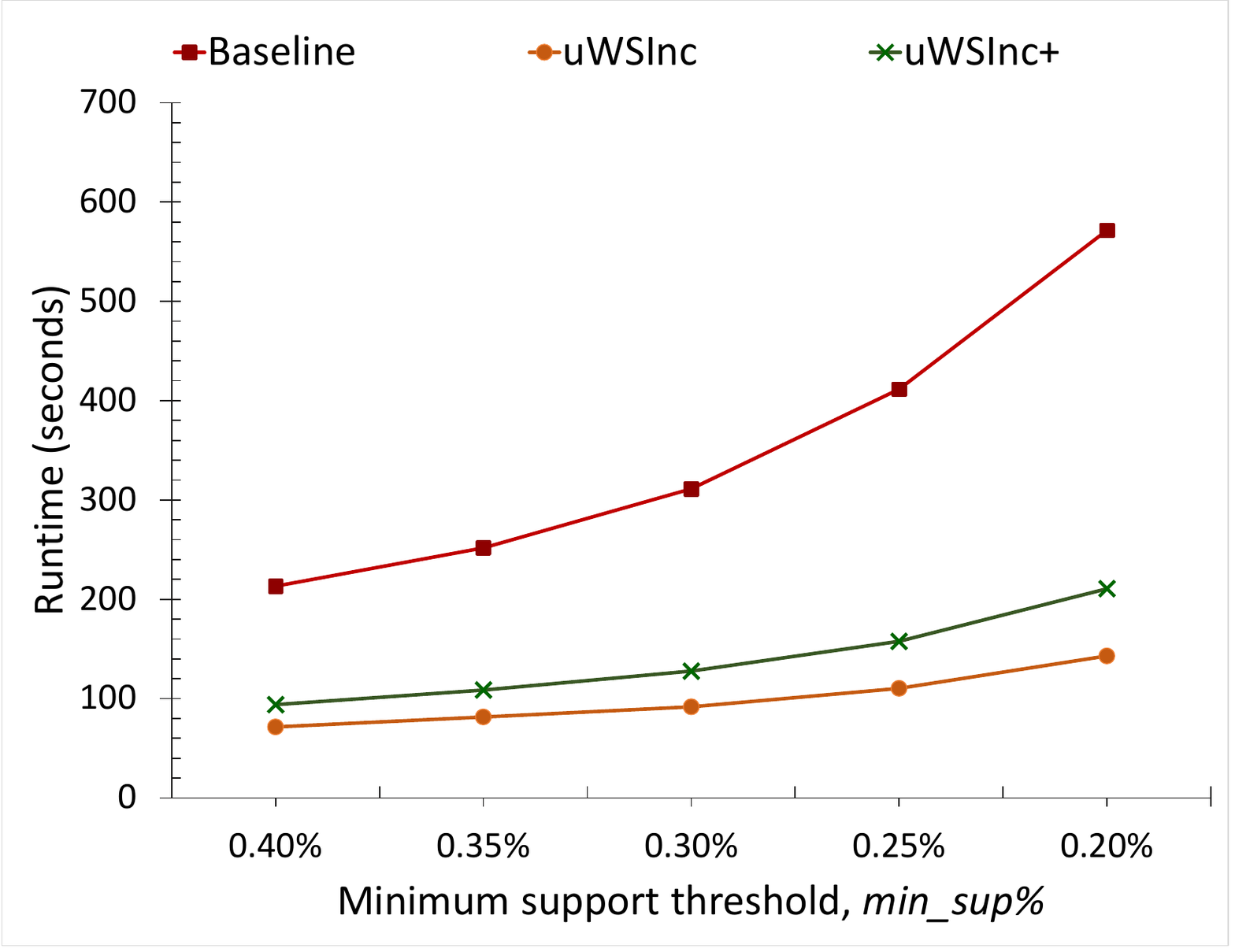}
    \caption{Runtime in \textit{Retail}}
    \label{fig:retail_rt_minsup}
    \end{subfigure}
    \hfil 
    \begin{subfigure}{0.45\textwidth}
      \includegraphics[width=\linewidth, height = .6\linewidth]{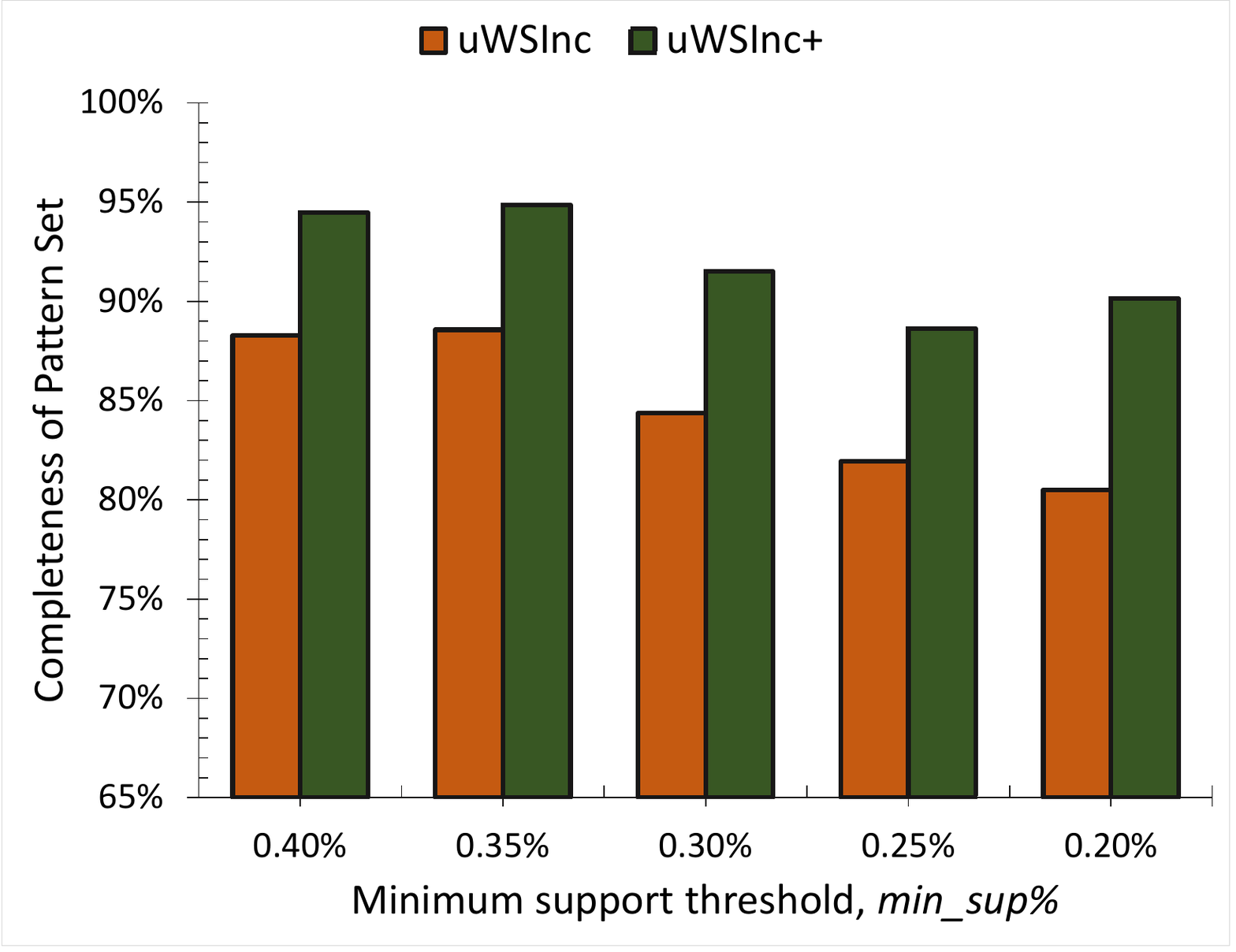}
      \caption{Completeness in \textit{Retail}}
      \label{fig:retail_comp_minsup}
    \end{subfigure}
    \medskip
    \begin{subfigure}{0.45\linewidth}
      \includegraphics[width=\linewidth, height = .6\linewidth]{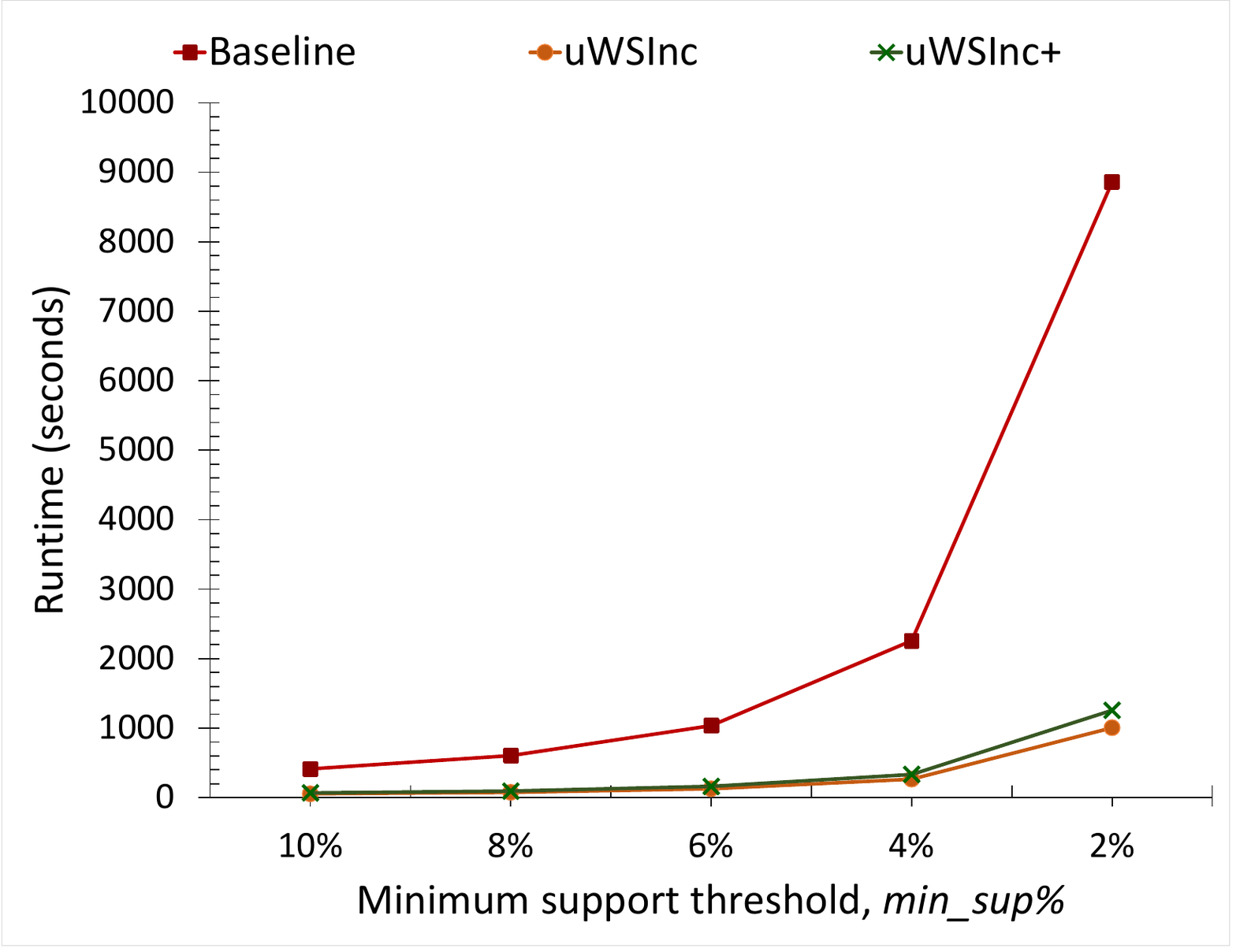}
      \caption{Runtime in \textit{MSNBC} dataset}
      \label{msn_rt}
    \end{subfigure}\hfil 
    \begin{subfigure}{0.45\linewidth}
      \includegraphics[width=\linewidth, height = .6\linewidth]{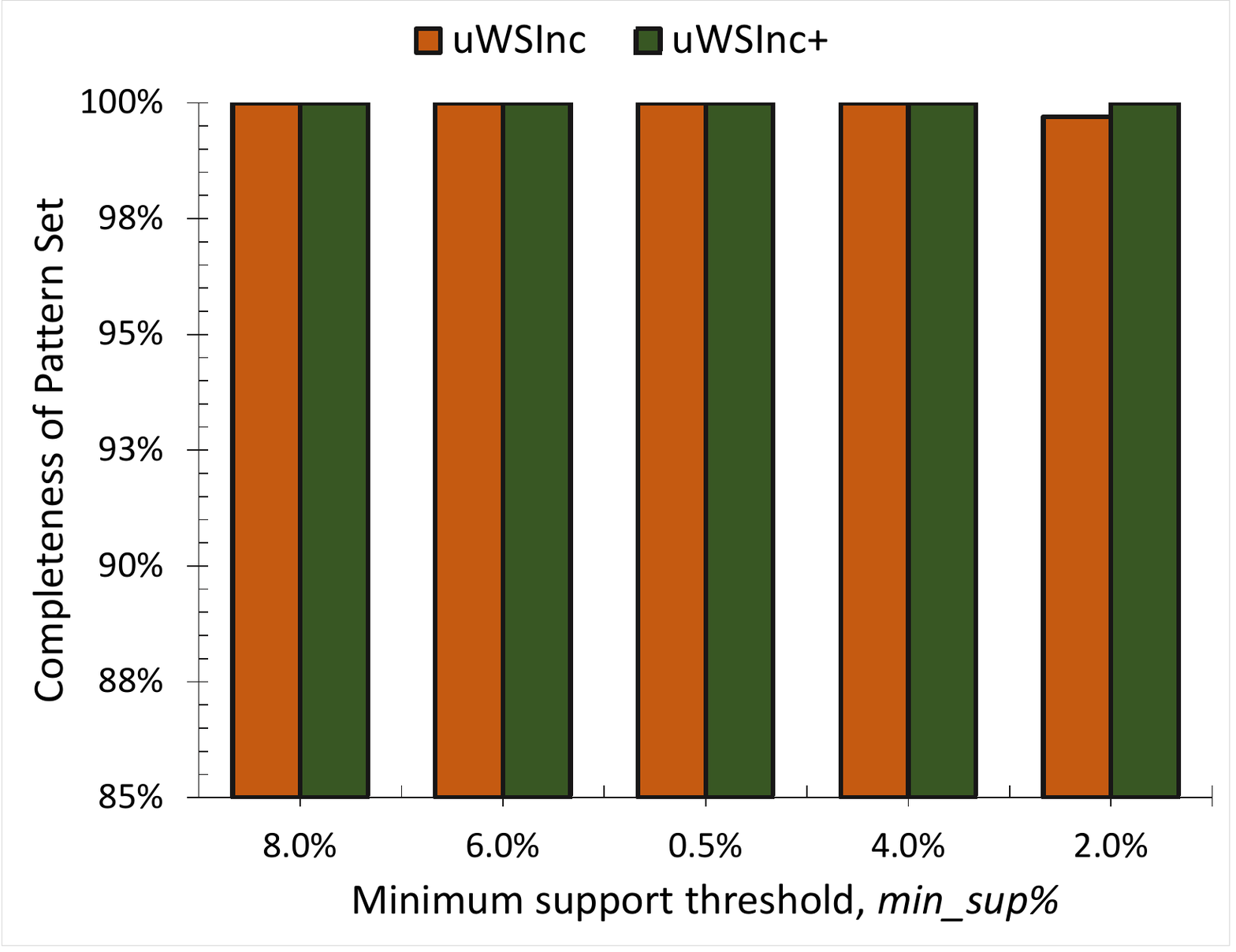}
      \caption{Completeness in \textit{MSNBC} dataset}
      \label{msn_comp}
    \end{subfigure}

\caption{Performance analysis of \textit{uWSInc} and \textit{uWSInc+} against various support threshold 
}
\label{inc_minsup}
\end{figure}

To analyze the runtime with respect to different support thresholds, we have run the baseline approach, \textit{uWSInc}, and \textit{uWSInc+} in a database several times and exhibited the average runtime for each support threshold. Total time required by an approach in the dataset for a single support threshold is the sum of the amount of time required after each increment plus the amount of time required for mining in the initial part.  
Figure \ref{fig:lev_rt_minsup} shows the result in the \textit{Leviathan} dataset for different support thresholds. As we can see, at 6\% support threshold, it requires 402.36 seconds in baseline approach whereas \textit{uWSInc} and \textit{uWSInc+} takes only 47.94 and 58.67 seconds, respectively.

\textit{uWSInc} takes 8.39 times less time than the baseline approach; for \textit{uWSInc+}, this number is 6.86. 
The difference between \textit{uWSInc} runtime and baseline is around 354.42 seconds and the difference between \textit{uWSInc} and \textit{uWSInc+} is around 10.73 seconds.
These differences with the baseline change very rapidly with a slight change in the support threshold. 
In the \textit{Retail} dataset, when the minimum support threshold is 0.4\%,
the difference between \textit{uWSInc} and the baseline is 159.73 seconds, the difference between \textit{uWSInc+} and the baseline is 137.27 seconds, and the difference between \textit{uWSInc} and \textit{uWSInc+} is 22.46 seconds as depicted in Figure~\ref{fig:retail_rt_minsup}.
When the threshold decreases to 0.2\%, the above differences rise to 428.51, 360.88, and 67.63 seconds respectively. 
Figure \ref{msn_rt} shows the runtime analysis with respect to minimum support threshold in \textit{MSNBC} dataset which is similar to the result of the \textit{Leviathan} dataset.

\subsubsection{Analysis of Completeness with respect to Different Support Thresholds}
\label{subsub:inc_min_sup_com}
\begin{figure}[thb]
    \centering 
    \begin{subfigure}{0.45\linewidth}
    \includegraphics[width=\textwidth,height=.6\textwidth]{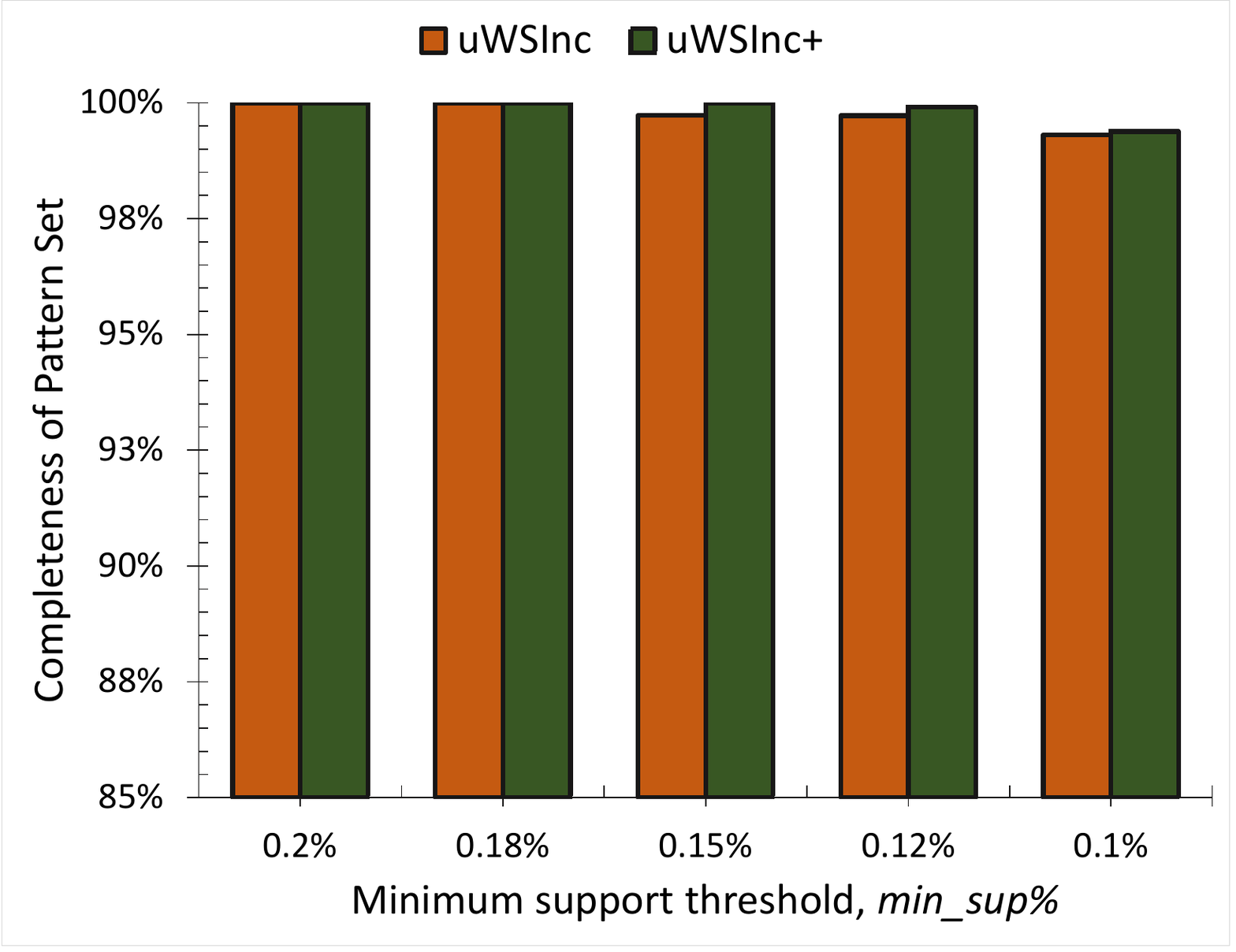}
    \caption{\textit{Kosarak} dataset}
    \label{kos_comp_minsup}
    \end{subfigure}
    \hfil 
    \begin{subfigure}{0.45\textwidth}
      \includegraphics[width=\linewidth, height = .6\linewidth]{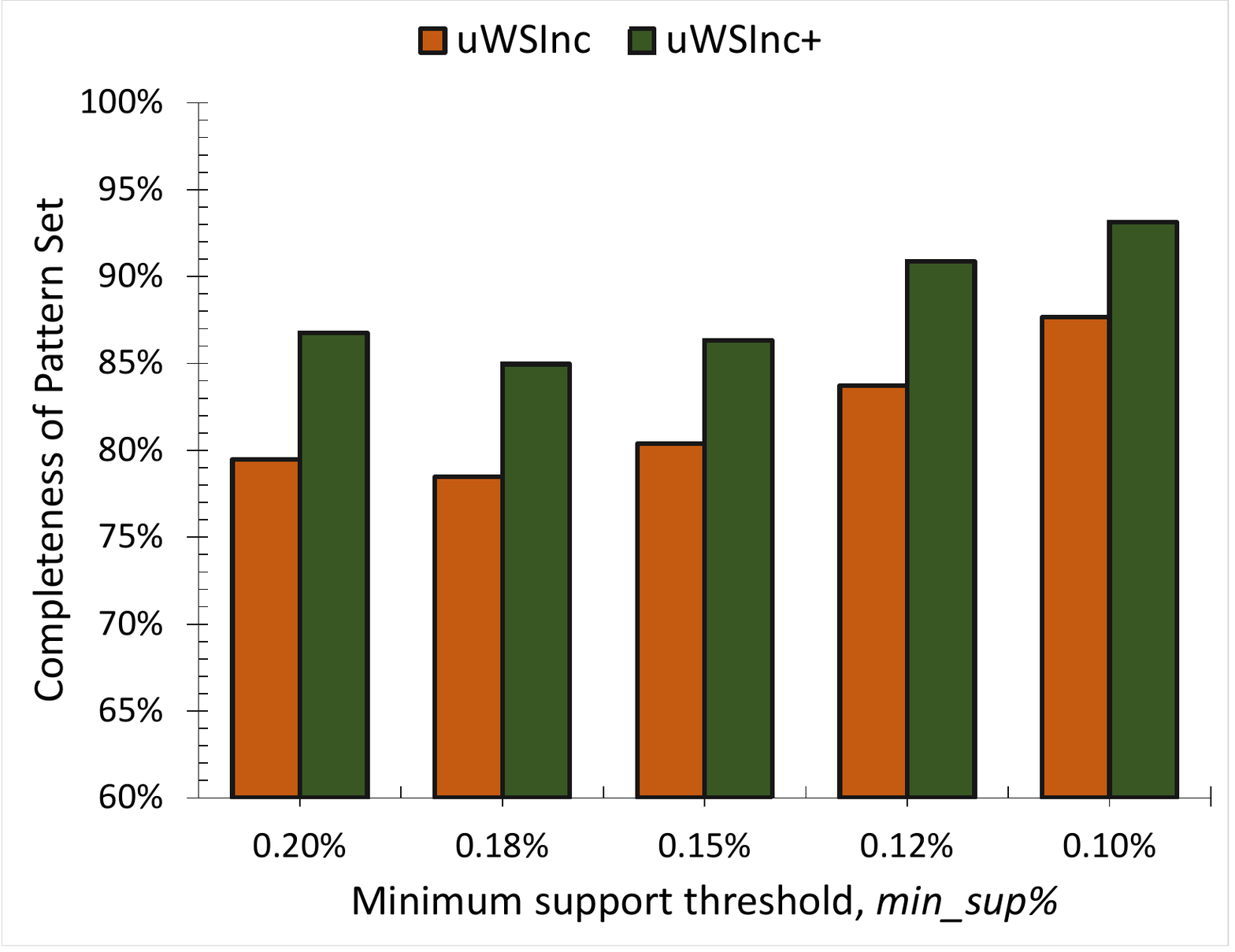}
      \caption{\textit{Foodmart} dataset}
      \label{food_comp_minsup}
    \end{subfigure}

\caption{Completeness of result 
in \textit{Kosarak} and \textit{Foodmart} datasets
against various support threshold 
}
\label{inc_minsup_2}
\end{figure}

Figure \ref{fig:lev_comp_minsup} shows the completeness comparison between \textit{uWSInc} and \textit{uWSInc+} in \textit{Leviathan} dataset for support threshold values ranging between 2\% and 6\%. As we can see, at any support threshold point, \textit{uWSInc+} gives better completeness.
Figure \ref{fig:retail_comp_minsup} shows the difference in completeness between \textit{uWSInc} and \textit{uWSInc+} in \textit{Retail} dataset for different support thresholds. The difference is more understandable than that was seen in \textit{Leviathan} dataset.
\textit{Foodmart} dataset has results similar to \textit{Retail} as shown in Figure \ref{food_comp_minsup}.
Datasets like \textit{Leviathan} and \textit{Bible} \footnote{Details of \textit{Leviathan}, \textit{Bible}, \textit{Retail}, and \textit{Foodmart} dataset can be found at \url{http://www.philippe-fournier-viger.com/spmf/index.php?link=datasets.php}} contain almost all the patterns (frequent word sequences) in their initial 50\% portion. Any new word that appears in future increments generally does not have enough support to become frequent. Thus, the completeness of our two approaches is very close in these cases for different min\_sup values.
On the other hand, market basket datasets like \textit{Retail} and \textit{Foodmart} have scenarios that any item that was initially infrequent or absent can come up in future increments with enough support to be frequent. Most of these new patterns can be found by \textit{uWSInc+} where \textit{uWSInc} can find none of them. Thus, a significant difference in completeness is found in these datasets.
Completeness of the result for both algorithm achieve completeness very close to 100\% in datasets like \textit{MSNBC} and \textit{Kosarak} as shown in Figures~\ref{msn_comp} and \ref{kos_comp_minsup}.

\subsubsection{Analysis with respect to Buffer Ratio}
\label{subsub:inc_buffer_ratio}

An incremental mining algorithm that consumes reasonably more time might be preferred to another only if it gives better completeness. Nevertheless, it is a matter of deciding what is an acceptable level of sacrifice. This decision may depend on many factors that vary from user to user. As we have seen runtime and completeness difference between \textit{uWSInc} and \textit{uWSInc+} for varying support thresholds in \textit{Leviathan}, \textit{Retail}, and \textit{Foodmart} dataset, let us discuss the effect of buffer ratio in them while using incremental approaches.
\begin{figure}[!htb]
    \centering 
    \begin{subfigure}{0.45\linewidth}
    \includegraphics[width=\textwidth,height=.6\textwidth]{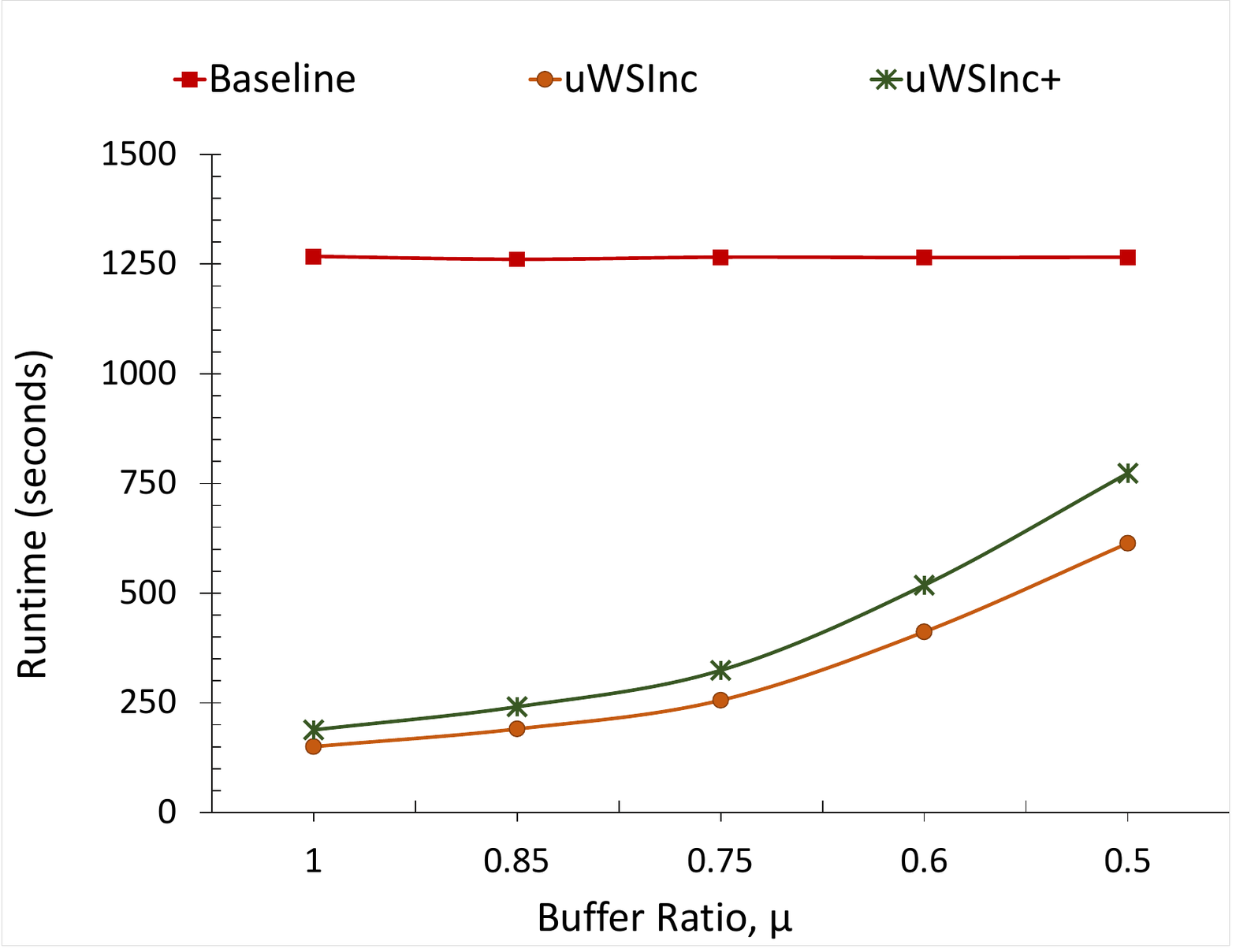}
    \caption{Runtime in incremental \textbf{Leviathan}, min\_sup 3\%}
    \label{fig:lev_rt_mu}
    \end{subfigure}\hfil 
    \begin{subfigure}{0.45\linewidth}
    \includegraphics[width=\textwidth,height=.6\textwidth]{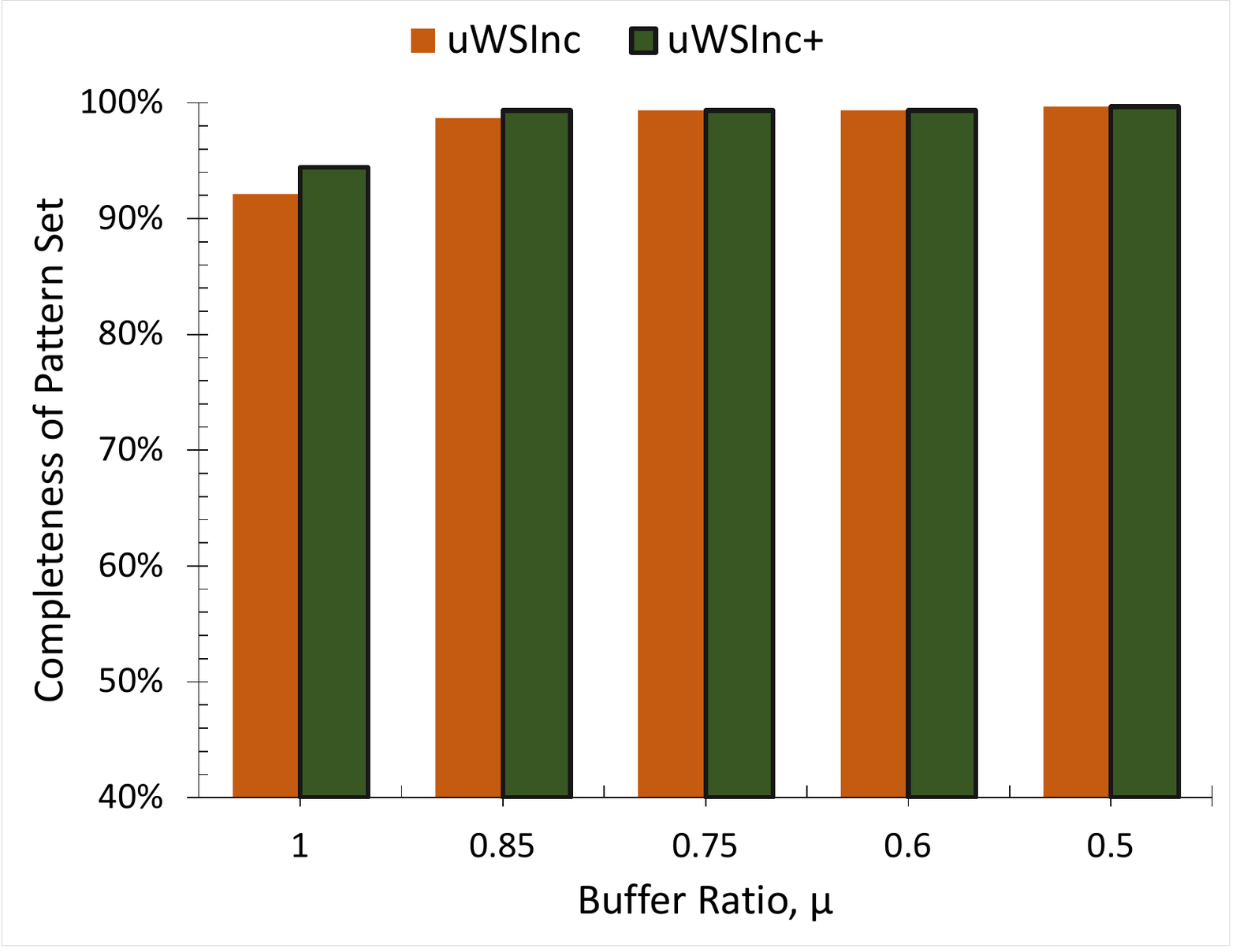}
    \caption{Completeness in incremental \textbf{Leviathan}, min\_sup 3\% }
    \label{fig:lev_comp_mu}
    \end{subfigure}
    \medskip
    \begin{subfigure}{0.45\linewidth}
    \includegraphics[width=\textwidth,height=.6\textwidth]{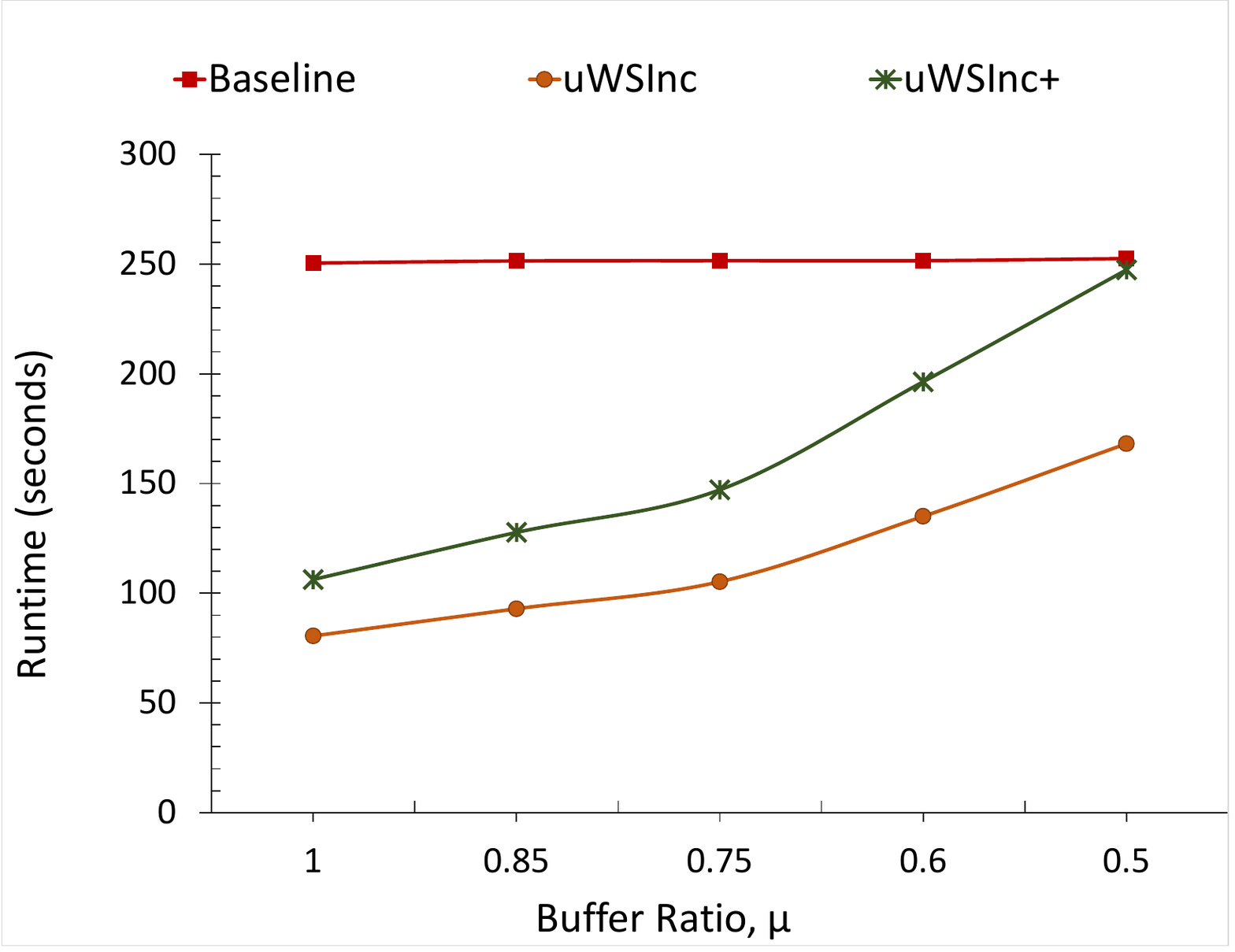}
    \caption{Runtime in incremental \textbf{Retail}, min\_sup 0.3\%}
    \label{retail_rt_mu}
    \end{subfigure}
    \hfil 
    \begin{subfigure}{0.45\textwidth}
      \includegraphics[width=\linewidth, height = .6\linewidth]{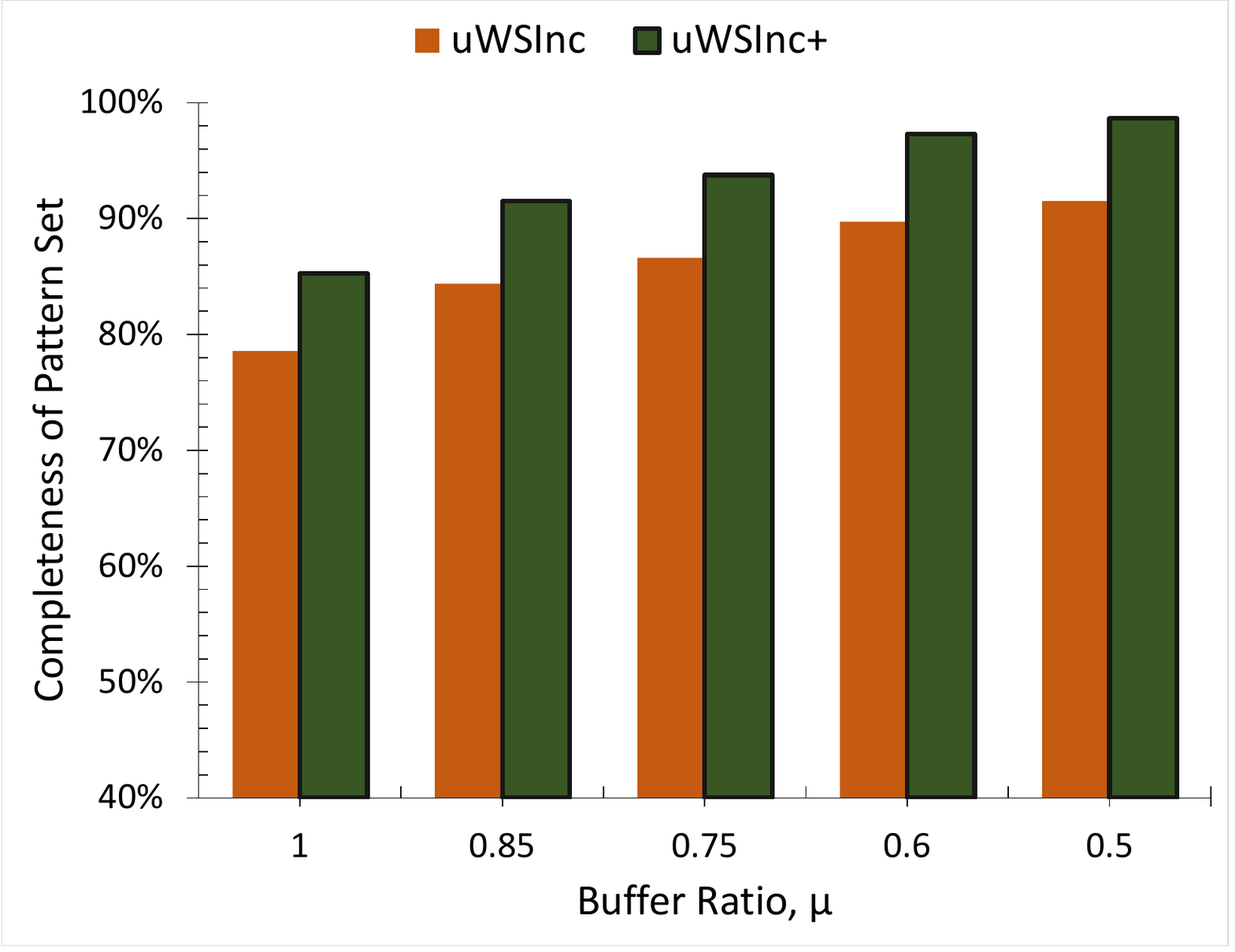}
      \caption{Completeness in incremental \textit{Retail}, min\_sup 0.3\%}
      \label{retail_comp_mu}
    \end{subfigure}
    \medskip
    \begin{subfigure}{0.45\linewidth}
      \includegraphics[width=\linewidth, height = .6\linewidth]{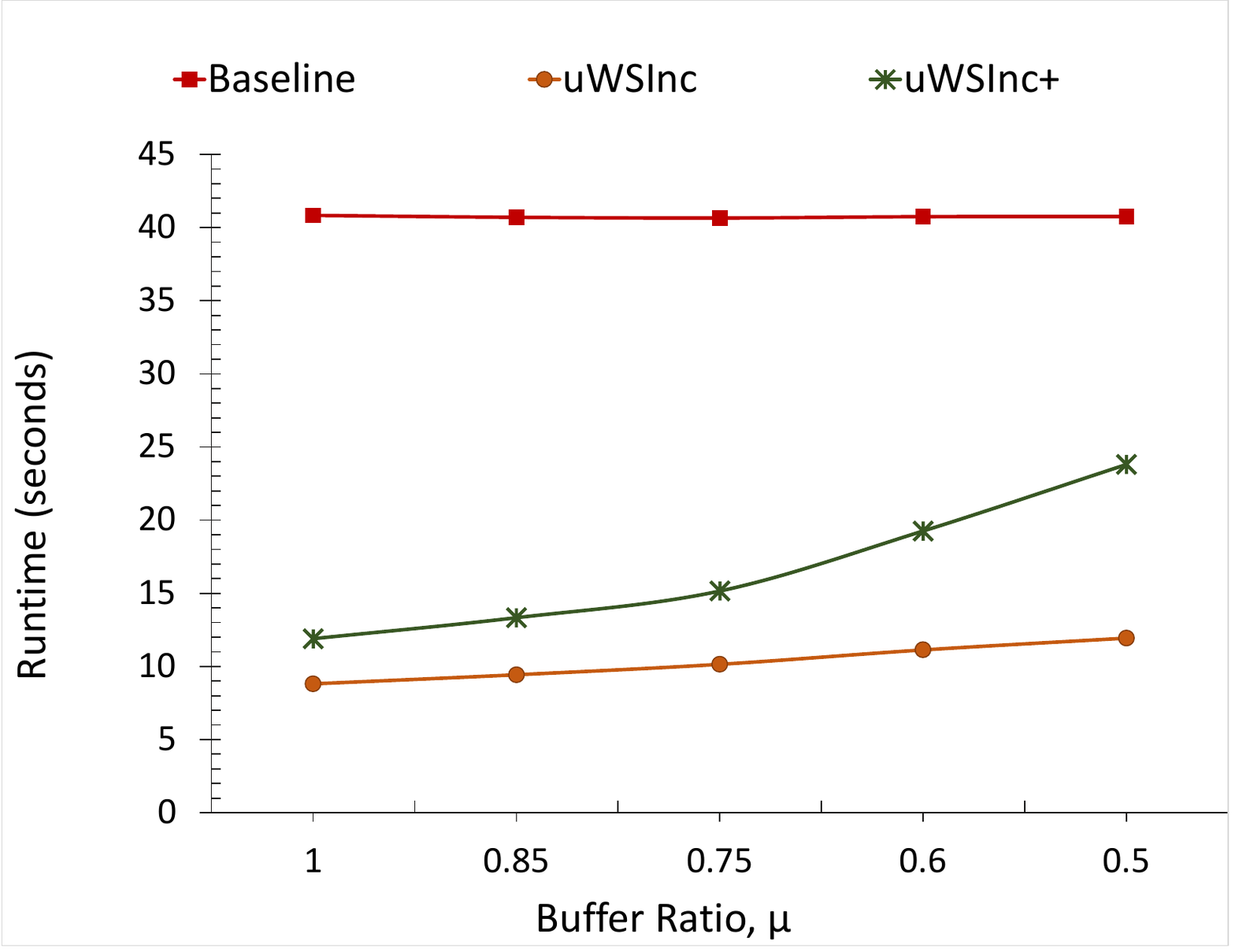}
      \caption{Runtime in incremental \textit{Foodmart}, min\_sup 0.2\%}
      \label{food_rt_mu}
    \end{subfigure}\hfil 
    \begin{subfigure}{0.45\linewidth}
      \includegraphics[width=\linewidth, height = .6\linewidth]{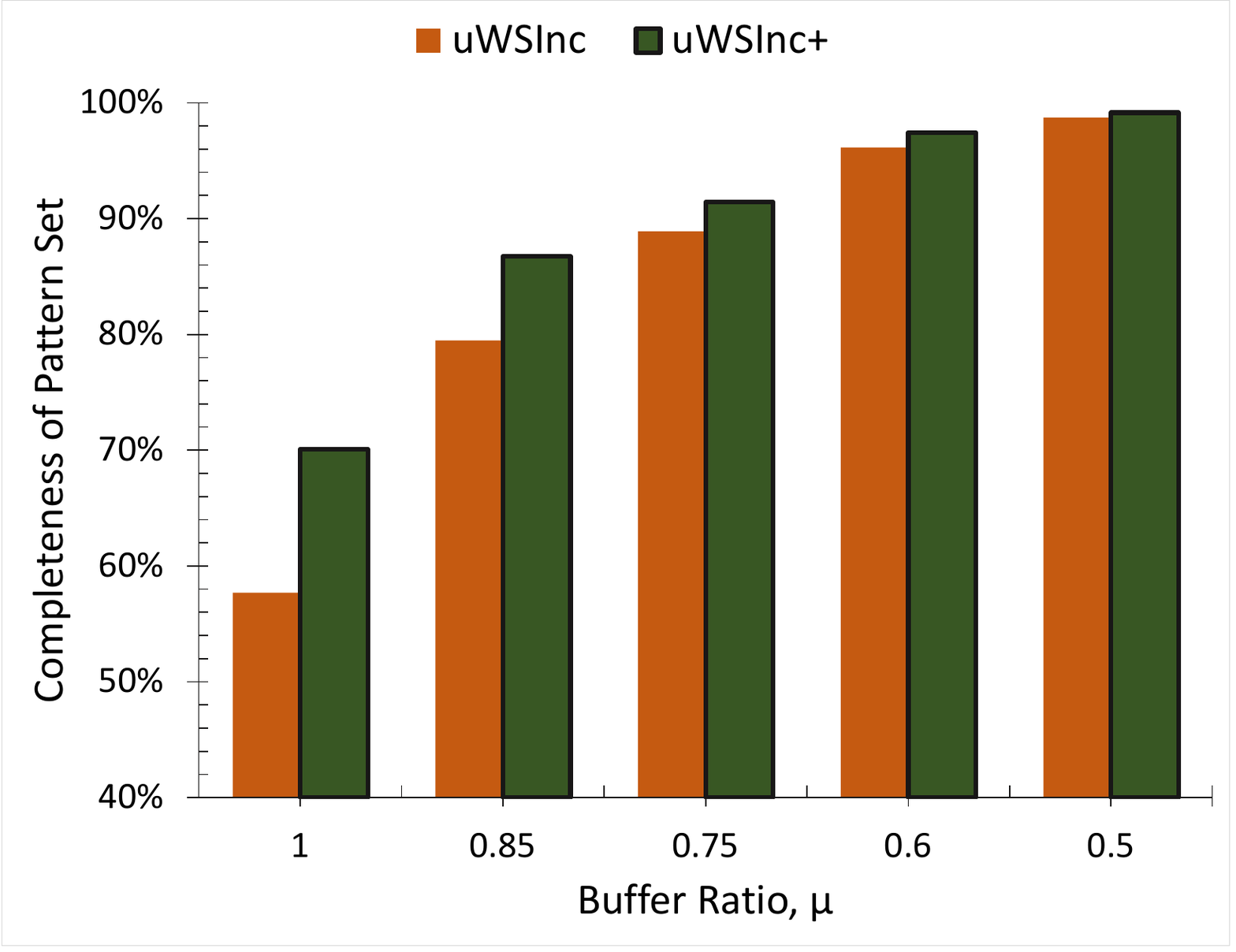}
      \caption{Completeness in \textit{Foodmart}, min\_sup 0.2\%}
      \label{food_comp_mu}
    \end{subfigure}

\caption{Performance analysis of \textit{uWSInc} and \textit{uWSInc+}
for different buffer ratio
}
\label{fig:inc_mu}
\end{figure}

The change in runtime between \textit{uWSInc} and \textit{uWSInc+} in \textit{Leviathan} is shown Figure \ref{fig:lev_rt_mu}. This figure validates our claim as stated above.
As we can see, when we use no buffer, i.e., $buffer\ ratio = 1.0$, both \textit{uWSInc} and \textit{uWSInc+} consume a very short amount of time compared To the baseline. 
However, there is a slight difference between their runtimes. \textit{uWSInc+} takes slightly more time because it has to run \textit{FUWS} locally in each increment that occurs and maintains promising frequent sequences. \textit{uWSInc} does not need this kind of step at all. It is easily observed that the runtime of both \textit{uWSInc} and \textit{uWSInc+} increases with the decrease in buffer ratio. At the same time, the decreased buffer ratio helps to achieve more completeness by maintaining extra sequences.
which can be observed in Figure \ref{fig:lev_comp_mu}.
 For datasets like \textit{Leviathan and Kosarak}, an average choice of buffer ratio = 0.85 gives a satisfactory result.

From Figure \ref{retail_rt_mu} and \ref{retail_comp_mu}, we can point out an interesting case. \textit{uWSInc} achieves around 91\% completeness using buffer ratio = 0.5 and it takes around 168 seconds. Whereas, a result of the same completeness can be achieved by \textit{uWSInc+} in around 127 seconds by using buffer ratio  = 0.85. 
\textit{Foodmart} is also a market basket dataset like \textit{Retail}. However, in this dataset, both the initial part and the increments are small in size.
Runtime and completeness analysis in \textit{Foodmart} for different choice of buffer ratio is shown in Figures \ref{food_rt_mu} and \ref{food_comp_mu} respectively.
In every case, both runtime and completeness for \textit{uWSInc+} is larger than that of \textit{uWSInc}.
Based on the results here, 
we can conclude that a lower value of buffer ratio gives better completeness, but at the same time, it also increases the amount of time required to generate the result. 

\begin{figure}[!thb]
    \centering 
    \begin{subfigure}{0.45\linewidth}
    \includegraphics[width=\textwidth,height=.6\textwidth]{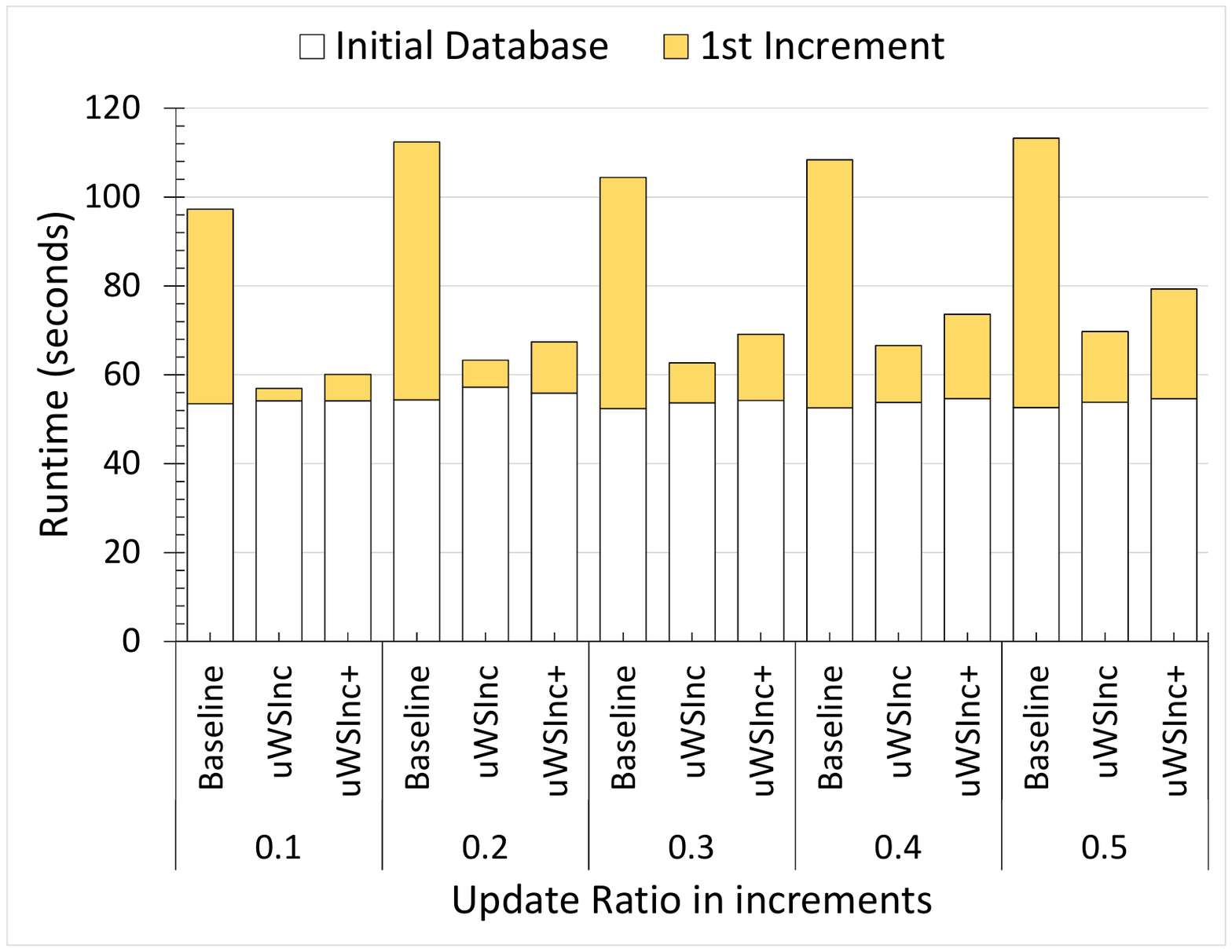}
    \caption{Upto first increment}
    \label{update_ratio_chain_one}
    \end{subfigure}
    \hfil 
    \begin{subfigure}{0.45\textwidth}
      \includegraphics[width=\linewidth, height = .6\linewidth]{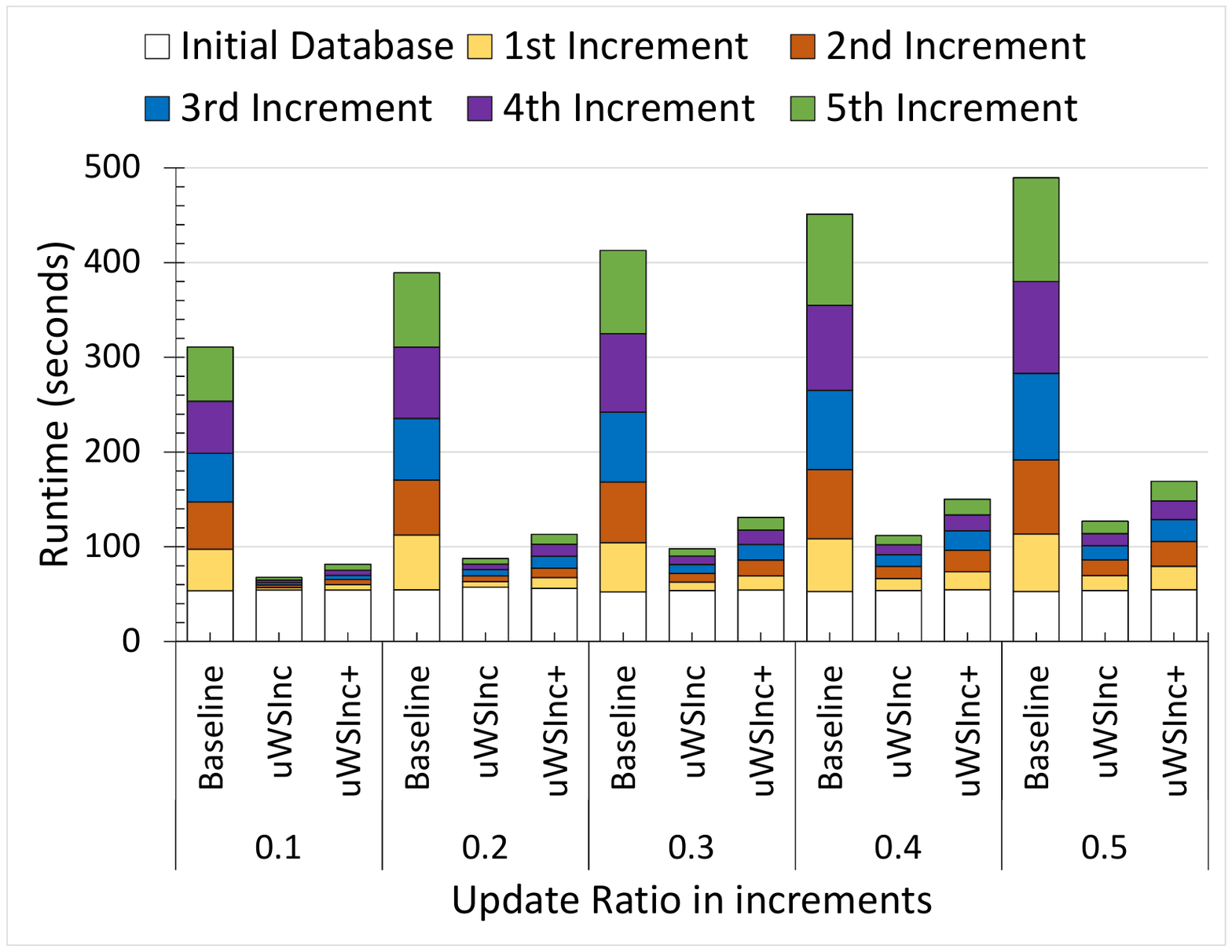}
      \caption{Upto five increments}
      \label{update_ratio_chain_five}
    \end{subfigure}

\caption{Performance of incremental solution for different increment size 
in \textit{Chainstore} dataset
with \textit{min\_sup} 0.01\%
}
\label{ur_chain}
\end{figure}

\begin{figure}[!thb]
    \centering 
    \begin{subfigure}{0.45\linewidth}
    \includegraphics[width=\textwidth,height=.6\textwidth]{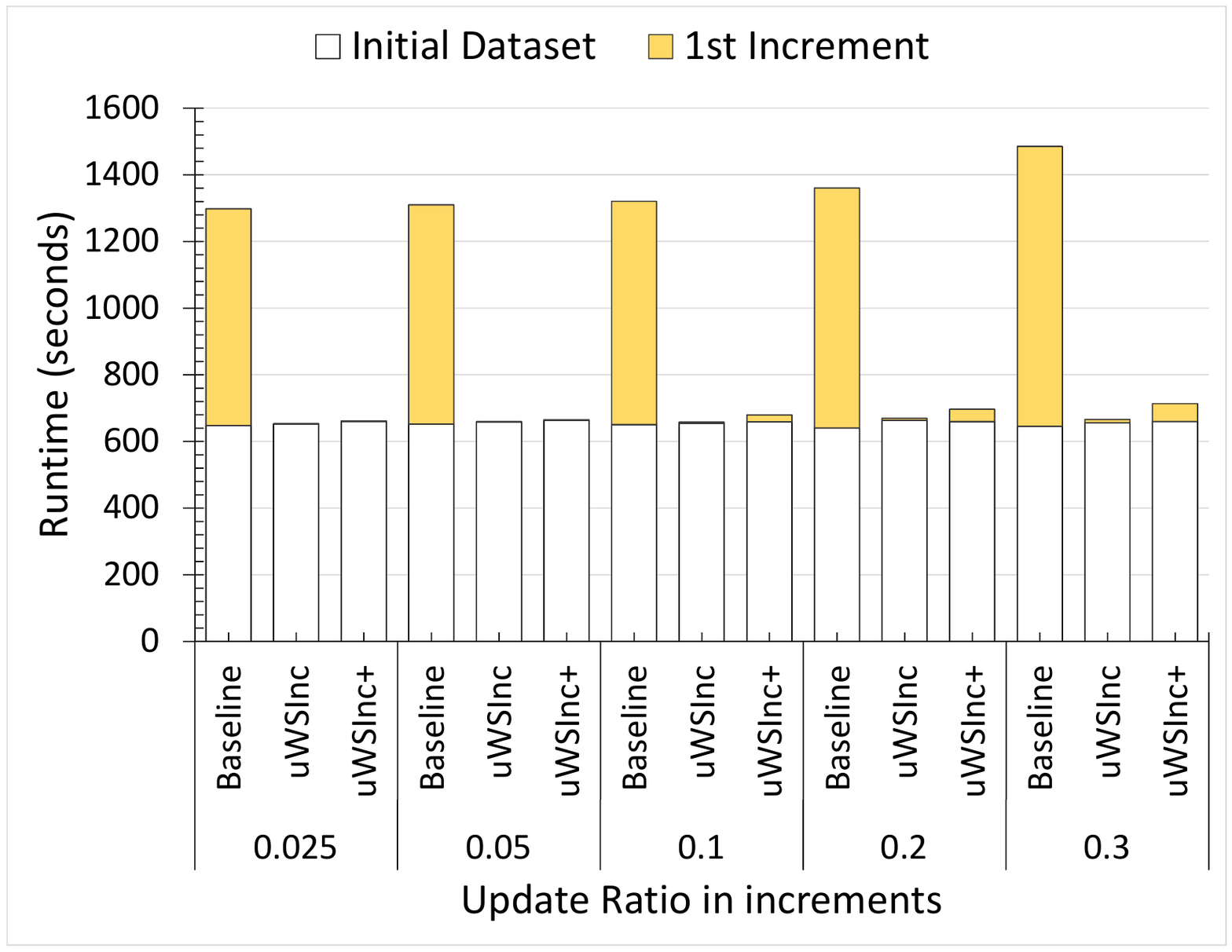}
    \caption{Upto first increment}
    \label{update_ratio_kos_one}
    \end{subfigure}
    \hfil 
    \begin{subfigure}{0.45\textwidth}
      \includegraphics[width=\linewidth, height = .6\linewidth]{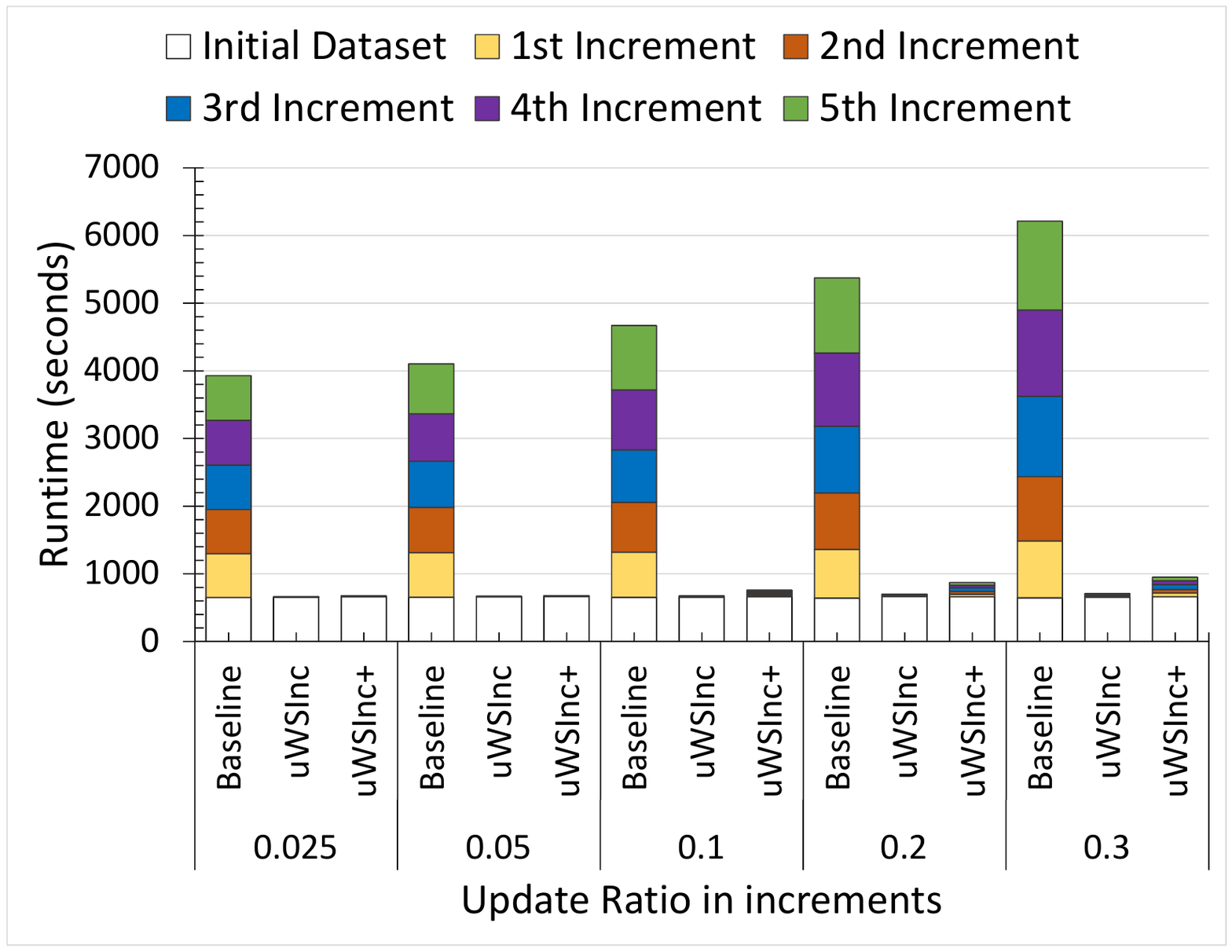}
      \caption{Upto five increments}
      \label{update_ratio_kos_five}
    \end{subfigure}

\caption{Performance of incremental solution for different increment size 
in \textit{Kosarak} dataset
with \textit{min\_sup} 0.03\%
}
\label{ur_kos}
\end{figure}

\subsubsection{Increment Size.}
\label{subsub:inc_inc_size}
To evalulate the effect of increment size, we have run our algorithms several times with different \textit{update ratios} ranging from 0.1 to 0.5 in the \textit{Chainstore} dataset and from 0.025 to 0.3 in the \textit{Kosarak} dataset. Here, the size of the initial dataset is 20,000 sequences. Hence, ${update\ ratio = 0.2}$ means an increment of $20000\times0.2=4000$ new sequences.
Results in Figures~\ref{ur_chain} and \ref{ur_kos} show that both of our incremental algorithms are very efficient not only for smaller increments but also for larger increments. 
However, the choice between \textit{uWSInc} and \textit{uWSInc+} can be made considering the trade-off between \textit{runtime} and \textit{completeness} as discussed in previous sections.
We highlight that the efficiency of our incremental mining algorithms is not limited by the update ratio or the total number of increments. Hence, they are highly scalable.

\begin{figure}[!bht]
    \centering 
    \begin{subfigure}{0.45\linewidth}
    \includegraphics[width=\textwidth,height=.56\textwidth]{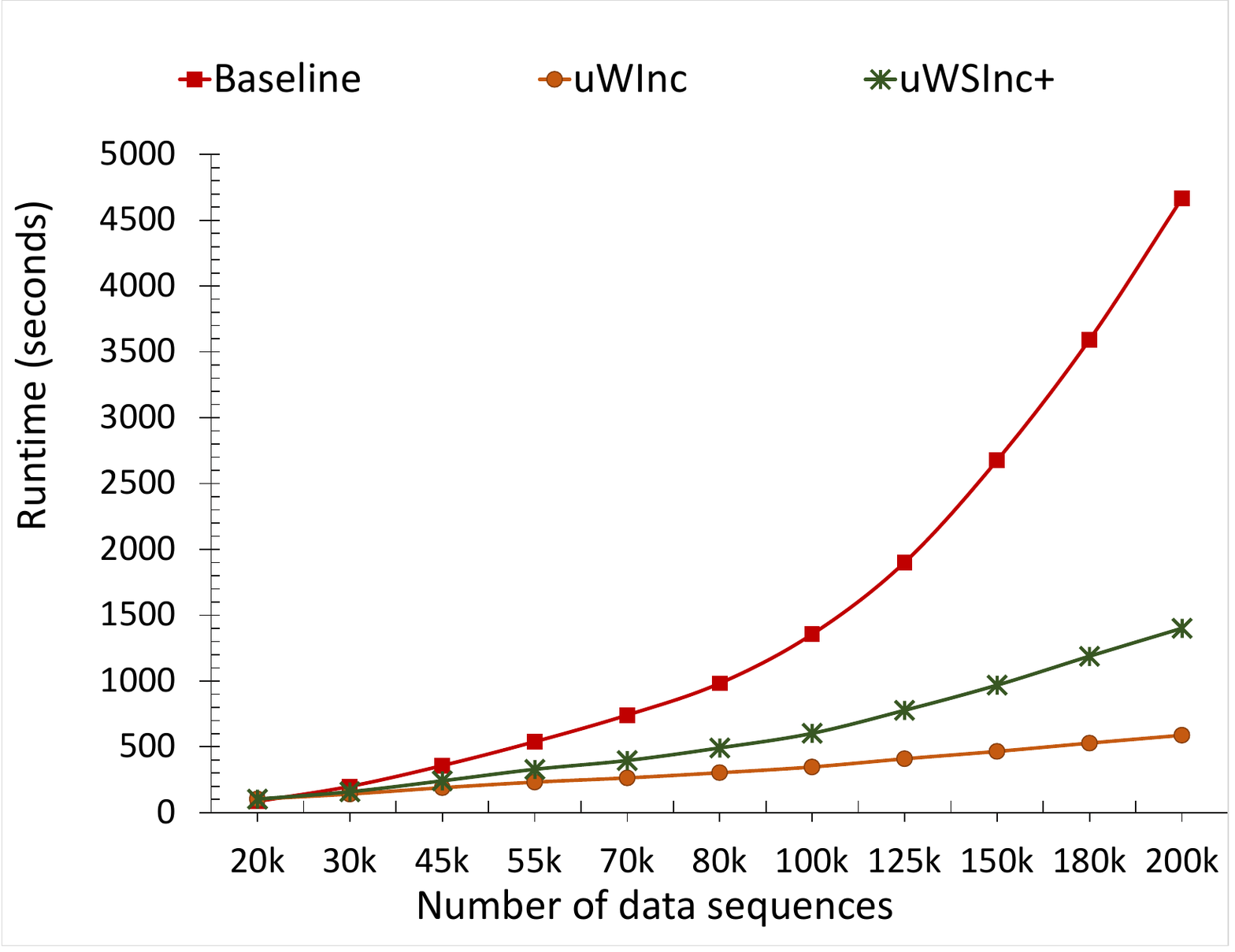}
    \caption{\textit{Chainstore} dataset with \textit{min\_sup} 0.05\%}
    \label{fig:scale_db_chain}
    \end{subfigure}
    \hfil 
    \begin{subfigure}{0.45\textwidth}
      \includegraphics[width=\linewidth, height = .56\linewidth]{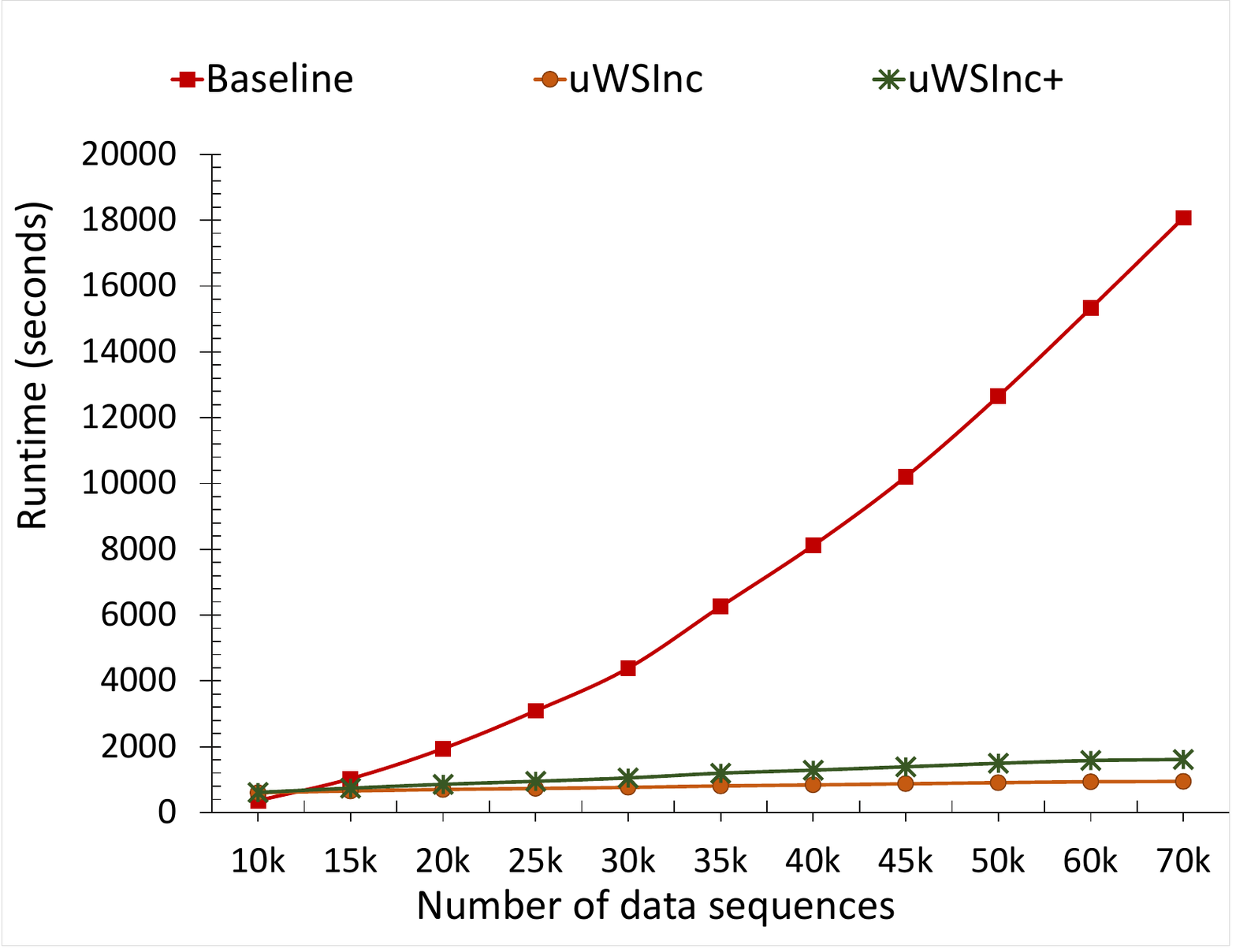}
      \caption{\textit{Kosarak} dataset with \textit{min\_sup} 0.1\%}
      \label{fig:scale_db_kos}
    \end{subfigure}

\caption{Performance of incremental solution for increasing database size 
}
\label{fig:scale_db}
\end{figure}

\subsubsection{Dataset Size}
\label{subsub:inc_dataset_size}
To test scalability against the dataset size, we have run our proposed algorithms and the baseline approach in several large datasets such as \textit{Chainstore} and \textit{Kosarak}.
 We have considered the first 10 thousand transactions as the initial dataset and then introduced several increments of varying sizes to use the full-length dataset.  

Figure \ref{fig:scale_db_chain} shows the performance analysis for the \textit{Chainstore} dataset with \textit{min\_sup} 0.05\%. 
Figure \ref{fig:scale_db_kos}  shows the scalability performance for the \textit{Kosarak} dataset with \textit{min\_sup} 0.1\%.
Both \textit{uWSInc} and \textit{uWSInc+} take an equal amount of time in the initial phase, and it is slightly higher than the baseline approach. The reason is that our proposed algorithms find and store additional patterns in the initial phase compared to the baseline approach for the same support threshold. Later on, for each increment, the baseline approach runs \textit{FUWS} from scratch in the updated database that leads to consuming huge time and memory.  In contrast, our proposed algorithms consume less time.
Thus, both our algorithms outperform the baseline approach in large datasets and are efficient in handling multiple updates. 

\begin{figure}[tbh]
    \centering 
    \begin{subfigure}{0.45\linewidth}
    \includegraphics[width=\textwidth,height=.6\textwidth]{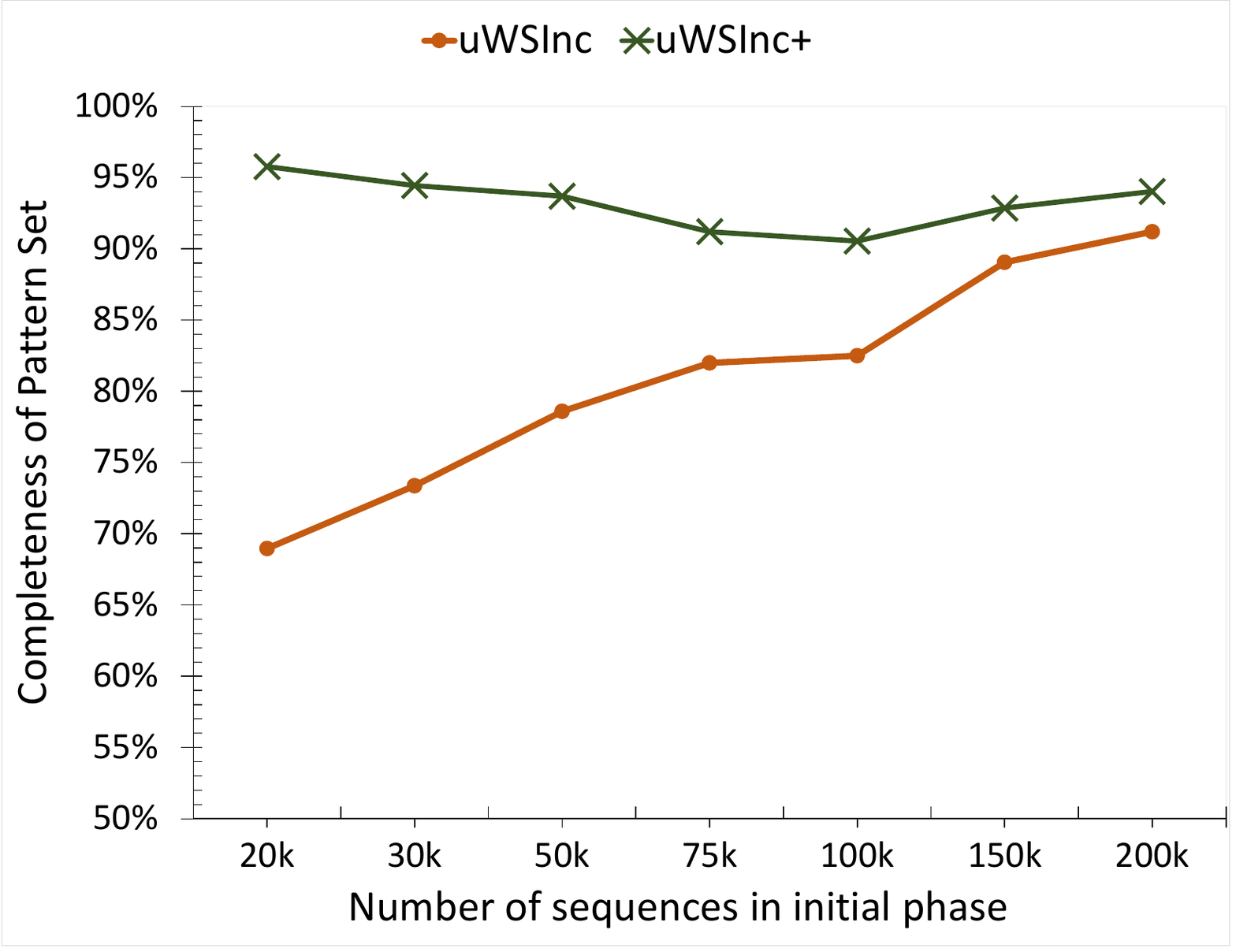}
    \caption{\textit{Chainstore} dataset with \textit{min\_sup} 0.05\%}
    \label{fig:chainstore init size}
    \end{subfigure}
    \hfil 
    \begin{subfigure}{0.45\textwidth}
      \includegraphics[width=\linewidth, height = .6\linewidth]{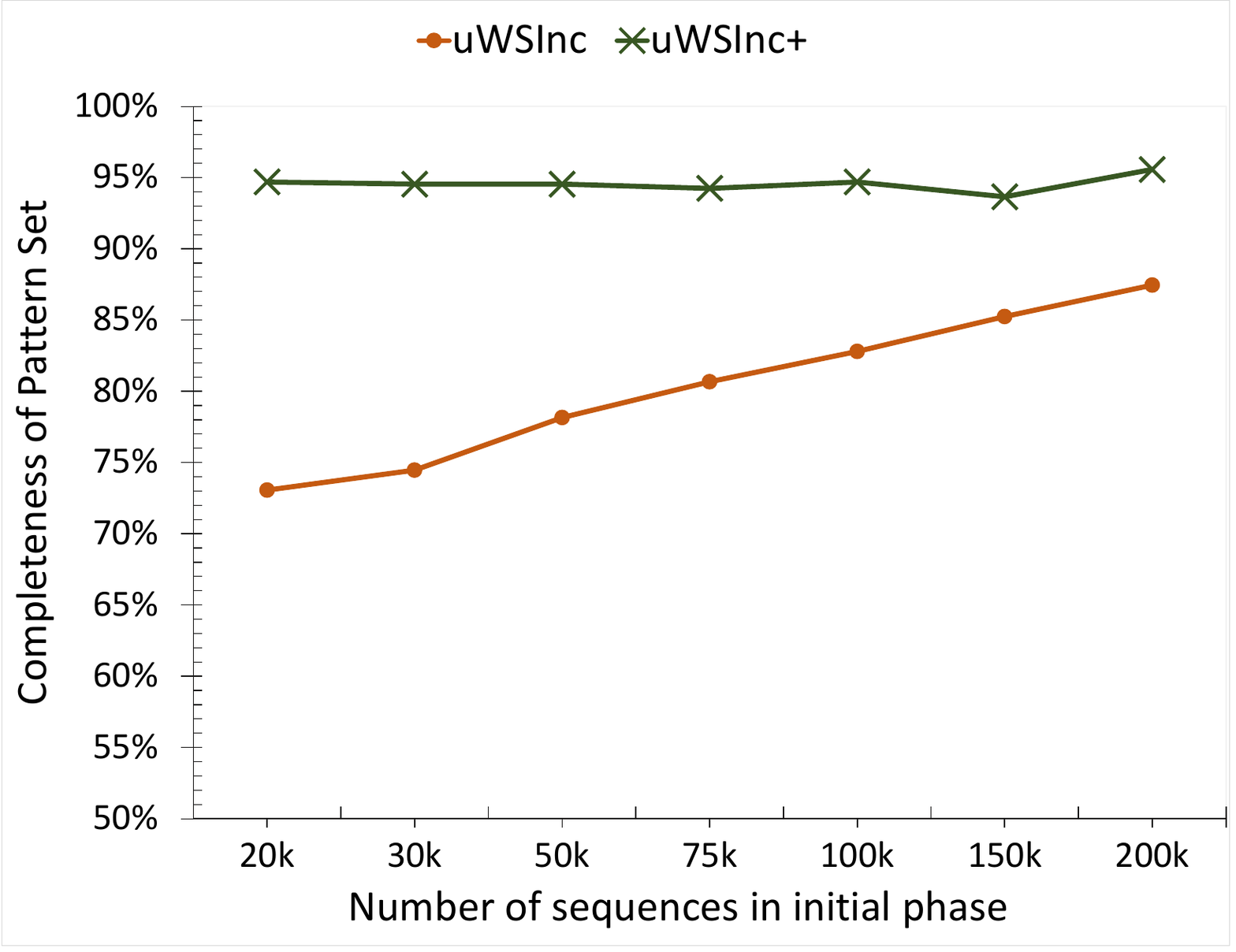}
      \caption{\textit{OnlineRetail} dataset with \textit{min\_sup} 0.05\%}
      \label{fig:online init size}
    \end{subfigure}

\caption{Performance of incremental solution for different dataset size of initial phase 
}
\label{fig:vary_init}
\end{figure}

\subsubsection{Completeness Analysis with Different Initial Size of Database}
\label{subsub:inc_initial_size}
The completeness of our proposed algorithms varies with the initial size of the datasets. 
The larger the initial size, the more patterns they can find and maintain to update in future increments. 
The smaller the initial size, the more initially infrequent sequences are found as new frequent patterns after successive increments. 
We have considered different sizes of the initial dataset, such as the first 10,000, first 20,000, first 30,000, and so on. 
After that, we have introduced a sufficient number of increments to cover the entire dataset. 
The size of increments was chosen randomly each time from the range of \textit{update ratio  0.4-0.8} to reflect the real-life use cases. 
Results for the \textit{Chaninstore} and \textit{OnlineRetail} datasets
are shown in Figure \ref{fig:vary_init}.
It can be seen that \textit{uWSInc} has a positive trend, which indicates greater completeness, with the larger initial dataset.
 Because \textit{uWSInc} updates the result after each increment based on only the semi-frequent patterns stored by mining in the initial phase, a larger initial dataset helps to find and store more potential patterns.

On the other hand, \textit{uWSInc+} is less dependent on the initial size as it also mines locally in the incremented portions, which helps to find new patterns. It also keeps track of patterns that have become infrequent recently and uses them as \textit{promising frequent sequences} in the next increment. Thus, the trend for completeness of \textit{uWSInc+} is somewhat neutral with respect to the initial dataset size, which means that the completeness of \textit{uWSInc+} is less affected by the size of initial datasets.
The completeness of incremental approaches also depends on the distribution of items among the increments.
However, for any initial dataset size, completeness values are always higher for \textit{uWSInc+} than for \textit{uWSInc}.\\

 Our extensive experimental analysis demonstrates that our proposed \textit{FUWS} outperforms the existing best solution \textit{uWSequence}~\cite{rahman2019mining_uWSeq} to mine frequent sequences in uncertain databases. The results also validate the efficiency of \textit{uWSInc} and \textit{uWSInc+}; that they can find the almost complete set of frequent patterns within a very short amount of time after each increment is introduced. Finally, one thing to highlight is that though \textit{uWSInc+} is better than \textit{uWSInc} in terms of completeness, \textit{uWSInc} is faster than \textit{uWSInc+}.

\section{Conclusions} 
\label{sec:concl}

In this work, we have developed an algorithm, \textit{FUWS}, to mine weighted sequential patterns in uncertain databases and proposed two new incremental mining approaches, \textit{uWSInc} and \textit{uWSInc+}, to mine weighted sequential patterns efficiently from incremental uncertain databases. 
The \textit{FUWS} algorithm applies $wExpSup^{cap}$ as an upper bound of weighted expected support to find all potential frequent sequences and then prunes false-positive sequences. We have used a hierarchical index structure named \textit{USeq-Trie} to maintain patterns and $SupCalc$ to calculate their support in a faster way. By using \textit{FUWS} as a tool and buffering semi-frequent sequences, the \textit{uWSInc} algorithm works efficiently in mining frequent sequences from incremental uncertain databases which have a uniform distribution of items. In the case of those datasets, the \textit{uWSInc} algorithm is very efficient because most of the frequent sequences are either found in the initial dataset or will come from semi-frequent sequences, and the appearance of new items after increments are sporadic. On the other hand, due to seasonal behavior, concept drifts, or different characteristics of datasets, new patterns can be largely introduced in some real-life datasets. In those cases, the \textit{uWSInc+} algorithm maintains promising sequences after each increment, additionally along with semi-frequent sequences to find new patterns effectively. 
We have tested them in many real-life and popular datasets by varying different parameters to prove their efficiency. 
These results show that our proposed techniques could be an excellent tool for many real-life applications that use uncertain sequential data, such as medical reports, sensor data, image processing data, social network data, privacy-preserving data, and so on.
In the future, this work can be extended to mine weighted sequential patterns in uncertain data streams. Furthermore, 
incremental mining of maximal and closed sequential patterns can be interesting for further research. 

\section*{Acknowledgements} 
We would like to express our deep gratitude to the anonymous reviewers of this paper. Their insightful comments have played a significant role in improving the quality of this work. This work is partially supported by NSERC (Canada) and the University of Manitoba.

\bibliographystyle{elsarticle-harv}
\bibliography{mybibfile}

\end{document}